# State and Memory is All You Need for Robust and Reliable AI Agents

Matthew Muhoberac[1], Atharva Parikh[2], Nirvi Vakharia[3], Saniya Virani[1], Aco Radujevic[4], Savannah Wood[4], Meghav Verma[4], Dimitri Metaxotos[4], Jeyaraman Soundararajan[4], Thierry Masquelin[4], Alexander G. Godfrey[4], Sean Gardner[4], Dobrila Rudnicki[4], Sam Michael[4], Gaurav Chopra[1,3,5]

[1]Department of Chemistry, Purdue University, 560 Oval Drive, West Lafayette, Indiana 47907, United States
[2]Department of Statistics, Purdue University, West Lafayette, Indiana 47907, United States
[3]Department of Computer Science, Purdue University, West Lafayette, Indiana 47907, United States
[4]National Center for Advancing Translational Sciences, National Institute of Health, 9800 Medical Center Dr, Rockville, MD 20850, United States
[5]Purdue Institute for Drug Discovery, Integrative Data Science Institute, Purdue Center for Cancer Research, Purdue Institute for Inflammation, Immunology, and Infectious Disease, Purdue Institute for Integrative Neuroscience

**Corresponding Author: gchopra@purdue.edu**

## Abstract

Large language models (LLMs) have enabled powerful advances in natural language understanding and generation. Yet their application to complex, real-world scientific workflows remain limited by challenges in memory, planning, and tool integration. Here, we introduce SciBORG (Scientific Bespoke Artificial Intelligence Agents Optimized for Research Goals), a modular agentic framework that allows LLM-based agents to autonomously plan, reason, and achieve robust and reliable domain-specific task execution. Agents are constructed dynamically from source code documentation and augmented with finite-state automata (FSA) memory, enabling persistent state tracking and context-aware decision-making. This approach eliminates the need for manual prompt engineering and allows for robust, scalable deployment across diverse applications via maintaining context across extended workflows and to recover from tool or execution failures. We validate SciBORG through integration with both physical and virtual hardware, such as microwave synthesizers for executing user-specified reactions, with context-aware decision making and demonstrate its use in autonomous multi-step bioassay retrieval from the PubChem database utilizing multi-step planning, reasoning, agent-to-agent communication and coordination for execution of exploratory tasks. Systematic benchmarking shows that SciBORG agents achieve reliable execution, adaptive planning, and interpretable state transitions. Our results show that memory and state awareness are critical enablers of agentic planning and reliability, offering a generalizable foundation for deploying AI agents in complex environments.

## Introduction

The ability of large language models (LLMs) to generate coherent, contextually rich language has transformed natural language processing and its applications across science, engineering, and society. These models, such as the GPT series[1,2], LLaMa[3,4], Gemini[5] and PaLM[6], excel in tasks ranging from summarization and translation to question-answering and scientific text generation. Despite their widespread success, LLMs are inherently static, in that, they are constrained by fixed training corpora, unable to dynamically update their knowledge, and prone to hallucination when queried on tasks outside their training distribution. These limitations pose significant challenges when LLMs are applied to domains requiring long-term memory, integration with external tools, or coordination across multiple agents and workflows.

In scientific settings, researchers frequently operate within complex, multi-step environments that require structured reasoning, access to instrumentation, and context-sensitive planning. While recent approaches such as domain-specific retrieval-augmented generation[7] (RAG) (i.e. BioRAG[8]) and fine-tuned domain-specific models (e.g., BioGPT[9], ChemGPT[10], BioBERT[11], ChemDFM[12], CRISPR-GPT[13], LegalBERT[14]) extend the capabilities of LLMs, they still fall short of enabling autonomous, goal-driven reasoning across physical and digital systems. Emerging work on LLM-powered agents offers promising direction: these agents combine language generation with decision-making, memory, and tool execution to autonomously perform tasks in iterative and dynamic workflows. However, current agentic systems often rely on rigid prompt engineering[15], lack modularity, or operate without persistent memory, thereby hindering their effectiveness in real-world scientific environments.

Here, we present Scientific Bespoke Artificial Intelligence Agents Optimized for Research Goals (SciBORG), a modular agentic framework designed to address these limitations. SciBORG constructs state-aware agents that plan, reason, and execute complex tasks by integrating memory architectures, tool-based control, and retrieval from structured knowledge sources (**Fig. 1a–e**). Agents are instantiated directly from documentation and instrument interfaces and operate using a finite-state memory model that tracks workflow progress over time. We demonstrate SciBORG across diverse use cases, including instrument control, data mining, prompt construction, and inter-agent collaboration. Through real-world deployment and systematic benchmarking, we show that SciBORG agents achieve robust, adaptive performance and offer a scalable foundation for LLM-driven scientific discovery.

## Agentic Infrastructure for Modular Planning and Execution with Memory Architectures for Agent Decision-Making

SciBORG AI agents (**Fig. 1a–e**) were developed using Python (v3.10.x) and the LangChain software development kit (SDK)[16] (v0.1.x), enabling structured LLM-based planning and tool interaction. The core agent infrastructure was constructed using modular components—parameters, commands, workflows, microservices, and command libraries—each implemented as BaseModel classes in Pydantic[17] (v2.6.x) to ensure robust type validation and JSON serializability. Agents were powered by large language models (LLMs) via LangChain-compatible chat-completion endpoints, including OpenAI's GPT-3.5-turbo, GPT-4, and internally hosted Azure OpenAI instances. These models served as the central reasoning engines for agentic planning, tool invocation, and prompt augmentation. To create operational agents, source code modules were parsed using custom LLM chains, which extracted function metadata and structured this information into callable tool definitions. These tools included schema-defined input parameters, operational descriptions, preconditions, postconditions, and expected outputs (**Fig. 1f**). High-level user goals were translated into actionable workflows using planning chains, and executable workflows were interpreted via command interpreters that interfaced directly with hardware or software APIs (**Fig. 2a**). Agent behavior was driven by ReAct-style[18] iterative reasoning, in which LLMs followed a structured sequence of thoughts, actions, and observations (**Fig. 1g**). To support long-horizon and stateful tasks, agents were equipped with one or more memory buffers, including: (i) **chat memory**, which retained conversational history with the user; (ii) **action summary memory**, which tracked tool usage summaries across interactions; and (iii) **pseudo–finite state automaton (FSA) memory**, which maintained a compressed, schema-driven representation of system state based on valid transitions between discrete states. Pseudo-FSA memory was particularly useful for long workflows involving instrumentation or multi-agent interaction, providing a scalable way to maintain operational context while avoiding prompt overflow. All memory buffers were dynamically updated after each agent execution and served as input to subsequent agent runs.

**Agent-mediated Prompt Construction for Adaptive Reasoning and Seamless Integration**

Prompt engineering includes manual tailoring of LLM inputs to direct generative behavior and has traditionally required human expertise, significant iteration, and fixed assumptions about task structure. While effective for narrow objectives, such static prompts lack generalizability across tools, domains, and evolving contexts. Automated approaches such as retrieval-augmented generation[7] (RAG) and LLM-driven prompt assembly offer improved relevance but are inherently constrained by the static nature of their retrieval sources or model priors. Moreover, these methods often require retraining or manual reconfiguration when tool interfaces or execution environments change. To overcome these limitations, we introduce a dynamic agentic prompt construction mechanism wherein LLM-powered agents autonomously generate and refine their own prompts at runtime (**Fig. 1h**). Rather than relying on static inputs, agents within the SciBORG framework iteratively construct context-aware prompts by incorporating operational memory, retrieved documentation, intermediate observations, and evolving tool states (**Fig. 2c**). These prompts are assembled in response to the agent's current reasoning process, tool feedback, and task requirements, allowing adaptive behavior as the problem space evolves. Central to this approach is the integration of structured memory, particularly a pseudo–finite state automaton (FSA) schema, which encodes agent and system state transitions (**Fig. 1i**). This allows agents to condition prompt construction not only on prior actions but also on anticipated state changes. Combined with a modular, service-agnostic base prompt (**Extended Data Fig. 1**) and schema-validated command hierarchies, this architecture enables agents to recover from execution failures, adapt to new tools, and operate across diverse computational and physical environments. By shifting prompt design from a static human-led process to a dynamic agentic operation, our framework facilitates robust, scalable, and interpretable AI agent design. This paradigm supports fine-grained control, system extensibility, and high-fidelity task execution across a range of scientific domains.

**Document-Embedded Retrieval Enhances Agent Domain Expertise for Context-Aware Decision Making**

Retrieval-augmented generation[7] (RAG) was implemented in SCIBORG to augment agent decision-making with context from unstructured documents. FAISS[19] (v1.8.0) vector databases indexed OpenAI-generated text embeddings from electronic lab notebooks (ELNs), protocols, and instructional manuals provided by the National Center for Advancing Translational Sciences (NCATS). Embedded documents were chunked semantically and indexed, enabling real-time retrieval of relevant content during agent operation. Agents interfaced with the embedding store via RAG tools, invoking queries in natural language. Retrieved content was summarized in real time and used to inform planning, fill knowledge gaps, or respond to user queries (**Fig. 2b**). This infrastructure supported both procedural decision-making (e.g., reaction setup) and educational use cases (e.g., technician training).

To extend domain-specific reasoning, embeddings were generated for two document sets: one consisting solely of ELN entries, and another incorporating the full protocol archive used in microwave synthesis training. In all cases, these embeddings functioned as opaque, interchangeable modules from the perspective of the agent thereby facilitating modularity in knowledge integration without necessitating changes in agent structure. In one representative use case, an agent was queried for, "recommended parameters for an N-alkylation reaction using 3-Bromoquinoline and morpholine", a specific reaction recorded in the ELN. Upon invocation, the embedding chain retrieved document chunks containing reaction conditions, reactants, solvents, and procedural details. The summarized output accurately specified a pressure of 1 bar, temperature of 120 °C, reactants including 3-Bromoquinoline, morpholine, and potassium 2-methylpropan-2-olate, solvents (dioxane, toluene), a PEPPSI catalyst (5 wt%), and a reaction time series (5–60 minutes) (**Supplementary Fig. 2**). These parameters were seamlessly incorporated into subsequent automated synthesis planning steps by the agent. Beyond experimental automation, we tested the same RAG-integrated agents on technician-facing training queries (**Supplementary Fig. 13**). When asked to explain the operational differences between microwave irradiation and conventional heating, the agent correctly retrieved excerpts from NCATS instructional documents, highlighting benefits such as uniform heating and reduced reaction times (**Extended Data Fig. 3**). Similarly, when queried on the function of the Initiator+ pressure setting, the agent identified documentation explaining that the setting allows for higher-temperature operation of low-boiling-point solvents via pressurization (**Supplementary Fig. 3**). These examples demonstrate how embedded knowledge retrieval enables agents to respond accurately to diverse queries, supporting both procedural execution and contextual education.

## Hardware Integration and Virtual Instrumentation

To evaluate real-world compatibility, agents were integrated with a Biotage Initiator+ microwave synthesizer using Python-based wrappers developed by NCATS. The wrappers exposed operational endpoints for session control, vial loading, lid manipulation, and thermal programming. A virtual clone of the instrument was also implemented in Python, simulating both current capabilities and future expansions (**Fig. 3a**). These virtual clones enabled agents to be tested and benchmarked without requiring physical hardware access, supporting safe and reproducible agent development. User-specified synthesis requests were processed using planning chains, which decomposed high-level goals into structured, executable command sequences. These workflows were then parsed and executed by an interpreter interfacing directly with the instrument (or its virtual twin). Reaction outcomes, such as conversion rates, were collected via simulated sensors and compared to human-executed controls.

**Automated Microwave Synthesis via Hierarchical Agentic Planning**

To evaluate hardware compatibility and real-world deployment of SciBORG, we integrated the framework with both a physical and virtual Biotage Initiator+ microwave synthesizer. These instruments automate synthetic chemistry reactions through sealed-vessel heating, stirring, and pressurization. The virtual clone, developed in Python, mirrors the operational endpoints of the physical hardware, enabling agents to prototype workflows *in silico* prior to deployment. This establishes a robust "two-pass" validation system for autonomous experimental control. In a representative demonstration, a human chemist issued a task to an AI agent to execute an N-alkylation reaction under benchmarking conditions: 1800 seconds (30 min) heating duration, a temperature of 100 °C, and a stir rate of 600 rpm. The agent was instructed not to release the session after heating and was provided with a username and password (**Fig. 3c**). This request was parsed by the agent's high-level planning chain, which generated a multi-step execution plan including session allocation, lid closure, and heating under the specified conditions (**Supplementary Fig. 4**). Notably, the user did not explicitly request session allocation or lid closure—these steps were inferred by the agent based on command prerequisites encoded in the infrastructure. The agent correctly reasoned that a session ID—absent in the initial request—would be generated by the session allocation command and subsequently required as input for the lid closure and heating commands. Moreover, although session release is typically called post-heating, the agent recognized the user's preference and omitted the post-requisite step. Internal validation routines ensured that any LLM-generated instructions conformed to expected schemas prior to execution. Following high-level planning, the proposed workflow was converted into a structured low-level plan in JSON format (**Supplementary Fig. 5**). This plan was passed to a command interpreter, which sequentially executed the session allocation, lid closure, and heating commands. The reaction proceeded under autonomous control, and after completion, the vial contents were analyzed by Ultra Performance Liquid Chromatography (UPLC) to determine conversion efficiency. The agent-operated reaction achieved 58.9% conversion, compared to 57.3% for the same procedure run manually by a chemist, thereby demonstrating equivalence in outcome and validating the agent's ability to autonomously interpret, plan, and execute multistep laboratory protocols.

**Finite-State Memory Improves Context Retention and Agentic Robustness in Instrument Workflows**

To assess the role of memory architectures in long-horizon scientific workflows, we deployed a series of large language model (LLM)-powered agents to control a virtual microwave synthesizer. This simulator mimicked operational features of the Biotage Initiator+ instrument, enabling automated execution of tasks such as vial loading, parameter setting, and heating (**Extended Data Fig. 4**). The agents were configured with varying memory modules: no memory, chat-only memory, action summary memory, combined chat and summary memory, and a hybrid chat and pseudo–finite-state automaton (FSA) memory. Agents were evaluated on a task requiring multistep execution with state dependencies, using both single-query and decomposed command sequences. In the baseline condition, a single compound instruction directed the agent to heat a vial to 100 °C for 50 minutes at 1 atm and recall specific user-provided context **(Supplementary Fig. 6)**. Agents without memory performed well on the baseline prompt but failed when the task was decomposed into four disjoint instructions. These agents repeatedly lost track of session identifiers, lid status, and heating parameters, leading to redundant actions and operational errors **(Supplementary Fig. 7)**. The addition of chat memory allowed agents to retain user interactions but failed to preserve system-level state, resulting in inconsistent tool usage **(Supplementary Fig. 8)**. Action summary memory improved tool tracking but often accumulated irrelevant or outdated content, reducing effective context resolution over time **(Supplementary Fig. 9)**. Agents equipped with both chat and summary memory showed moderate improvements, particularly in preserving user-specific variables such as conversational entities **(Supplementary Fig. 10)**. The most robust performance was observed in agents with pseudo-FSA memory, which used a predefined state schema and transition rules to represent and update instrument status **(Supplementary Figs 11 and 12; Extended Data Figs 5 and 6)**. This memory class retained a minimal, task-relevant state representation across multiple interactions, enabling agents to correctly infer dependencies, avoid redundant actions, and adapt to new instructions **(Fig. 3b,d)**. For example, the FSA-equipped agents consistently remembered whether the lid was open, or a session had been initialized, enabling seamless execution of follow-up tasks. Quantitatively, agents with FSA memory achieved the highest success rate across all benchmark scenarios. Their memory buffers remained compact (mean size: 197 characters) compared to action summary logs (mean: 756 characters), preserving valuable prompt space and reducing token overflow. In contrast to summary memory, FSA memory filtered out extraneous details such as secret phrases or non-instrumental inputs, improving signal-to-noise ratio in LLM prompting. These results demonstrate that schema-driven, state-aware memory architectures substantially enhance the robustness of agentic systems operating in scientific workflows. This approach is particularly advantageous in instrument-rich environments or distributed pipelines where state tracking is essential but not natively available from drivers or sensors.

### PubChem Agent and Cross-Agent Communication

To enable scientific data mining, a dedicated PubChem agent was constructed using a suite of RESTful API endpoints from the PubChem database[20,21]. Supported operations included compound identification, assay filtering, citation retrieval, and metadata extraction. An external tool was also integrated from the National Cancer Institute's (NCI) Cactus API[22] to support name-to-SMILES[23] conversion for molecular structures. Cross-agent communication was implemented by exposing one agent's interface as a callable tool to another agent. For example, the synthesis agent invoked the PubChem agent to look up molecular weights or retrieve SMILES strings. The delegated agent processed queries autonomously and returned structured responses. This configuration supports a modular, decentralized architecture in which task-specific agents collaborate to fulfill broader scientific goals while maintaining domain boundaries and reducing hallucination risk.

### Multi-step planning, reasoning and execution in chemical information retrieval

To evaluate the ability of agents to autonomously reason across web-based databases, we integrated PubChem REST API[20] endpoints and the NCI Chemical Identifier Resolver[22] (CACTUS) API into SciBORG agents. These endpoints provide access to curated chemical structures, bioactivity data, assay metadata, and identifiers, enabling execution of multi-step information retrieval[24] workflows. In an initial benchmarking experiment, we tested whether agents could plan PubChem data mining tasks using planning chains without performing any API calls. Given the input query, "What is the IC50 of 1-[(2S)-2-(dimethylamino)-3-(4-hydroxyphenyl)propyl]-3-[(2S)-1-thiophen-3-ylpropan-2-yl]urea to the Mu opioid receptor, cite a specific assay in your response," the agent generated a plan involving sequential execution of five API tools: compound name-to-CID conversion, filtered bioassay retrieval, assay filtering by receptor, and extraction of assay name and description **(Extended Data Fig. 2; Supplementary Fig. 1).** Upon real execution of a similar query using the agent, the plan was dynamically adjusted: the assay name retrieval step was skipped because the information was already returned by the bioassay endpoint, highlighting flexible adaptation to API output. To test multi-step execution and reasoning, we queried the agent with "What is the Ki of PZM21 to the Mu opioid receptor, cite a specific assay in your response." The agent successfully identified multiple compound IDs for PZM21, retrieved bioassay results filtered by Ki values, selected the most relevant assay based on target gene annotations, and extracted the assay description and citation (**Fig. 4**). The correct inhibitory constant (0.0011 μM) was returned alongside the PubChem assay ID, title, and a DOI-based reference **(Extended Data Fig. 7; Supplementary Fig. 14).**

To test integration with embedded experimental protocols, we equipped the agent with access to an electronic lab notebook (ELN) via a RAG tool. When prompted to identify reactants and products in a specific ELN, the agent invoked the RAG tool, which retrieved and summarized chemical entities from embedded documentation. Upon follow-up, the agent was asked to convert these entities to SMILES using the CACTUS API. The agent initially failed by submitting all entities at once, then retried with multiple variations (e.g., combining names and identifiers), encountering repeated 500 errors. Eventually, it deduced that a valid input required one identifier at a time and resolved most entries (**Supplementary Fig. 15**). For entries not found in the database, the agent returned informative fallback suggestions.

These results demonstrate that agents using SciBORG infrastructure can reason over multi-step data mining tasks, recover from structured API errors, and resolve ambiguous entities, thereby displaying capabilities that exceed static pipeline approaches. The PubChem integration illustrates the ability of modular agents to query domain-specific knowledge resources with resilience and adaptability.

**Agent-to-agent communication enables coordination and trusted information sharing**

To evaluate inter-agent coordination within the SciBORG framework, we configured a virtual microwave synthesis agent to communicate with a PubChem API[20] agent via a delegated tool. The PubChem agent, integrated with a curated chemical knowledge base, serves as a trusted retrieval endpoint for molecular identifiers, properties, and assay data. This architecture represents a modular retrieval-augmented generation[7] (RAG) strategy, wherein factual queries are delegated to domain-specific agents to minimize hallucinations and enforce provenance. In a representative task, a user instructed the synthesis agent to heat a vial containing 3-bromoquinoline and morpholine at 100 °C for 30 minutes under atmospheric pressure, using 5 mg of each reactant. To calculate molar quantities, the agent queried the PubChem agent with the prompt: "What are the molecular weights of 3-bromoquinoline and morpholine?" The PubChem agent retrieved property data from compound-specific endpoints, returning molecular weights of 208.05 g/mol and 87.12 g/mol, respectively. The synthesis agent followed up with a second query: "How many moles are in 5 mg of each compound?" The PubChem agent applied the appropriate conversions and returned calculated values of $2.40 \times 10^{-5}$ mol for 3-bromoquinoline and $5.74 \times 10^{-5}$ mol for morpholine. This information was used by the synthesis agent to complete the reaction protocol and report the percent conversion as 51.16%, derived from virtual instrument feedback. This example illustrates that SciBORG agents can dynamically delegate complex sub-tasks to specialized agents while maintaining task-specific autonomy. The ability to chain inter-agent reasoning steps, grounded in trusted data, provides a flexible and transparent mechanism for multi-agent scientific collaboration. Full trace shown in **Supplementary Fig. 16**.

**Benchmarking Framework for Agent Robustness and Reliability**

To evaluate agent performance and reproducibility, we implemented a benchmarking suite with three validation strategies: (i) **Path-based benchmarking**: Validated whether agents followed correct action sequences to reach the goal. Paths were compared against reference workflows defined by tool invocation order and argument structure. (ii) **State-based benchmarking**: Assessed whether agents reached predefined system states starting from a known initial configuration. States were represented as JSON objects and validated against target schemas. (iii) **Output-based benchmarking (regex and schema validation)**: Verified correctness of final outputs against either regular expression patterns (for textual responses) or JSON schemas (for structured data). Each benchmark was run for 20 independent trials per agent configuration. Agent memory configurations were systematically varied, including ablation of memory buffers and state initializations. Outputs, action traces, and final states were recorded and analyzed for consistency, error frequency, and recovery behavior.

**Benchmarking AI agents reveals robustness of memory architectures and retrieval pipelines**

To assess the robustness and reproducibility of agent behavior across memory architectures and retrieval contexts, we developed a benchmarking framework for structured evaluation of planning, execution, and information retrieval tasks. Benchmark types included path-based, state-based, and regular expression (regex)-based validation across multiple agents operating in simulated scientific environments (**Fig. 5**).

We first evaluated the necessity of state augmentation using the microwave synthesis agent. Path-based benchmarks were conducted using the prompt, "Heat vial 3 to 100 °C for 50 min at 3 atm." Agents either received or omitted initial state information, which included session status, vial loading state, and heating parameters. Without initial state augmentation, agents completed the correct execution path in 13 of 20 runs (65%), frequently failing to open the lid prior to vial loading **(Extended Data Fig. 8; Supplementary Figs 17–19).** When provided with a JSON-formatted initial state buffer, agents completed the correct path in 17 of 20 runs (85%). Failures in this setting were linked to incorrect handling of NoneType session IDs, underscoring the importance of schema-aware input validation **(Supplementary Figs 20–22)**. We next compared memory architectures by benchmarking agents equipped with either action summary or pseudo–finite state automaton (FSA) memory. In both cases, agents were asked to "heat the vial" after prior tool usage had left the lid open. Successful execution required the agent to close the lid before initiating heating. Agents with pseudo-FSA memory completed this path in 18 of 20 runs (90%), while those using summary memory succeeded in only 10 of 20 runs (50%) (**Supplementary Figs 23–28**). These results demonstrate that focused, schema-based memory representations enhance the agent's ability to reason about stateful prerequisites during multi-step workflows.

To benchmark retrieval-augmented generation[7] (RAG) workflows, we used regex-based evaluation of agent responses generated from embedded documentation. Agents accessing an embedding trained on electronic lab notebook (ELN) protocols for cross-coupling reactions were asked to return pressure, temperature, and duration values in a specific format. The agent succeeded in 19 of 20 runs (95%), with one failure due to formatting inconsistency **(Extended Data Fig. 10; Supplementary Figs 29–31).** Agents equipped with microwave synthesis protocol embeddings were similarly tested. When asked to explain the difference between microwave irradiation and conventional heating, agents generated responses containing required keywords in 20 of 20 trials (100%), often elaborating with comparisons of heating efficiency and reaction rates (**Supplementary Figs 32–34**). A second query on the pressure setting of the Biotage Initiator+ microwave synthesizer also yielded 20 of 20 successful matches, with responses referencing solvent boiling points, operational ranges, and instrument-specific safety mechanisms (**Supplementary Figs 35–37**).

Finally, we conducted a path-based benchmark on the PubChem agent by instructing it to retrieve the *Ki* value of PZM21 for the mu opioid receptor along with an assay citation and description. The correct action sequence involved retrieving the compound ID (CID), filtering assay results for *Ki* activity, and fetching assay metadata. The agent succeeded in 17 of 20 runs (85%). One error involved redundant retrieval of the assay name, while the remaining two failures were due to input parsing errors **(Extended Data Fig. 9; Supplementary Figs 38–40).**

Collectively, these benchmarks demonstrate that agent reliability improves with schema-guided memory, state-aware prompting, and targeted retrieval frameworks. Such modular benchmarking pipelines provide essential tools for evaluating and validating autonomous scientific systems under conditions of non-deterministic model behavior.

**Discussion**

LLMs have revolutionized natural language understanding and generation, but their deployment in complex scientific workflows is hindered by several structural limitations. Chief among these are hallucinations, static knowledge bases, brittle prompt dependencies, and lack of task-specific memory. These challenges become acute in iterative real-world workflows that require - reliable context tracking, adaptive reasoning, and modular knowledge integration - features essential for real-world laboratory and computational environments. We introduce SciBORG, a modular framework for building AI agents that autonomously plan, reason, and execute scientific tasks by operationalizing domain-aware agents built from modular infrastructure, enriched by structured memory including state-tracking and agent-to-agent communication. By integrating memory-aware planning, embedded document retrieval, and tool execution, these agents exhibit capabilities beyond those of standalone LLMs. The use of agentic prompt engineering[25] allows agents to dynamically construct input representations aligned with system constraints and goals, eliminating the need for brittle manual prompt design. Through real-world integration with laboratory instrumentation and web-based databases, we demonstrate that SciBORG agents can operate reliably in both physical and digital environments including tasks often required in drug discovery pipelines. Using a Biotage Initiator+ microwave synthesizer, we show that agents autonomously plan and carry out an N-alkylation reaction with yields comparable to human chemists, differing by only 1.6% in percent conversion. When deployed in virtual hardware simulations, agents exhibited reliable tool orchestration and state-awareness across disjoint command inputs. When queried, agents retrieved relevant conditions, reactants, and parameters from ELNs and instructional documentation, as well as, provided accurate responses to technician training questions based on multi-source documentation, demonstrating their ability to synthesize and apply context-rich procedural knowledge. Through interaction with a PubChem-querying agent, they retrieved and computed molecular properties and bioassay information. Specifically, these agents performed complex information retrieval tasks, including compound-to-assay mapping, assay filtering, and citation extraction. Agents composed multi-step workflows to retrieve activity data (e.g., Ki values), handle ambiguous compound identifiers, and adapt to tool errors through retry logic and input reformulation. This enabled robust execution of data mining queries from natural language. The seamless use of inter-agent communication for fact delegation demonstrates a scalable mechanism for domain-specific RAG, one that enhances factual accuracy and restricts LLM hallucination by bounding access to trusted tools and data sources.

At the heart of SciBORG is a core infrastructure that abstracts computational and instrumental systems into standardized representations, including parameters, commands, microservices, and workflows, all formatted in hierarchical, human-readable JSON. These components are dynamically generated from source code and integrated into the agent's runtime prompt. This eliminates the need for hand-crafted prompts and supports rapid agent creation with minimal manual engineering. Agents constructed using SciBORG are tool-aware, memory-augmented, and capable of accessing domain-specific knowledge through retrieval-augmented generation (RAG) and inter-agent communication. Unlike static LLM pipelines, SciBORG agents maintain operational continuity through custom memory architectures. A central innovation of SciBORG is its use of pseudo–finite state automaton (FSA) memory. This mechanism enables scalable and interpretable tracking of a system's operational state using a discrete schema of allowed variables and transitions. By encoding agentic memory as schema-driven state transitions rather than long free-text chat histories, SciBORG preserves relevant context over extended sequences without overwhelming the LLM's context window. This allows agents to reason more effectively about system preconditions, reduce redundant tool calls, and adapt flexibly across disjoint steps within scientific workflows.

Benchmarking is central to establishing agent robustness. We developed three benchmarking modalities—output, state, and path-based—to assess agent performance and identify sources of error across configurations. Agents provided with structured state initialization performed better (85% success rate) than those without (65%), highlighting the value of encoding system state prior to execution. Similarly, memory-augmented agents using pseudo-FSA buffers outperformed summary memory counterparts (90% vs 50%) in tasks that required attention to tool prerequisites. These quantitative metrics provide a foundation for reproducible agentic development and deployment. Benchmarking confirms that agents equipped with state-based memory and structured reasoning pipelines outperform alternatives across planning, retrieval, and execution tasks. Specifically, agents with state-based memory outperform those with only summary or chat-based memory in path accuracy and tool use consistency. Moreover, benchmarking on output correctness, action trace fidelity, and system state progression demonstrates SciBORG's ability to diagnose failure modes and quantify robustness across diverse use cases. Inter-agent communication further expands this capability, allowing agents to delegate sub-tasks and query domain-specific resources effectively. We implemented inter-agent communication by enabling a synthesis agent to delegate factual queries to a PubChem agent. In a representative task, the synthesis agent retrieved molecular weights and computed molar quantities using information obtained from the PubChem agent. This collaboration model supports division of labor among specialized agents and improves factual consistency through trusted query delegation. SciBORG allows agents to invoke one another as tools thereby delegating sub-tasks such as data retrieval, compound resolution, or metadata enrichment. Such decentralized and modular approach to problem-solving supports domain-specific agents to be composed into a network that mimics the collaborative workflow of scientific teams.

By design, SciBORG contrasts with prior agentic frameworks such as ChemCrow[26], CRISPR-GPT[27], TAIS[28], and Coscientist[29]. These systems, while pioneering in their domains, are limited by fixed toolsets, manual prompt engineering, lack of persistent memory, or reliance on web search tools that reduce reproducibility. Specifically, ChemCrow[26], while integrating multiple chemistry tools, lacks memory support and modular extensibility for domain-specific protocols. CRISPR-GPT[27] and TAIS[28] showcase role-based delegation in gene editing and bioinformatics pipelines, respectively, but rely on prompt engineering and static toolsets without robust memory tracking. Coscientist[29] explores experimental planning but depends on prompt engineering, web search tools and code interpreters that may limit reproducibility and control. Compared to existing communication frameworks such as the Message Communication Protocol[30] (MCP), which emphasizes stateless message exchange between autonomous components, SciBORG offers a memory-augmented model of agent operation. While MCP provides a low-level infrastructure for message routing and inter-agent communication, it does not inherently support task state retention or schema-driven planning. In contrast, SciBORG agents retain internal memory of prior actions, utilize finite-state schemas to track workflow progression, and reason dynamically about their operational environment. This enables more robust multi-step execution, improved coordination between agents, and greater fault tolerance across distributed workflows.

SciBORG promotes modularity by enabling agents to be constructed from well-documented source code and deployed for task-specific operations. This design supports rapid agent generation, robust role specialization, and seamless substitution or expansion of agent capabilities that are key features for scalable scientific automation. Moreover, the framework mirrors real-world scientific collaboration, such as the interplay between computational and experimental chemists. Task-specific agents (e.g., synthetic chemist or cheminformatician) can coordinate, exchange critical information, and execute interdependent subtasks in pursuit of a shared objective. These interactions are mediated through structured memory, tool interfaces, and minimal abstraction layers, all essential for robust scientific delegation. Agents within SciBORG can operate independently yet communicate effectively, exchanging task-relevant information and advancing a shared scientific objective. The framework is LLM-agnostic and supports integration with general-purpose models alongside curated RAG sources, avoiding the need for expensive domain-specific fine-tuning. It enables agents to ground their actions in structured APIs and curated documents, learning iteratively through exploratory operation, not retraining. This design supports evolving toolchains and workflows, allowing SciBORG agents to adapt to new systems and scientific protocols without re-engineering.

Looking forward, further development of the SciBORG framework may include incorporation of uncertainty estimation, active learning from user or task-specific agent feedback, and support for multimodal inputs such as spectral data or images. As autonomous agents assume larger roles in experimental planning, data analysis, and literature mining, infrastructure such as SciBORG will be crucial for maintaining trust, interpretability, and reproducibility in scientific AI systems supporting dynamic orchestration of computational and experimental processes. Future work includes enhancing collaborative coordination among agents, extending support for multi-modal reasoning, and integrating uncertainty estimation to improve trust and transparency in agentic systems.

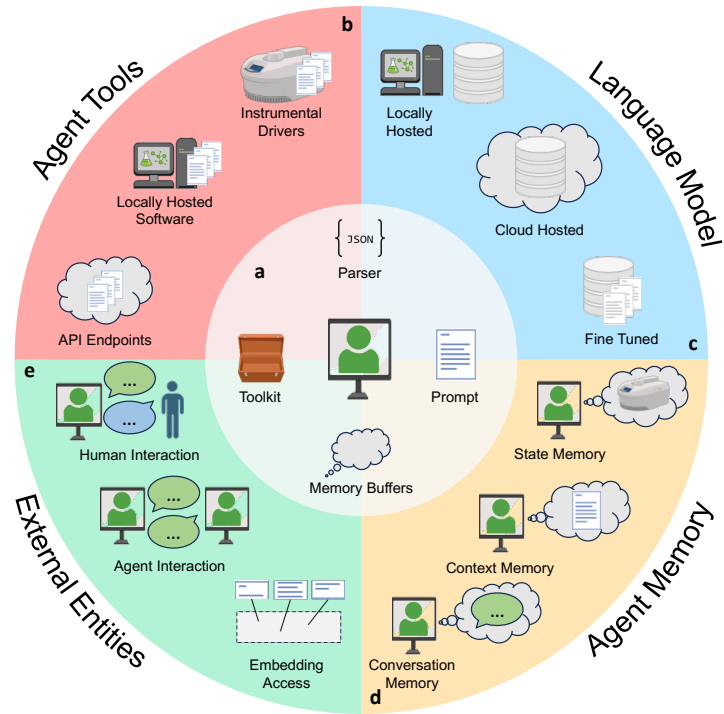

**General Agent Tool Categories**

**f** Internal Scope

API Endpoints    Instrumental Drivers

Locally Hosted Software

External Scope

Agent Interaction    Human Interaction

Document Embeddings

**Agent Thought-Action-Observation (TAO) Process**

**g**
**Thought:**
What do I need to do to satisfy the request based my tools and domain knowledge?

**Action:**
How should I use the tool which I have chosen?

**Observation:**
What have I learned from using this tool?

Request    Thought    LLM

Observation    Action

Prompt Augmentation

**Dynamic Augmentation of Agent Prompt**

**h**
```
You are an agent controlling {service}
which can perform the following actions
{tools}.

You goal is to fulfill the user's request
as accurately as possible. If you are
confused about anything...

Previous conversation history:
{history}
Current system state:
{state}
User request:
{request}
Agent scratchpad
{scratchpad}
```

{service}    {tools}

{history}    {state}

{request}    {scratchpad}

...

**Update of State Memory Based on Agent Actions**

**i**
State N    Agent Operation    State Update    State N+1

Augment Prompt with State    Iterative TAO Process    LLM Conversion of Log to State    Write to State Memory Buffer

**Fig. 1 | State and Memory based AI agent infrastructure. a,** The core internal infrastructure of an agent including a command parser, toolkit, dynamic prompt, and set of memory buffers. The command parser is used to interpret LLM output into tool usage, the toolkit contains links to tools

which the agent can use, the dynamic prompt is augmented by various entities including user requests, tool, and memory and is sent off to the LLM for interpretation, and the memory buffers save off key outputs for future use. **b,** The agent's tools are programmatic functions which interact with external services where the agent is generally aware of the input format, scope, and expected output format of the tool. The agent uses a LLM to decide when and how to use these tools and the tool's use is handled by the command parser. **c,** The language model provides the agent with the ability to process complex requests by utilizing its domain knowledge, analyzing the dynamic prompt augment with requests, tools, and memory, and formatting output for the command parser to operationalize its thoughts. **d,** The agent's memory which is generally comprised of a buffer or set of buffers which are designed to save specific agentic outputs for future use. State memory keeps track of the operational state of a system or set of systems, context memory keeps track of the context learned from external entities (i.e. embedding access), and conversation memory keeps track of a conversation between the agent and a human or another agent. **e,** External entities are presented to the agent as tools, but they differ from tools as the agent is generally told to call upon them for direction, clarification, or action that is outside the current scope of the agent. **f,** Agentic tools both inside and outside the contextual scope of the agent. **g,** The agent's iterative thought-action-observation[18] (TAO) process controlled by the LLM and used to explore the problem space. **h,** The dynamic prompt augmentation process using specific entities relating to the service and state of the service. **i,** The agent's state-based memory which is updated at the termination of the iterative TAO process by using an LLM chain to analyze the action log and update the state in a controlled manner.

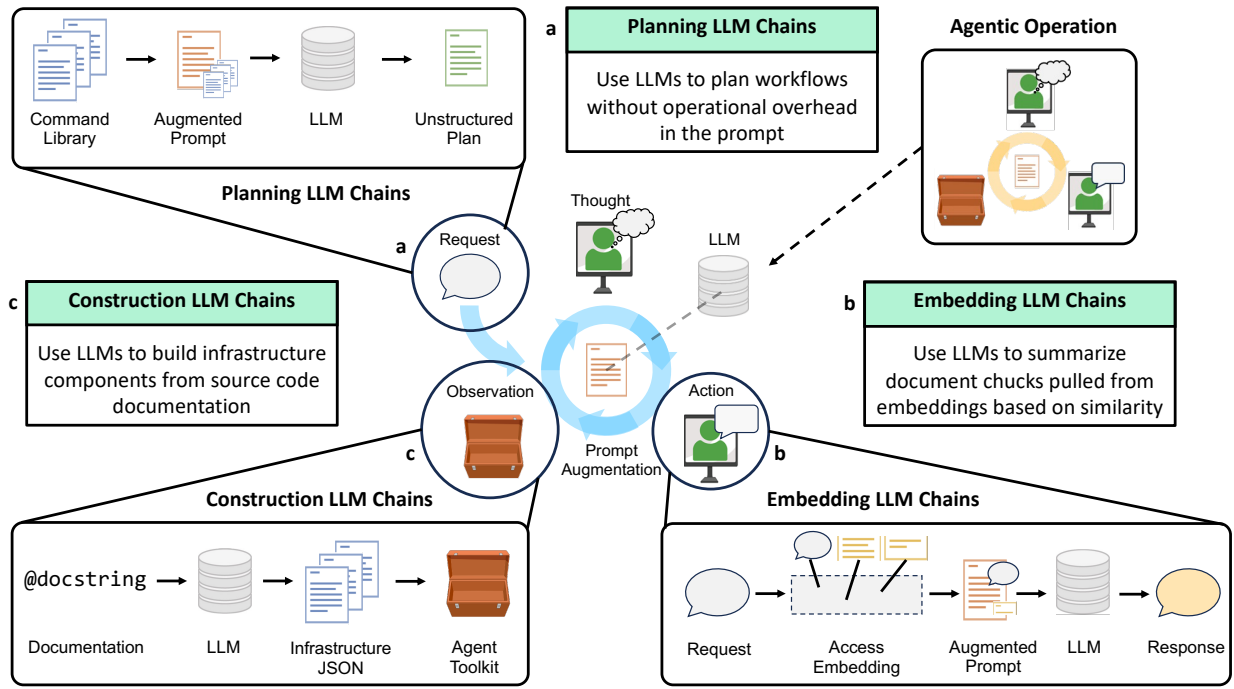

**Fig. 2 | Integration of external LLMs chains for agentic planning, construction and operation. a,** LLM chains for supplementing agentic requests with high-level workflow planning without overhead of operational formatting. **b,** LLM chains which act as tools and can supplement an agent's prompt with knowledge from external document embeddings. **c,** LLM chains which are used to build core infrastructure components dynamically at runtime including parameters, commands, microservices, etc. These chains (**a-c**) all have modular prompts which are augmented with key information at runtime including operational endpoints (**a**), document embedding access (**b**), and driver code documentation (**c**).

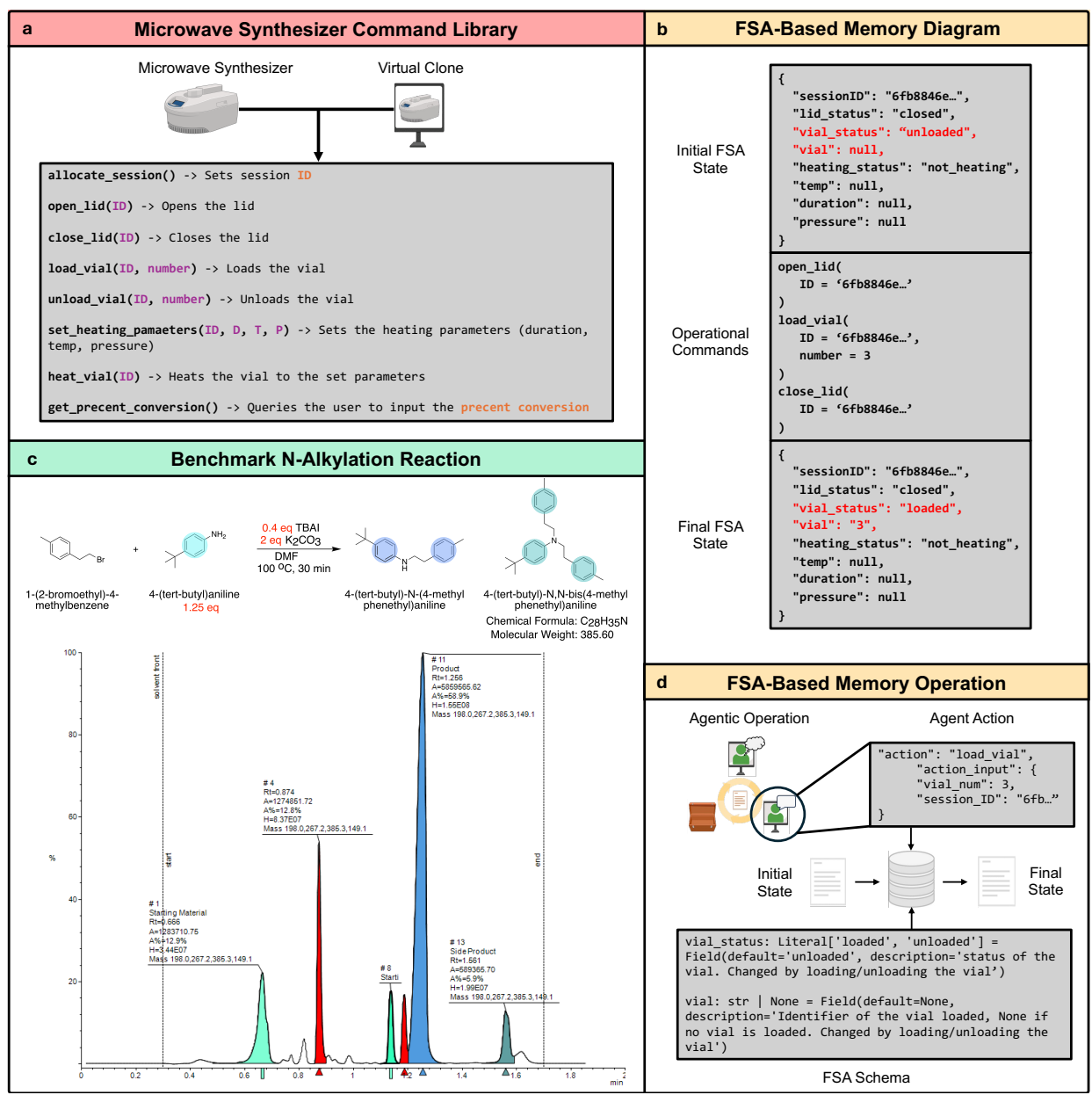

**Fig. 3 | Finite state memory integration in AI Agents is required for robust operation of microwave synthesis protocols. a.** Virtual clone of microwave synthesizer instrument includes a command library designed for both experimental planning and execution. This library includes commands for physical tasks (such as open_lid, close_lid, etc.) that are directly linked to the operations of the physical instrument. **b.** The finite state automata (FSA) schema of the microwave synthesizer is edited as the agent operates the tasks on the instrument. **c.** The benchmark N-alkylation reaction (light blue starting material, dark blue product) was planned and executed on a physical Initiator+ microwave synthesizer instrument using a two pass AI planning chain (high-level and low-level plan) and worklist executor. The conversion efficiency of the product (blue peak, 58.9% conversion) was determined by UPLC. **d.** Schematic of FSA based memory in operation that is updated with defined transition rules that the agent autonomously executes based on iterative actions and observations. Finite state memory provides robust operational execution.

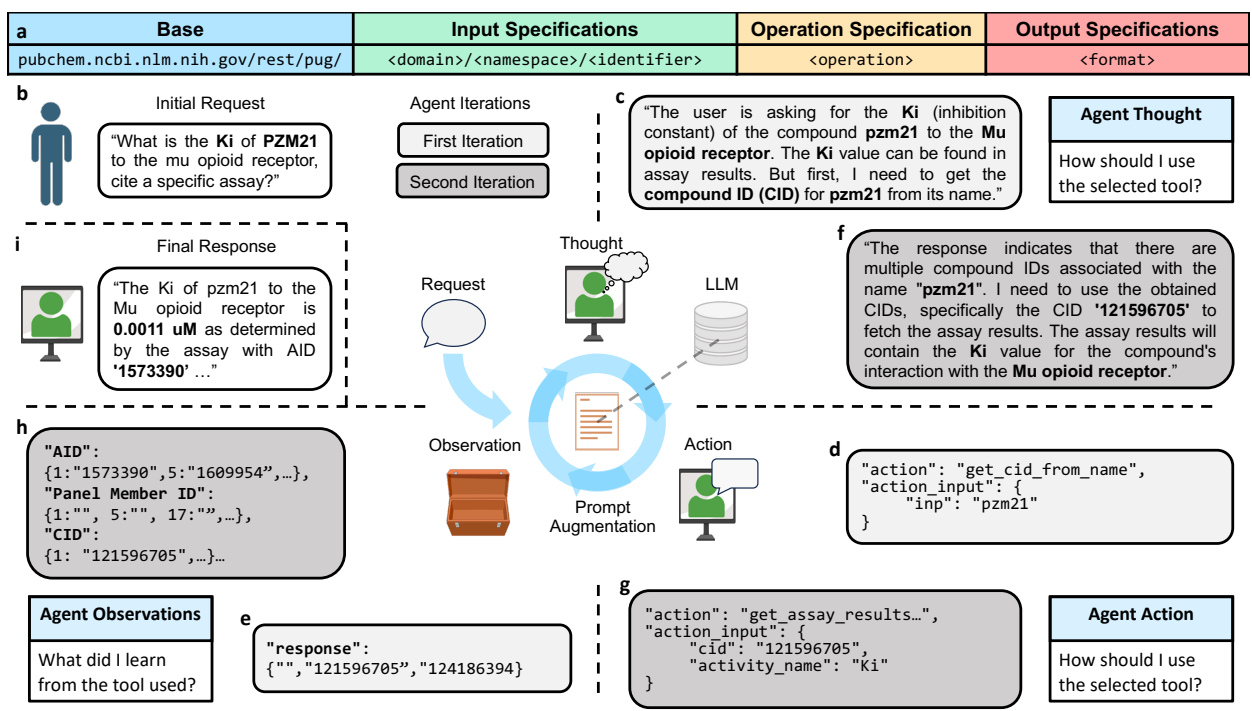

**Fig. 4 | SciBORG AI agent plan, reason, and execute complex data mining tasks using PubChem database. a.** The core structure of the API URL is used in custom functions to access relevant data from different sections of the PubChem database. **b.** The initial request provided to the agent by the human in natural language. **c.** The initial thought of the agent about accessing the drug targets compound ID (CID) without any prompt engineering. **d.** The initial action of the agent to get the CID from the targets name. **e.** The initial observation the agent on the CIDs associated with the drug target. **f.** The second thought of the agent is based on the initial request and initial observation on accessing the Ki. **g.** The second action of the agent to access the Ki using one of the drug targets CIDs. **h.** The second observation the agent makes regarding the Ki assays retrieved for the drug target. **i.** The final output of the agent after additional iterations (not shown) were performed. The iterative planning, reasoning and execution of these tasks are without any prompt engineering by human user to fulfill the requested task.

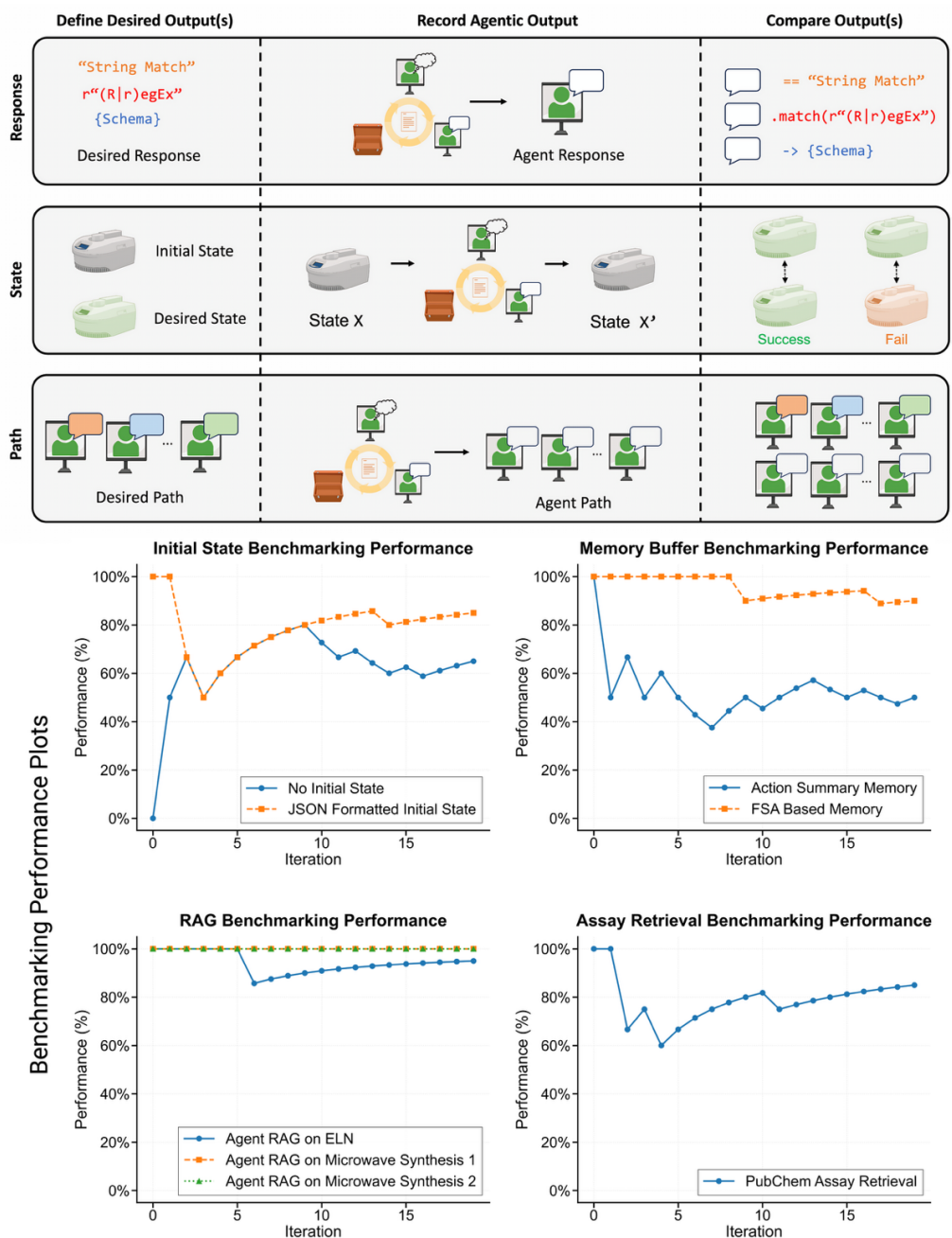

**Fig. 5 | Agentic benchmarks are essential for robustness and reliability of output, state, and planned workflow operation. a,** Output-based benchmarking takes agentic output at the conclusion of iterative operation and compares it to a string, regular expression or validation schema. **b,** State-based benchmarking access the final state of a system which the agent interacts with after operation and compares it to a schema representing the desired state(s). **c,** Path-based benchmarking which takes the agents action path after operation and compares the actions and action inputs with the desired path. Each benchmark regardless of type is broken down into three common steps where **1)** the user defines the benchmarking goals, **2)** the agent undergoes agentic operation and relevant output is logged, and **3)** the output is compared with the user defined goal to assess success or failure.

**METHODS**

**Software, Hardware, and Documentation**

All AI agents were developed in Python (3.10.X) using the Langchain AI[16] (0.1.X) software development kit (SDK) for building context-aware applications. Classes of the core infrastructure were built by extending the Pydantic[17] (2.6.X) `BaseModel` class for robust validation and JSON serializability. All LLM interaction occurs using Langchain that allows for our infrastructure to support any LLM supported by Langchain and including internally deployed LLMs (i.e. Azure OpenAI) as well as fine-tuned LLMs. During internal development and testing, LLMs that were used include GPT 3.5 turbo, GPT 4, and OpenAI embeddings. LLMs which work with planning chains and agents use chat completion endpoints. Additionally, we have utilized FAISS[19] library (v1.8.0) from Meta for efficiently searching in the vector-space of document embeddings. Hardware used for instrumental integration testing was a Biotage Initiator+ microwave synthesizer modified by engineers at the National Center for Advancing Translational Science (NCATS), a division of the NIH. The modifications that were used in testing include a set of Python wrappers which control core instrumental components such as opening and closing the lid and heating the vial. A virtual clone of this instrument was implemented in Python based on current and future modifications which NCATS plans to integrate with the physical hardware. This virtual clone had similar endpoints and descriptions as the physical instrument but was designed with planned integrations such as automated vial loading already incorporated. Documentation used included protocols for microwave synthesis operation and electronic lab notebook (ELN) entries for microwave synthesis reactions provided by the National Center for Advancing Translation Science (NCATS) automation chemistry team.

**SciBORG SDK Core Infrastructure**

Our core infrastructure is designed to create a modular, standardized framework to represent both theoretical and operational information to an AI agent back by an LLM, creating a context aware, operational agent which can interact with a computational or instrumental "microservice" in directed yet autonomous manner. A microservice is defined as a script, software package, API, physical instrument, or any combination of these which has a relatively small number of endpoints in a certain scope which can be built into AI ready tools. The AI agent can then use these tools to dynamically interact with the microservice to accomplish tasks. In general, microservice endpoints revolve around a relatively specific set of tasks but there are no formal restrictions on the scope. Endpoints which interact with computational services can reside in the same microservice as endpoints which interact with physical instrumentation, creating the ability for hybrid services such as instrumental operation and data analysis to coexist in the same agent scope.

To best structure this concept of a microservice to an AI agent, we have created a Python object hierarchy which includes parameters, commands, workflows, microservices, and libraries. Parameters contain information on values that are passed into specific commands including data types, data ranges, allowed values, units, context, etc. Commands contain information on python functions which either perform operational tasks or wrap other functions which perform operational tasks including parameters, description, and expected output structure. When run operationally, commands undergo internal validation on inputs, providing an additional layer of robustness if an AI agent incorrectly uses a command. Workflows contain information on the operational order of command and contain explicit instructions on how information will be transferred between commands. These are quite useful tasks which involve a high degree of planning and abstraction of certain predefined processes (i.e. running a set protocol) at agent runtime. Microservices contain information on the components and scope of an individual service including operational commands, a high-level description, and access to an embedded document space which can be queried for context. Finally, libraries include a set of operational commands that span across multiple microservices which can be combined to accomplish tasks that require more than one type of service. Many of these core infrastructure components, including parameters, commands, and microservices can be directly build from Python source code functions and modules, using LLMs to analyze source code documentation and structure that information into infrastructure components. This allows for users to build agents directly from source code with any "prompt engineering" coming in at the source code documentation level **(Supplementary Tables 5–9).**

**JSON Serializability of AI Agent Components**

All core infrastructure components are fully serializable in JavaScript Object Notation (JSON) format. This means that python objects representing core infrastructure can be easily written to a JSON string or file and then that string or file can be read back into a python dictionary and passed to the object constructor to recreate the object. This rebuilt object maintains the properties of the original object. The JSON format was chosen for reasons relating to both core infrastructure and LLM interaction. JSON objects are inherently hierarchical, which works well with the hierarchical core infrastructure. This allows for JSON objects for components lower in the hierarchy (i.e. parameters) to naturally be included in objects higher in the hierarchy (i.e. commands). Additionally, JSON objects are easily transferable between machines, allowing for core infrastructure components to be moved from one system to another. Finally, JSON objects are highly human readable allowing for core infrastructure components to augment input prompts to LLMs as well as be generated by LLMs as output.

**LLM Chains for Infrastructure Creation, Planning, and Embedding Access**

In our work, we develop custom LLM chains (structured, sequential calls to an LLM) for various context dependent applications including core infrastructure construction, high- and low-level operational planning, and construction and access of LLM based document embeddings. Langchain provides three major chain concepts which we utilize heavily in our custom chains: dynamic prompts, output parsers, and memory. LLM prompts are instructional text input that is provided to the LLM which define a request or task and include any additional conditions or context around that task which may benefit the LLM during output generation. We dynamically augment prompts at runtime to include contextual information pertaining to specific requests or operational information pertaining to specific tasks. Output parsers are objects which provide context on output format and then subsequence enforce this format on generated output. We employ Langchain's built in JSON output parser frequently in our custom chains to structure LLM output into core infrastructure components. Output parsers formatting instructions are included in the prompt. Finally, memory refers to ways to keep track of important information between chain queries. In the context of LLM chains, this generally refers to keeping track of past input and output in some format. We utilize memory in our work with LLM chains to keep track of previously generated infrastructure, operational plans, and external context and use this to augment generation of future output in a more targeted manner. This memory can contain a buffer (context window) and/or use summarization to reduce the character count of retained information to prevent the memory becoming a computing overhead.

The first type of LLM chain which we employ are chains that generate core infrastructure components based on user input, command drivers, or operational documentation. These chains are always equipped with Langchain's built in JSON output parser which instructs the LLM via the prompt to generate a JSON formatted string that defines the infrastructure object in question. This JSON formatted output can then be directly converted into a Python dictionary and then passed into the infrastructure object's constructor to create an object that fulfills the request. This type of chain has been implemented for most infrastructure components including parameters, commands, workflows, microservices. These chains can include memory, but it is generally only needed for iterative debugging of inaccurate or incomplete object generation. A practical use case of this type of chain is where a python module which represents operational endpoints of a specific software package, API service, or scientific instrument driver is converted into a structure microservice object. The overall description of the service is taken either from the modules docstring or external documentation and then each python function defined in the module is converted to an operational command. For each function in the module, an LLM prompt is dynamically augmented with the microservice definition, function signature, function docstring, and formatting instructions for a command object JSON. The LLM then uses this information and its internal domain of knowledge to determine and format the command's description, parameters (with data types, units, ranges, etc.), and output signature. This allows for a simple, one line function to easily convert operational code into a structured microservice.

The second type of LLM chain employed in our system are chains that assist humans or AI agents in operational planning of complex workflows. The prompts that are provided to these chains define the overall goal as workflow planning and can be augmented with both information about operational commands which can be included in the workflow and additional scientific or procedural context related to workflow planning. Operational planning chains are separated into two main categories, high-level planning and low-level planning. High-level planning chains plan workflows without worrying about the structure of the output. The main use of these chains is assisting humans or AI agents in reducing abstract tasks to clearly defined step-by-step workflows. These chains can be implemented with memory of past queries and their corresponding generated workflows to engage in iterative "debugging" of incomplete or inaccurate workflow. Low-level planning chains plan workflow with output structured in an executable format that can then be passed directly to a command interpreter to execute the AI generated workflow. Their prompts include formatting instructions from Langchain's built in JSON output parser, set to generate a JSON that can be directly converted to a workflow object. This formatting step introduces additional overhead which is not present in high-level planning and thus can cause more issues when generating workflows from a more abstract request. To remedy this, we can combine both high- and low-level planning chains to convert an abstract user request to a high-level operational plan, and then convert the resulting high level operational plan to a low-level operational plan which is ready for execution **(Supplementary Table 1)**

The final type of LLM chain which is used in our system are chains which construct and interact with text embeddings. Each task-specific chain is augmented with relevant context in the form of a vectorstores, acting as a knowledge base. Various models can be utilized to curate these embeddings, alongside different methods for their storage. We primarily utilize OpenAI embeddings to curate the embeddings of contextual documents and employ FAISS and Chroma DB for their storage. This approach of "grounding" the LLM to a specific knowledge base that comprehensively covers all pertinent information for a given task aims to mitigate hallucinations. Furthermore, as the LLM searches mainly within the most relevant sources, which are provided after peer review by experts, the responses generated are supported with cited sources used as references.

**Development of AI Agents with Heightened Context Awareness**

LLM integrated agents are entities which allow the LLM to perform autonomous or semi-autonomous problem solving in an iterative and directed manner. These agents are generally provided with a toolkit which corresponds to a set of code functions that the agent can use to accomplish tasks or access information outside of the LLMs domain of knowledge and an output parser which converts structured LLM text output to function calls. During operation, the agent undergoes an iterative exploratory process based on the ReAct[18] framework where it works through the problem in a step-by-step manner. The primary agent which we use in our work is a modified version of Langchain's internal structured chat agent. Structure chat agents can use tools (python functions) which receive multiple inputs, iterate through an internal though/action/observation process, and provide output to humans in a conversational manner. The though/action/observation process is the key component of the agent's problem-solving ability, and it is where a specific tool is selected based on the current goal (thought), the tool is used with appropriate inputs (action), and the output of the tool is fed back to the agent (observation). This process can be repeated until the agent achieves its overall goal.

We make several modifications to the base form of this agent to heighten their context awareness. These agents are designed to be structured around a single or small set of microservices but can operate in a broader context by communicating with other agents in a network. This agent's prompt is defined in a highly modular and dynamic manner, with information on context, tools, user input, embeddings, chat history, current thought/action/observation, system state, and additional retrieved context all being provided to the agent at runtime. An initial toolkit is created by accessing commands of each microservice provided to the agent and converting those commands into Langchain tools. Each command provides the description, pre and post requisites, parameters, and expected output format to a tool object which then serves as a function wrapper to the command. The prompt contains instructions to check for operational prerequisites prior to using commands, to assist in robust planning and operation. Additional tools which are separate from the microservice command set include human and external agent interaction, operational planning, and document embedding access. These tools can be included at build time to improve the agent's ability to reason and solve complex tasks.

Providing the current agent with the ability to communicate with a human or external agent allows for the agent to access information outside the LLMs domain of knowledge in a targeted manner. Interaction with humans and external agents can be accomplished in two ways: using a tool or acting as if the current agent is being used as a tool. Using a human or external agent as a tool refers to creating a Langchain tool which queries the human or external agent for input when the current agent decides to use that tool. The input to the tool will be determined by the LLM and it will define information or clarification which the agent needs to accomplish the task at hand. The key here is that the current agent knows the context of what it is communicating with and makes an active decision to query the human or external agent. This can even be customized to include multiple humans or external agents as tools. Our current implementation uses Langchain's built in human tool for human interaction and any agent or chain's invoke method can be wrapped by a tool decorator to use an external agent as a tool. In contrast, having the current agent being used as a tool removes a level of abstraction from the agent decision making process as the agent no longer must decide if it needs to access a human or external agent or which one to access. Instead, the agent operates autonomously while it can and then once it reaches an impasse, it stops iteration and returns details of what additional information it needs to continue operating. The human or external agent which is operating the current agent can then provide additional information and restart the iterative process.

Another way to provide the agent with access to additional information outside of the LLMs domain of knowledge is to create a tool that calls an embedding chain, allowing for an agent to call and view the response of that chain. If the agent is provided with a request with incomplete information for a specific task, the agent can decide to query the embedding prior to asking a human or external agent to see if any documentation in the embedding can fill in the information gaps. Since the chain is accessed as a tool, the agent can iteratively use this tool in a question-and-answer style manner.

The final modification that we make is a more robust memory system for the agent which retains chat history, information on past agentic actions which affect the state of the system, and external context access as separate memory entities. There are several chat-based memories which are natively built into Langchain, including two which we used in our implementation, chat history and chat summary. Chat history retains the direct conversation text between the human and AI agent whereas conversation summary summarizes this conversation. There is also an option to introduce a buffer to either memory type which serves as a context window that overwrites older chat history with newer chat history. For retaining information on past agentic actions, we modify Langchain's built in chat summary memory with a new prompt which dynamically receives a trace of the thought/action/observation process and summarizes actions and observations correspond to internal (microservice driver) tools in the trace. This is key to retaining information about the state of the microservice (or microservices) between agent calls. If a set of values which define the agent's state is provided, we create a separate implementation of this memory which is designed to track the agent's state between operation calls based on the state structure. This in turn treats the microservice as a finite state automaton (FSA) and provides the LLM with this information as the agent operates. Finally, we retain information on external context by using summary memory with a modified prompt which takes in the internal thought/action/observation trace and filters out and summarizes tool calls that are not a part of the microservice driver. This includes communication with humans, external agents, or document embeddings. These three memory categories are combined using Langchain's built in combined memory, which supports the use of multiple memory objects in a single agent **(Supplementary Table 2).**

**Agentic Benchmarking Infrastructure**

We implement agentic benchmarking in our infrastructure using three methodologies: output, state, and path-based benchmarking **(Fig. 5)**. Output-based benchmarking refers to the ability of an AI agent to generate desired results at the end of its internal thought-action-observation process. The actions which the agent takes and the final state of the system which the agent interacts with (if applicable) is not considered in this benchmarking method. This is implemented by providing an individual or set of strings, regular expressions, or Pydantic validation schemas to compare against one or more agent outputs. The benchmarking object also requires a python function which builds an AI agent, key word arguments provided to that function which defines agent specifications, and an initial question which is provided to the agent. During each benchmarking run a new agent is created, the agent is run with the provided initial question, the agent is allowed to operate until completion, and then the agent's output(s) are compared against the desired outputs provided. For direct string-based comparisons, a successful benchmark run matches one of the provided strings for each output key character-for-character to the corresponding output from the agent. For regular expression-based comparisons, a successful benchmark run matches one of the provided regular expressions for each output key to the corresponding output from the agent. For JSON schema-based validation, a successful benchmark run successfully validates JSON formatted agent output against one of the provided Pydantic schema for each output key. The benchmarking object performs this run for n iterations and reports the number of runs which are successful as m. This provides the user with a verifiable success rate for a specific system, agent, and question paring and can help gauge agent robustness despite LLM non-determinism.

State benchmarking refers to the ability of an AI agent to perform a set of actions which takes a system it can interact with from the provided initial state to a desired final state or states. The actions which the agent takes and the order in which those actions are taken are not considered. This is implemented by providing a python object or function which tracks the system's state, a python function which builds an AI agent which interacts with a system via its tools, keyword arguments provided to that function which defines the agent's specifications, an initial question, a Pydantic schema for the initial state, and one or more Pydantic schema(s) for the desired final state(s). During each benchmarking run, the benchmarking object ensures that the system is in its initial state, runs the agent with the initial question, allows the agent to operate until completion, and then checks to see if the system's final state matches the desired final state. If the states match, the run is considered a success and if the states do not match the run is considered a failure. From a benchmarking statistic perspective, the state benchmarking object behaves in the exact same manner as the output benchmarking object.

Path benchmarking differs from output and state benchmarking whereas instead of consisting of the final output or state of a system, the actions which the agent takes, the order of those actions, and the inputs provided (if applicable) are considered by the benchmark. The is implemented by providing one or more lists of actions which constitute the desired operational path, a python function which builds an AI agent which interacts with a system via its tools, keyword arguments provided to that function to define agent specifications, and an initial question. Each action in the desired path list can be either a name of a specific tool or a tuple with the name and a schema to validate against the input for the specific tool. During each benchmarking run, the agent is run with the initial question and the actions and action inputs which the agent takes from state to finish are recorded. If the actions and inputs which the agent took during operation match one of the desired paths, the run is considered a success, otherwise the run is considered a failure. From a benchmarking statistic perspective, the path benchmarking object behaves in the exact same manner as the other benchmarking object **(Supplementary Table 3).**

**Synthetic Benchmark Reaction on Microwave Synthesizer**

The benchmark reaction run on the Initiator+ was the synthesis of 4-(tert-butyl)-N-(4-methylphenethyl)aniline. The starting materials 1-(2-bromoethyl)-4-methylbenzene (1 eq.) and 4-tert-butylaniline (1.2 eq) were added to a solution of potassium carbonate (2.0 eq.), TBAI (0.2 eq.), and DMF (solvent) in a microwave vial. The vial was loaded to the initiator plus and the temperature was set at 100 degrees Celsius and run for 30 minutes. The only difference between the human initiated run and AI initiated run was that human run solution was pre-stirred for 30 seconds. This feature was not integrated on the automated microwave synthesizer. See figure 5c and SI figure S6 for additional information on the benchmarking reaction.

## CODE AND DATA AVAILABILITY

The code will be available for the specific commit corresponding to the manuscript's publication timeframe and will be released on GitHub https://github.com/chopralab/sciborg_manuscript_repo. All scientific examples shown in this work will be released as notebooks at https://github.com/chopralab/sciborg/tree/master/notebooks/SI.

**ACKNOWLEDGEMENTS.** We thank the members of Chopra laboratory and NCATS staff for their input and assistance. This work was supported by the National Center for Advancing Translational Sciences (NCATS) U18 TR004146, NCATS ASPIRE Challenge and Reduction-to-Practice awards to G.C. who is the James Tarpo Jr. and Margaret Tarpo Professor.

**COMPETING INTERESTS.** G.C. is the Director of Merck-Purdue Center funded by Merck Sharp & Dohme, a subsidiary of Merck and co-founded Meditati Inc., BrainGnosis Inc. and LIPOS BIO Inc. M.M cofounded Meditati Inc. and BrainGnosis Inc. A.P cofounded BrainGnosis Inc. The remaining authors declare no competing interests.

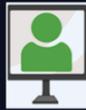

## Agent Prompt Template

```
BASE_LINQX_CHAT_PROMPT_TEMPLATE = """You are an AI agent in control of the following microserivce - {microservice}

A short description of the microservice is as follows:
{microservice_description}

Respond to the human as helpfully and accurately as possible. You have access to the following tools:
{tools}
```

**Input Variables**

```
Use a json blob to specify a tool by providing an action key (tool name) and an action_input key (tool input).

Valid "action" values: "Final Answer" or {tool_names}
{additional_instructions}
You may not need to use a tool to answer certian questions. You can instead refer to your domain of knowledge or the chat
history and then return your "Final Answer."

Provide only ONE action per $JSON_BLOB, as shown:
```
{{
    "action": $TOOL_NAME,
    "action_input": $INPUT
}}
```
```

**Output Format**

```
Follow this format:

Question: input question to answer
Thought: consider previous and subsequent steps
Action:
```
$JSON_BLOB
```

Observation: action result
... (repeat Thought/Action/Observation N times)
Thought: I know what to respond
Action:
```
{{
    "action": "Final Answer",
    "action_input": "Final response to human"
}}
```
```

**Example response**

```
Begin! Reminder to ALWAYS respond with a valid json blob of a single action. Use tools if necessary. Respond directly if
appropriate. Format is Action:```$JSON_BLOB```then Observation'''

Chat histroy:
{chat_history}

Past Action Summary:
{past_action_log}

human = '' {input}

{agent_scratchpad}
(reminder to respond in a JSON blob no matter what)
"""
```

**Enforcing JSON-only response**

**Input Variables**

*Extended Data Fig. 1* | **Structure of the** *dynamic* **base prompt** *template* ***used for making SciBORG Agents.*** Input variables appear in yellow bounding boxes; they will be filled in at runtime to enable flexible agent creation. The output format is marked by a blue bounding box; it defines how the agent's response must be structured. A generic example response is included for clarity; it is not tied to any specific domain but demonstrates the required format.

Planning Chain on PubChem REST API Endpoints

Input Prompt

```
What is the IC50 of 1-[(2S)-2-(dimethylamino)-3-(4-
hydroxyphenyl)propyl]-3-[(2S)-1-thiophen-3-ylpropan-2-yl]urea
to the Mu opioid receptor, cite a specific assay in your response?
```

Agent Output

```
1. Use the `get_cid_from_name` function to retrieve the CID for the chemical name
   "1-[(2S)-2-(dimethylamino)-3-(4-hydroxyphenyl)propyl]-3-[(2S)-1-thiophen-3-ylpropan-2-yl]urea".
   - Parameter:
     - `inp`: "1-[(2S)-2-(dimethylamino)-3-(4-hydroxyphenyl)propyl]-3-[(2S)-1-thiophen-3-ylpropan-2-yl]urea"
   - Save the returned CID as `compound_cid`.

2. Use the `get_assay_results_for_compound` function to get the assay results for the retrieved CID,
   and specifically filter for the IC50 activity.
   - Parameters:
     - `cid`: `compound_cid`
     - `activity_name`: "IC50"
   - Save the returned assay result table as `assay_results`.

3. From the `assay_results`, find the assay result related to the Mu opioid receptor
   and save the assay ID as `assay_id`.

4. Use the `get_assay_name_from_aid` function to retrieve the name of the assay for the saved assay ID.
   - Parameter:
     - `aid`: `assay_id`
   - Save the returned assay name as `assay_name`.

5. Use the `get_assay_description` function to retrieve the description, protocol, and comments for the saved assay
   ID.- Parameter:
     - `aid`: `assay_id`
   - Save the returned assay description as `assay_description`.

The IC50 of the compound to the Mu opioid receptor is found in the `assay_results`,
and the specific assay used is described by `assay_name` and `assay_description`.
```

The response lists the exact steps the agent planned to answer the prompt

*Extended Data Fig. 2* | **Response of a planning chain with access to PubChem driver file containing PubChem REST API Endpoints.** The input prompt sits in a yellow box and asks for the IC50 of a specified compound at the Mu opioid receptor, with a request to cite an assay. The agent's output appears in a blue box and breaks the task into ordered steps. Each step names the tool to call, lists its parameters under an input key, and explains where to save the returned data. The closing line points to the assay results table for the IC50 value and notes which assay name and description were retrieved.

## Document Embedding Chain Trained on Microwave Synthesizer Protocols and Operational Manuals

How does microwave irradiation influence reaction mechanisms differently compared to conventional heating methods?

**Input Prompt**

Thought: The user is asking for parameters for a specific chemical reaction. This is not information that the MicrowaveSynthesizer microservice can provide directly because it doesn't contain specific chemical reaction parameters. However, I can use the call_RAG_agent tool to query relevant sources of information.

**Thought**

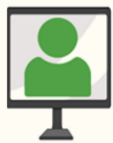

get_answer_from_information

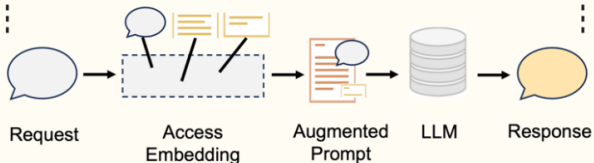

Request     Access     Augmented     LLM     Response
            Embedding    Prompt

**Action**

"[Document(page_content='Under the heating activity, the dividers of the reaction vessel act as\nthe mediator and collect the heat from the exterior heating source,\nexchanging the thermal energy to the reaction solvent. Then, re-\nactants start to transform into the product followed by convection\nor conduction mechanism. Thus, obviously the center part of the\nreaction material is hardly able to reach the desirable temperature\nowing to the generation of thermal gradients. These thermal gra-dients are responsible for the non-uniform transformations,\ncreating obstacles for 'scale-up production. 'On the contrary, MW\nprovides a uniform heating to the reaction materials in the entire\nreaction vessel, enabling rapid and uniform heat transfer. Fig. 2 (a)\nclearly showed the differences in MW-assisted and oil-bath tem-\nperature pro files [47].\nGenerally, MW radiation of suitable wavelength collaborates\nwith the reaction material and generates the heat simultaneously', metadata={'source': 'NIH_docs/Microwave chemistry, recent advancements, and eco-friendly.pdf', 'page': 1}),

**Observation**

Microwave irradiation provides a more uniform heating to the reaction materials in the entire reaction vessel, which allows for rapid and uniform heat transfer. This is different from conventional heating methods where thermal gradients can lead to non-uniform transformations and issues in scale-up production. Moreover, microwave radiation collaborates with the reaction material and generates heat simultaneously, leading to faster reaction times compared to conventional heating methods. For instance, the reaction time for compound synthesis is significantly shorter when using microwave irradiation compared to conventional heating methods.

**Agent Output**

*Extended Data Fig. 3* | **Demonstration of a document embedding chain process that has access to microwave synthesizer protocols and operational manuals.** The user's prompt occupies a yellow box and asks how microwave irradiation influences reaction mechanisms compared to conventional heating methods. In the thought box the agent acknowledges that it lacks enough context to answer on its own. It then performs a retrieval action against the vectorized documents and logs the resulting observation. The blue box contains the agent's final answer, which integrates the retrieved passages to explain uniform heating and faster reaction rates. This example shows how embedding can augment an LLM with specialized content to improve its responses.

## Microservice JSON for microwave synthesizer

```json
{
  "name": "MicrowaveSynthesizer",
  "uuid": "890a8f31-5847-488b-b834-aecb8b1d9748",
  "desc": "Module for a virtual microwave synthesizer.",
  "commands": {
    "allocate_session": {
      "name": "allocate_session",
      "microservice": "MicrowaveSynthesizer",
      "uuid": "890a8f31-5847-488b-b834-aecb8b1d9748",
      "desc": "Allocates a session on the microwave synthesizer. Must be called prior to any other action.",
      "parameters": {},
      "fn": "allocate_session",
      "module": "sciborg.testing.models.drivers.MicrowaveSynthesizer",
      "package": "sciborg.testing.models.drivers",
      "has_return": true,
      "return_signature": {
        "session_ID": "the id of the allocated session"
      }
    },
    "close_lid": {
      "name": "close_lid",
      "microservice": "MicrowaveSynthesizer",
      "uuid": "890a8f31-5847-488b-b834-aecb8b1d9748",
      "desc": "Closes the lid on the microwave synthesizer. Must be run prior to running heating.",
      "parameters": {
        "session_ID": {
          "name": "session_ID",
          "data_type": "str",
          "precision": -1,
          "upper_limit": null,
          "lower_limit": null,
          "allowed_values": [],
          "is_optional": false,
          "is_list": false,
          "default": "None",
          "from_var": false,
          "var_name": "",
          "desc": "the id of the current session"
        }
      },
      "fn": "close_lid",
      "module": "sciborg.testing.models.drivers.MicrowaveSynthesizer",
      "package": "sciborg.testing.models.drivers",
      "has_return": true,
      "return_signature": {
        "status": "is a status string that provides the result of the operation"
      }
    },
    "get_precent_conversion": {...},
    "heat_vial": {...},
    "load_vial": {...},
    "open_lid": {...},
    "unload_vial": {...},
    "update_heating_parameters": {...}
  }
}
```

Driver commands for the microwave synthesizer that become tools for AgentOS Agents

Extended Data Fig. 4 | **Structure of Microservice JSON drive file for Microwave Synthesizer.** The JSON defines driver-level commands for operating a virtual microwave synthesizer. These commands are exposed as tools for AgentOS Agents. Each command (e.g., allocate_session, close_lid) includes metadata such as function name, parameters, and return signature. The user issues high-level prompts, and the agent determines which microservice tools to invoke through a structured thought-action-observation loop. When the agent lacks sufficient context or has to perform an action on the instrument, it can retrieve and utilize these commands to ensure precise and valid operation execution.

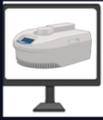

**Microwave Synthesizer as system with allowed states and transition rules between states**

```python
class MicrowaveSynthesizerFSA(BaseModel):
    sessionID: Optional[str] = Field(
        default=None,
        description="ID of the session allocation or None if no session allocated. Changed by allocating the session"
    )
    lid_status: Literal['open', 'closed'] = Field(
        default='closed',
        description="Status of the lid. Changed by opening or closing the lid"
    )
    vial_status: Literal['loaded', 'unloaded'] = Field(
        default='unloaded',
        description="Status of the vial. Changed by loading/unloading the vial"
    )
    vial: Optional[str] = Field(
        default=None,
        description="Identifier of the vial loaded, None if no vial is loaded. Changed by loading/unloading the vial"
    )
    heating_status: Literal['not_heating', 'heating'] = Field(
        default='not_heating',
        description="Status of heating. Only changed by heating the vial, not by updating the parameters"
    )
    temp: Optional[int] = Field(
        default=None,
        description="Set temperature to heat at, None if currently set. Changed by updating the heating parameters"
    )
    duration: Optional[int] = Field(
        default=None,
        description="Set duration to heat for, None if not currently set. Changed by updating the heating parameters"
    )
    pressure: Optional[float] = Field(
        default=None,
        description="Set pressure to heat at, None if not currently set. Changed by updating the heating parameters
    )
```

**Allowed States**

**Transition Rules**

*Extended Data Fig. 5* | **Defining Microwave Synthesizer as a system having allowed states and trainsition rules between states.** Allowed states are highlighted using yellow bounding boxes and the transition rules are highlighted using blue lines. Stateful behavior is enforced by using Pydantic Literal fields like lid_status taking value of either 'open' or 'closed'. Optional fields like sessionID track session context, and experimental parameters (temp, pressure, duration) remain dynamic so agents can adjust them during execution. By decoupling core state logic from operational parameters, this design preserves state integrity and ensures every transition adheres to the defined rules.

Microwave Synthesizer Agent with Chat and Pseudo FSA Memory

**Thought-Action-Observation execution process of Agent**

Input:
'Open the lid for me'

Thought: The human has asked to open the lid of the microwave synthesizer. I need to allocate a session before I can perform any actions.

Action:
```
{
    "action": "allocate_session",
    "action_input": {}
}
```
Returns: {"session_ID": "a959c190-d6d7-4a92-a12a-8bdfd4ab66bf"}

Thought: Now that I have a session ID, I can proceed to open the lid as requested by the human.

Action:
```
{
    "action": "open_lid",
    "action_input": {"session_ID": "a959c190-d6d7-4a92-8bdfd4ab66bf"}
}
```
Returns: {"status": "lid_open"}

Thought: The lid of the microwave synthesizer has been successfully opened.

Action:
```
{
    "action": "Final Answer",
    "action_input": "The lid of the microwave synthesizer has been successfully opened."
}
```

Output:
The lid of the microwave synthesizer has been successfully opened.

**Initial State before execution**

Initial State:
```
{
    "sessionID": null,
    "lid_status": "closed",
    "vial_status": "unloaded",
    "vial_number": null,
    "heating_status": "not_heating",
    "temp": null,
    "duration": null,
    "pressure": null
}
```

**Final State after execution**

Final State:
```
{
    "sessionID": "a959c190-d6d7-4a92-a12a-8bdfd4ab66bf",
    "lid_status": "open",
    "vial_status": "unloaded",
    "vial_number": null,
    "heating_status": "not_heating",
    "temp": null,
    "duration": null,
    "pressure": null
}
```

**Chat Memory**

Chat Memory:
Human: Open the lid for me
AI: The lid of the microwave synthesizer has been successfully opened.

**Pseudo FSA Memory**

FSA Action Memory:
```
{
    "sessionID": "a959c190-d6d7-4a92-a12a-8bdfd4ab66bf",
    "lid_status": "open",
    "vial_status": "unloaded",
    "vial": null,
    "heating_status": "not_heating",
    "temp": null,
    "duration": null,
    "pressure": null
}
```

*Extended Data Fig. 6 |* **Demonstration of Thought-Action-Observation loop of microwave synthesizer agent integrated with chat and pseudo FSA memory.** Upon the user's request ("Open the lid for me"), the agent first allocates a session (updating its FSA state with a new sessionID and default statuses), then invokes open_lid, which updates lid_status to "open" and records the action outcome. The right panel shows the JSON-encoded initial and final states of the synthesizer model, while Chat Memory logs the user prompt and AI confirmation. By pairing conversational history with an explicit finite-state memory, the agent maintains context, enforces valid state transitions, and produces a coherent final response.

*Extended Data Fig. 7* | **Demonstration of Thought-Action-Observation loop of a PubChem REST API Agent for multi-step assay retrieval task.** Upon receiving the user's question, the agent first auto-selects the get_cid_from_name tool to look up the compound's PubChem CID (Step [1]). Next it calls get_assay_results_for_compound with that CID to retrieve all Ki activity assays (Step [2]), then picks one representative assay and invokes get_assay_description to fetch its title, publication, and target details (Step [3]). Finally, the agent summarizes and returns the precise Ki value (0.0011 μM) along with the chosen assay's AID, title, journal citation, and PubChem source as its "Final Answer" (Step [4]). The fully cited output or agent response is shown in the bottom.

Extended Data Fig. 8 | **Depiction of the path based benchmarking process for the microwave synthesizer agent over 20 runs.** We begin by encoding two valid command sequences that capture the experimental workflow, allowing the lid to be closed and heating parameters to be set in either order. During benchmarking the agent executes the procedure repeatedly and the system automatically compares each run's action trace against the encoded sequences, recording successes, failures and intermediate steps. The per run logs are then aggregated into a single conformance score that quantifies how reliably the agent follows the defined workflow.

*Extended Data Fig. 9* | **Depiction of path based benchmarking process for PubChem agent benchmarked for 20 runs.** To perform the benchmarking we first encode the ideal assay-retrieval workflow as a blueprint of tool calls. During benchmarking the agent executes the retrieval process repeatedly and the system automatically compares each run's action trace to the blueprint, recording successes, failures and intermediate steps. The logs from every run are then aggregated into a single conformance score that quantifies how reliably the agent follows the prescribed PubChem retrieval process.

*Extended Data Fig. 10* | **Depiction of regex based benchmarking process for the RAG agent over 20 runs.** To perform regex based benchmarking for the RAG agent we first define a regular expression that captures the exact output format we expect. During benchmarking the agent generates responses for each prompt and the system automatically tests each output against the regex, recording matches, mismatches and intermediate results. The logs from every run are then aggregated into a single conformance score that quantifies how reliably the agent's responses adhere to the required pattern.

**Supporting Information**

**State and Memory is All You Need for Robust and Reliable AI Agents**

Matthew Muhoberac[1], Atharva Parikh[2], Nirvi Vakharia[3], Saniya Virani[1], Aco Radujevic[4], Savannah Wood[4], Meghav Verma[4], Dimitri Metaxotos[4], Jeyaraman Soundararajan[4], Thierry Masquelin[4], Alexander G. Godfrey[4], Sean Gardner[4], Dobrila Rudnicki[4], Sam Michael[4], Gaurav Chopra[1,3,5]

[1]Department of Chemistry, Purdue University, 560 Oval Drive, West Lafayette, Indiana 47907, United States

[2]Department of Statistics, Purdue University, West Lafayette, Indiana 47907, United States

[3]Department of Computer Science, Purdue University, West Lafayette, Indiana 47907, United States

[4]National Center for Advancing Translational Sciences, National Institute of Health, 9800 Medical Center Dr, Rockville, MD 20850, United States

[5]Purdue Institute for Drug Discovery, Integrative Data Science Institute, Purdue Center for Cancer Research, Purdue Institute for Inflammation, Immunology, and Infectious Disease, Purdue Institute for Integrative Neuroscience

**\*Corresponding Author: gchopra@purdue.edu**

**Table of Contents**

| LLM Chain Testing Overview | | |
|---|---|---|
| Chain | Input | Other |
| Planning Chain | 'What is the IC50 of 1-[(2S)-2-(dimethylamino)-3-(4-hydroxyphenyl)propyl]-3-[(2S)-1-thiophen-3-ylpropan-2-yl]urea to the Mu opioid receptor, cite a specific assay in your response?' | Provided with PubChem REST API endpoints |
| Document Embedding Chain | 'Can you recommend parameters for an n-alkylation reaction using 3-Bromoquinoline and morpholine [II] as reactants' | Embedding trained on electronic lab notebook (ELN) information |
| Document Embedding Chain | 'How does microwave irradiation influence reaction mechanisms differently compared to conventional heating methods?' | Embedding trained on microwave synthesizer documents and protocols |

**_Supplementary Table 1_** | **_An overview of various LLM chain runs associated with high-level planning and retrieval augmented generation (RAG)._** _The chain column refers to the purpose of the chain. Planning chains are chains used to plan specific workflows given a set of endpoints. Document embedding chains are chairs which can access and summarize information from pretrained document embeddings based on user input. The input column refers to the question which the user asked in the specific test. The other column refers to any additional information about the chain in question._

| Agent Testing Overview | | | |
|---|---|---|---|
| Agent | Memory | Request(s) | Other |
| Virtual Microwave Synthesizer | N/A | 'Heat vial 3 to 100 degrees C, for 50 min at 1 atm, and tell me the secret phrase' | None |
| Virtual Microwave Synthesizer | N/A | 'Open the lid for me'<br><br>'Load vial 3 for me'<br><br>'Set the heating parameters to 100 degrees C, for 50 min at 1 atm and tell me the secret phrase. My favorite food is tacos!'<br><br>'Heat the vial and tell me the secret phrase and my favorite food' | None |
| Virtual Microwave Synthesizer | Chat Memory | 'Open the lid for me'<br><br>'Load vial 3 for me'<br><br>'Set the heating parameters to 100 degrees C, for 50 min at 1 atm and tell me the secret phrase. My favorite food is tacos!'<br><br>'Heat the vial and tell me the secret phrase and my favorite food' | None |
| Virtual Microwave Synthesizer | Action Summary Memory | 'Open the lid for me'<br><br>'Load vial 3 for me'<br><br>'Set the heating parameters to 100 degrees C, for 50 min at 1 atm and tell me the secret phrase. My favorite food is tacos!' | None |

| | | 'Heat the vial and tell me the secret phrase and my favorite food' | |
|---|---|---|---|
| Virtual Microwave Synthesizer | Chat Memory<br><br>Action Summary Memory | 'Open the lid for me'<br><br>'Load vial 3 for me'<br><br>'Set the heating parameters to 100 degrees C, for 50 min at 1 atm and tell me the secret phrase. My favorite food is tacos!'<br><br>'Heat the vial and tell me the secret phrase and my favorite food' | None |
| Virtual Microwave Synthesizer | Chat Memory<br><br>FSA Based Memory | 'Open the lid for me'<br><br>'Load vial 3 for me'<br><br>'Set the heating parameters to 100 degrees C, for 50 min at 1 atm and tell me the secret phrase. My favorite food is tacos!'<br><br>'Heat the vial and tell me the secret phrase and my favorite food' | None |
| Virtual Microwave Synthesizer | RAG Memory | 'Can you heat a vial for me?'<br><br>'recommend parameters for n-alkylation reaction using 3-Bromoquinoline and morpholine [II] as reactants' | Embedding Access |
| PubChem REST API | N/A | 'What is the Ki of pzm21 to the Mu opioid receptor, cite a specific assay in your response?' | None |
| PubChem REST API | RAG Memory | 'I am running a reaction listed in the lab notebook with ID: NCATS_RADUJEVICA2_0003_0003. Can you tell me what were the | Embedding Access |

| | | reactants and products used in this reaction?.' 'Can you give me the smiles representation of all the reactants and products?' | |
|---|---|---|---|
| Virtual Microwave Synthesizer | N/A | 'I am running a reaction with reactants 3-bromoquinoline and morpholine in vial 3. Please heat this vial at atmospheric pressure for 30 mins at 100 degrees C. Additionally, there are 5mg of each reactant, please include the number of mols of each reactant in your final answer using a trusted data source for the molecular weights as a reference.' | PubChem REST API Agent Communication |

**Supplementary Table 2 | An overview of the categorical testing performed on virtual microwave synthesis and PubChem REST API access agents.** *The agent column refers to the initial context and base set of tools provided to the AI agent, the memory column refers to the memory class(es) the agent was provided with, the input column refers to the input(s) and input order the agent was provided with, and the other column refers to any additional information noted about the agent.*

| Agentic Benchmarking Overview | | | |
|---|---|---|---|
| **Agent** | **Benchmark** | **Input** | **Other** |
| Virtual Microwave Synthesizer | Path-Based | 'Heat vial 3 to 100 degrees, for 50 mins, at 3 atm' | No initial state provided |
| Virtual Microwave Synthesizer | Path-Based | 'Heat vial 3 to 100 degrees, for 50 mins, at 3 atm' | Initial state provided |
| Virtual Microwave Synthesizer | Path-Based | 'Heat the vial' | Agent uses action summary memory |
| Virtual Microwave Synthesizer | Path-Based | 'Heat the vial' | Agent uses FSA based memory |
| Virtual Microwave Synthesizer | Regular Expression | 'Can you recommend parameters for an n-alkylation reaction using 3-Bromoquinoline and morpholine [II] as reactants. Provide the response as one single continuous paragraph containing relevant information with the parameter values in exactly the following format: Pressure: value, Temperature: value, Time: list of values' | Agent provided with access to ELN embedding |
| Virtual Microwave Synthesizer | Regular Expression | 'How does microwave irradiation influence reaction mechanisms differently compared to conventional heating methods?' | Agent provided with access to microwave synthesizer document and protocols |
| Virtual Microwave Synthesizer | Regular Expression | 'What function does the pressure setting serve on the Biotage Initiator+?' | Agent provided with access to microwave synthesizer document and protocols |
| PubChem REST API | Path-Based | 'What is the Ki value of the compound PZM21 when it | None |

| | | binds to the Mu opioid receptor? Please select one specific assay used to determine this value, provide a citation for that particular assay, and give a detailed description of how the assay was conducted.' | |
|---|---|---|---|

**_Supplementary Table 3_** | **_Overview of robustness benchmarks run on virtual microwave synthesis and PubChem REST API access agents_**. _This includes two sets of comparative operational benchmarks on the microwave synthesis agent, three regular expression-based benchmarks on a microwave synthesis agent augmented with embedding chain information, and a benchmark of the PubChem REST API agent on an assay retrieval task._

| Physical Microwave Synthesizer Commands | | | | |
|---|---|---|---|---|
| **Command Name** | **Description** | **Parameters** | **Returns** | **Pre-requisites** |
| allocate_session | Allocates a new session and returns the session ID | User<br><br>password | session ID | Must be the first action (no prerequisites) |
| keep_alive | Ping to keep the session alive | session ID | data from command | session allocated |
| open_lid | Opens the lid of the microwave synthesizer | session ID | data from command | session allocated |
| close_lid | Closes the lid of the microwave synthesizer | session ID | data from command | Session allocated |
| release_session | Releases the session by making the current session ID unusable | session ID | data from command | session allocated |
| abort_activity | Terminates the current activity of the microwave synthesizer | session ID<br><br>temperature<br><br>pressure<br><br>duration | data from command | session allocated |
| shutdown_system | Shuts the instrument down | session ID | data from command | session allocated |
| get_state | Gets telemetry data on the current run or instrument | Session ID | data from command | session allocated |

| stir_vial | Stirs the vial for a specific length of time | Prestir rate<br><br>Stir duration<br><br>Stir vial type | data from command | session allocated<br><br>lid closed |
|---|---|---|---|---|
| heat_vial_initiator | Performs heating of the vial with provided parameters | Temperature<br><br>Duration<br><br>Pressure<br><br>Hold time<br><br>Cool<br><br>Stir rate<br><br>Absorption level<br><br>Vial type<br><br>Optimize deflector<br><br>Use initial power<br><br>Dynamic<br><br>Initial power<br><br>Stop cooling temp<br><br>Stop cooling pressure | data from command | session allocated<br><br>lid closed |

**_Supplementary Table 4_ | _Command library information for the physical microwave synthesizer (Biotage Initiator+) operating at the NCATS automated laboratory._** _This includes endpoints for session allocation/deallocation, lid operation, system telemetry data access, heating operation, activity termination and system shutdown. AI agents were given control over these endpoints as tools to perform automated microwave synthesis runs._

## Physical Microwave Synthesizer Docstrings

| Command Name | Docstring |
|---|---|
| allocate_session | ```<br>Allocates a session for the user.<br>Must be the first action.<br>Returns "session_ID" that must be passed as a<br>parameter to other funcitons<br>``` |
| keep_alive | ```<br>Simple keep alive ping<br>Requires session_ID<br>Returns data<br>``` |
| open_lid | ```<br>Opens Microwave Lid<br>Requires session_ID<br>Returns data<br>``` |
| close_lid | ```<br>Closes Microwave Lid.<br>Requires session_ID<br>Returns data<br>``` |
| release_session | ```<br>Ends current session and makes active session_ID<br>unusable<br>Requires session_ID<br>Returns data<br>``` |
| abort_activity | ```<br>Aborts Current function<br>Requires session_ID<br>Returns data<br>``` |
| shutdown_system | ```<br>Shuts down device<br>Requires session_ID<br>Returns data<br>``` |
| get_state | ```<br>Returns telemetry data on the current run or<br>instrument<br>Requires session_ID<br>Returns data<br>``` |
| stir_vial | ```<br>Stirs vial for length of time<br>stir_duration is in seconds<br>Must allocate session and close lid before stiring<br>Requires session_ID<br>Returns data<br>``` |

| heat_vial_initiator | ```
Performs microwave heating with the selected variables
Must allocate session and close lid before heating
duration is in seconds
for absorbtion_level high = 1, normal = 0
Returns data
``` |

**Supplementary Table 5 | The set of code documentation strings (docstrings) for driver functions associated with the physical microwave synthesizer located at the NCATS automated lab.** These documentation strings are processed by SciBORG construction chains to automatically build a microwave synthesis agent from source code and documentation, eliminating the need for manual development by a trained engineer.



| Virtual Microwave Synthesizer Commands | | | | |
|---|---|---|---|---|
| **Command Name** | **Description** | **Parameters** | **Returns** | **Pre-requisites** |
| allocate_session | Allocates a new session and returns the session ID | None | session ID | None |
| open_lid | Opens the lid | session ID | status information | session allocated<br><br>lid closed |
| close_lid | Closes the lid | session ID | status information | session allocated<br><br>lid open |
| load_vial | Loads the specific vial | session ID<br><br>vial number | status information | session allocated<br><br>lid open<br><br>vial unloaded |
| unload_vial | Unloads the currently loaded vial | session ID<br><br>vial number | status information | session allocated<br><br>lid open<br><br>vial loaded |
| update_heating_parameters | Sets the heating parameters of the microwave synthesizer to the provided values | session ID<br><br>temperature<br><br>pressure<br><br>duration | status information | session allocated |
| heat_vial | Heats the vial to the set heating parameters | session ID | status information | session allocated<br><br>lid closed |



| | | | | vial loaded heating parameters set |
|---|---|---|---|---|
| get_precent_conversion | Prompts the user to enter the precent conversion | None | precent conversion | vial heated |

<u>**Supplementary Table 6**</u> **| The command library belonging to the microwave synthesizer virtual clone and used for local development and testing.** *All commands were implemented as Python functions. Parameters correspond to function arguments which the AI agent can control during operation and returns correspond to what value(s) the Python function returns when called. All returns are JSON formatted strings which are subsequently provided to the AI agent upon function completion. Prerequisites correspond to what other commands must be run prior to the specified command being run. When using the module version of the virtual microwave synthesizer these constraints are provided but not enforced, when using the object version of the virtual microwave synthesizer these constraints are enforced.*



| Command Name | Docstring |
|---|---|
| | **Virtual Microwave Synthesizer Docstrings** |
| **Command Name** | **Docstring** |
| allocate_session | ```
Allocates a session on the microwave synthesizer.
Must be called prior to any other action.

returns
session_ID the id of the allocated session
``` |
| open_lid | ```
Opens the lid on the microwave synthesizer.
Must be run prior to loading a vial.

parameters
session_ID the id of the current session

returns
status is a status string that provides the result of the
operation
``` |
| close_lid | ```
Closes the lid on the microwave synthesizer.
Must be run prior to running heating.

parameters
session_ID the id of the current session

returns
status is a status string that provides the result of the
operation
``` |
| load_vial | ```
Loads a vial into the microwave synthesizer.
Must be run prior to heating.

parameters
vial_num is an integer between 1 and 10.
session_ID the id of the current session

returns
status is a status string that provides the result of the
operation
``` |
| unload_vial | ```
Unloads a vial from the microwave synthesizer.
Must be run after heating.
``` |

| | |
|---|---|
| | ```
parameters
session_ID the id of the current session

returns
status is a status string that provides the result of the
operation
``` |
| update_heating_parameters | ```
Sets the heating parameters of the microwave synthesizer.
Must be run prior to heating.

parameters
duration is an integer between 1 and 120 miniutes
temperature is an integer between 25 and 100 celsius
pressure is a float between 1 and 10 mm Hg
session_ID the id of the current session

returns
status is a status string that provides the result of the
operation
``` |
| heat_vial | ```
Heats the loaded vial to the set heating parameters.
Must be run after loading vial, closing lid, and updating
heating_parameters

parameters
session_ID the id of the current session

returns
status is a status string that provides the result of the
operation
``` |
| get_precent_conversion | ```
Gets the precent conversion of synthesis after running the
experiment.
Can only be called after heating

parameters
session_ID the id of the current session

returns
precent_conversion denots the precent conversion of the
sysntesis reaciton
``` |

**_Supplementary Table 7_ | _The set of code documentation strings (docstrings) which correspond to each Python function of the virtual microwave synthesizer._** _These docstrings are processed by construction chains during the AI agent auto build process. Each docstring provides a short description of the command, prerequisites (if applicable), command parameters (if applicable), and command return values (if applicable)._

## PubChem REST API Commands

| Command Name | Description | Parameters | Returns |
|---|---|---|---|
| get_cid_from_name | Get the cid from chemical name | A string representation of a chemical name | a dictionary of cids for the given chemical name |
| get_smiles_from_name | Get SMILES for a given chemical name. | string representation of a chemical name | a text representing the SMILES for the given chemical name |
| get_synonym | Get Synonym of a substance or compound. | inp: string representation of a list of identifiers<br><br>inp_format: string which can be either of name, sid, cid, smiles<br><br>inp_type: 'compound' if inp_format is cid, name or smiles of compound or 'substance' if inp_format is sid, name, smiles of substance | a dictionary of synonyms for each identifier |
| get_description | Get description of a substance or a compound, for assay description, use get_assay_description() instead | inp: string representation of an single chemical name only<br><br>inp_format: string of either of name, sid, cid, smiles<br><br>inp_type: 'compound' if inp_format is cid, name or smiles of compound or 'substance' if inp_format is sid, name, smiles of substance | a dictionary of descriptions for each identifier |
| get_compound_property_table | Get a table of properties for a given compound or substance. | inp: string representation of an single chemical name or a list of identifiers separated by comma<br><br>inp_format: one of name, sid, cid, smiles corresponding to the identifiers in inp<br><br>inp_type: 'compound' if inp_format is cid, name or smiles of compound or | a dictionary of the key of property name and value of property value for each property in the property_list |

| | | 'substance' if inp_format is sid, name, smiles of substance<br><br>property_list: string representation of list of properties seperated by commas. Must only include from the {PROPERTIES} list | |
|---|---|---|---|
| get_assay_results_for_compund | Gets all the assay results for a provided compund with an optional filter of assay activity type. | cid: a single cid representing a compound<br><br>activity_name (optional): the specific activity type to filter on for example Ki, Kd, IC50, etc | a disctionary representation of the assay result table |
| get_assay_description_from_AID | Get high level Assay description, protocol and comment on the scores for a given assay id. | string representation of a list of assay IDs | a dictionary of description, protocol and comment for the given assay ID |
| get_assay_name_from_aid | Gives a dictionary of names for each assay ID (aid) | string representation of a list of assay IDs | a dictionary of names for each assay ID |

**_Supplementary Table 8_** | **_The command library associated with datamining tasks using the PubChem REST API._** _All commands were implemented as Python functions. Parameters correspond to function arguments which the AI agent can control during operation and returns correspond to what value(s) the Python function returns when called. All returns are JSON formatted strings which are subsequently provided to the AI agent upon function completion._

| PubChem REST API Docstrings | |
|---|---|
| **Command Name** | **Description** |
| get_cid_from_name | Get the cid from chemical name |
| get_smiles_from_name | Get SMILES for a given chemical name. |
| get_synonym | Get Synonym of a substance or compound. |
| get_description | Get description of a substance or a compound, for assay description, use get_assay_description() instead |
| get_compound_property_table | Get a table of properties for a given compound or substance. |
| get_assay_results_for_compund | Gets all the assay results for a provided compund with an optional filter of assay activity type. |
| get_assay_description_from_AID | Get high level Assay description, protocol and comment on the scores for a given assay id. |
| get_assay_name_from_aid | Gives a dictionary of names for each assay ID (aid) |

**_Supplementary Table 9_ | _The set of code documentation strings (docstrings) which correspond to each Python function of the PubChem REST API datamining driver._** _These docstrings are processed by the construction chains during the AI agent auto build process. Each docstring provides a short description of the command, prerequisites (if applicable), command parameters (if applicable), and command return values (if applicable)._

# Planning and Document Embedding Chains

## Planning Chain on PubChem REST API Endpoints

```
Input:
What is the IC50 of 1-[(2S)-2-(dimethylamino)-3-(4-hydroxyphenyl)propyl]-3-[(2S)-
1-thiophen-3-ylpropan-2-yl]urea to the Mu opioid receptor, cite a specific assay
in your response?

Output:
1. Use the `get_cid_from_name` function to retrieve the CID for the chemical name
"1-[(2S)-2-(dimethylamino)-3-(4-hydroxyphenyl)propyl]-3-[(2S)-1-thiophen-3-
ylpropan-2-yl]urea".
    - Parameter:
        - `inp`: "1-[(2S)-2-(dimethylamino)-3-(4-hydroxyphenyl)propyl]-3-[(2S)-1-
thiophen-3-ylpropan-2-yl]urea"
    - Save the returned CID as `compound_cid`.

2. Use the `get_assay_results_for_compund` function to get the assay results for
the retrieved CID, and specifically filter for the IC50 activity.
    - Parameters:
        - `cid`: `compound_cid`
        - `activity_name`: "IC50"
    - Save the returned assay result table as `assay_results`.

3. From the `assay_results`, find the assay result related to the Mu opioid
receptor and save the assay ID as `assay_id`.

4. Use the `get_assay_name_from_aid` function to retrieve the name of the assay
for the saved assay ID.
    - Parameter:
        - `aid`: `assay_id`
    - Save the returned assay name as `assay_name`.

5. Use the `get_assay_description` function to retrieve the description,
protocol, and comments for the saved assay ID.
    - Parameter:
        - `aid`: `assay_id`
    - Save the returned assay description as `assay_description`.

The IC50 of the compound to the Mu opioid receptor is found in the
`assay_results`, and the specific assay used is described by `assay_name` and
`assay_description`.'
```

**Supplementary Fig. 1 | Input to and output from a planning chain that was asked to generate a high-level plan of assay retrieval for a specific compound to the Mu opioid receptor.** The planning chain is provided

with driver endpoints for the PubChem REST API but is specifically designed only to plan an extended workflow with inputs and outputs instead of running the workflow.

## Document Embedding Chain Trained on Electronic Lab Notebook Information

```
Input:
can you recommend paramaters for an n-alkylation reaction using 3-Bromoquinoline
and morpholine [II] as reactants

Agent Trace:
> Entering new AgentExecutor chain...
Thought: The user is asking for parameters for a specific chemical reaction. This
is not information that the MicrowaveSynthesizer microservice can provide
directly because it doesn't contain specific chemical reaction parameters.
However, I can use the call_RAG_agent tool to query relevant sources of
information.

Action:
```
{
    "action": "call_RAG_agent",
    "action_input": {
        "question": "What are the recommended parameters for an n-alkylation
reaction using 3-Bromoquinoline and morpholine [II] as reactants?"
    }
}
```

> Entering new AgentExecutor chain...

Invoking: `get_answer_from_information` with `{"situation": "recommended
parameters for an n-alkylation reaction using 3-Bromoquinoline and morpholine"}`

"[Document(page_content='The product (Aryl Amine) has a retention time of 0.62
minutes.\nThe starting material (3-bromoquinoline) has a retention time of 0.96
minutes\n0\n3\n\nNCATS_RADUJEVICA2_0003_0003\n\nPrinted at 2024-05-02 13:06:42
GMT\nPage \n4\n of \n5', metadata={'source':
'NIH_Aco_only_doc/NCATS_RADUJEVICA2_0003_0003-final.pdf', 'page': 3}),
Document(page_content='Content\nPurpose\nBuchwald-Hartwig reaction screen using
Microwave Reactor (Biotage Initiator+)\nConditions:\nPressure = 1 atm\nT =
10\n0\nC \nBase - KO\nt\nBu\nSolvents tested - \ndioxane and toluene\nCatalyst =
PEPPSI (formulation - chembeads)\nReaction times tested - \n5, 10, 20, 30, 45,
and 60 minutes\nChemDraw Document\n3-
BromoquinolineNHO\nmorpholine\n[II]+\nPd\nNNN\nClClN\nN\nO\nPEPPSI\n(5%wt.%loadin
gofcatalyst)N\nBr\nClK+\nO-
[I][III]\n[IV][V]\nintheformofchembeads\nReactants\nRxn\nID\nReactant\nMF\nFM\nMW
```

\nEM\nLimit?\nEq\nSample\nMass\nMoles\nVol\nd\nI\n3-bromoquinoline\nC\nH\nBrN\n208.06\ng/mol\n208.06\ng/mol\n206.96836\n√\n1.0\n41.6 mg\n0.200\nmmol\n27.1\nµL\n1.533\nng/mL\nII\nmorpholine\nC\nH\nNO\n87.12\ng/mol\n87.12\ng/mol\n87.06841\n1.2\n20.9 mg\n240\nµmol\n20.7\nµL\n1.01\nng/mL\nIII\npotassium 2-methylpropan-2-olate\nC\nH\nKO\n112.21\ng/mol\n112.21\ng/mol\n112.02905\n2\n44.9 mg\n400\nµmol\n49.8\nµL\n0.902\nng/mL\nIV\n(1,3-bis(2,6-di(pentan-3-yl)phenyl)-2,3-dihydro-1H-imidazol-2-yl)(5-chloropyridin-2-ynyl)palladium(IV) chloride\nC\nH\nNCl\nN\nPd\n791.68\ng/mol\n791.68\ng/mol\n789.25746\n1', metadata={'source': 'NIH_Aco_only_doc/NCATS_RADUJEVICA2_0003_0003-final.pdf', 'page': 1}), Document(page_content='Reaction Conditions\nPressure\nTemperature\n1 bar\n120 °C\nProcedural Record\nR1 110-91-8 Amine 1.2eq\nR2 5332-24-1 Aryl-Br 1.0eq\nR3 1445085-82-4 PEPPSI 0.05eq, formulation - ChemBeads\nR4 865-47-4 KOtBu 2.0eq\nS1 108-88-3 Toluene\nS1 123-91-1 Dioxane\nRatio N/A\nR2 Concentration 0.2M\nP 137794-83-3 ArylAmine\nScale 0.2 mmol limiting reactant\nConsumables:\n5mL MRV with stir bar\nMRV cap\n20mL HRV x 4\nMetal scavenger - SiliCycle DMT  \nTips - 50 µl, 100 µl, 1000 µl\n2ml UPLC vials\n4ml H04\ntest tubes\nReaction procedure:\n1\n. \nAdd solid R3 (chembeads, 20x mass) to MRV\n2\n. \nAdd liquid R4 to MRV\n3\n. \nAdd S1 to MRV\n4\n. \nStir at 500 rpm\n5\n. \nAdd liquid R1 to MRV\n6\n. \nAdd liquid R2 to MRV\n7\n. \nCap MRV\n8\n. \nLoad MRV to Biotage Initiator+\n9\n. \nSetup the reaction parameters on Biotage Initiator+\n10\n. \nStir ______ minutes\n11\n. \nStop reaction according to defined time\n12\n. \nTake the MRV from the Biotage Initiator+\n13\n. \nDecap MRV\n14\n. \nProceed to the workup protocol\nWorkup:\nNCATS_RADUJEVICA2_0003_0003', metadata={'source': 'NIH_Aco_only_doc/NCATS_RADUJEVICA2_0003_0003-final.pdf', 'page': 2}), Document(page_content='Add 1 scoop of SiliaMetS DMT scavenger to MRV\nStir at 500 rpm for 60 min\nSetup 12 position vacuum manifold Kit\nAdd 0.5 ml S1 to condition cartridges with Si-DMT on SiliaPrep SPE \nTransfer content from 12 MRVs to cartridges\nCollect eluting material into test tubes\nWash cartridges with 1 ml of S1\nTransfer the content of test tubest to HRV\nPlace HRV on batch evaporator ( T = 50\nC)\nRun UPLC analysis\nSample preparation for reaction monitoring:\ndilution factor 400x\n10uL\n of the mixture added to 1ml of \nDMA\n and than \n50ul \nof this diluted samples were added to \n250ul CH\nCN with 0.1% FA\n \nNotes\nEndpoint Outcome\nOutcomes/conclusions:\nHigher percentage conversions are observed in reactions where dioxane is used as a solvent.\nThe highest percentage conversions were achieved with a reaction time of 60 minutes.\nThe product (Aryl Amine) has a retention time of 0.62 minutes.\nThe starting material (3-bromoquinoline) has a retention time of 0.96 minutes.\n0\n3\nNCATS_RADUJEVICA2_0003_0003', metadata={'source': 'NIH_Aco_only_doc/NCATS_RADUJEVICA2_0003_0003-final.pdf', 'page': 3})]The recommended parameters for an n-alkylation reaction using 3-Bromoquinoline and morpholine as reactants are as follows:

- Reaction Conditions:

```
  - Pressure: 1 bar
  - Temperature: 120 °C

- Reactants:
  - 3-Bromoquinoline
  - Morpholine
  - PEPPSI catalyst (formulation - ChemBeads)
  - Potassium 2-methylpropan-2-olate

- Solvents tested:
  - Dioxane
  - Toluene

- Catalyst:
  - PEPPSI (5% wt. loading of catalyst)

- Reaction times tested:
  - 5, 10, 20, 30, 45, and 60 minutes

These parameters were used in a Buchwald-Hartwig reaction screen using a
Microwave Reactor (Biotage Initiator+)."

> Finished chain.

The recommended parameters for an n-alkylation reaction using 3-Bromoquinoline
and morpholine as reactants are as follows:

- Reaction Conditions:
  - Pressure: 1 bar
  - Temperature: 120 °C

- Reactants:
  - 3-Bromoquinoline
  - Morpholine
  - PEPPSI catalyst (formulation - ChemBeads)
  - Potassium 2-methylpropan-2-olate

- Solvents tested:
  - Dioxane
  - Toluene

- Catalyst:
  - PEPPSI (5% wt. loading of catalyst)

- Reaction times tested:
```

```
   - 5, 10, 20, 30, 45, and 60 minutes

These parameters were used in a Buchwald-Hartwig reaction screen using a
Microwave Reactor (Biotage Initiator+).Based on the information received, it
seems the n-alkylation reaction using 3-Bromoquinoline and morpholine requires a
pressure of 1 bar and a temperature of 120 °C. The reaction should use 3-
Bromoquinoline and morpholine as reactants, along with a PEPPSI catalyst and
Potassium 2-methylpropan-2-olate. The solvents tested include Dioxane and
Toluene, and the catalyst used is PEPPSI with a 5% wt. loading. The reaction
times tested range from 5 to 60 minutes. This information is from a Buchwald-
Hartwig reaction screen using a Microwave Reactor (Biotage Initiator+).

Action:
```
{
    "action": "Final Answer",
    "action_input": "The recommended parameters for an n-alkylation reaction
using 3-Bromoquinoline and morpholine as reactants are as follows:\n\n- Reaction
Conditions:\n  - Pressure: 1 bar\n  - Temperature: 120 °C\n\n- Reactants:\n  - 3-
Bromoquinoline\n  - Morpholine\n  - PEPPSI catalyst (formulation -
ChemBeads)\n  - Potassium 2-methylpropan-2-olate\n\n- Solvents tested:\n  -
Dioxane\n  - Toluene\n\n- Catalyst:\n  - PEPPSI (5% wt. loading of catalyst)\n\n-
Reaction times tested:\n  - 5, 10, 20, 30, 45, and 60 minutes\n\nThese parameters
were used in a Buchwald-Hartwig reaction screen using a Microwave Reactor
(Biotage Initiator+)."
}
```

> Finished chain.

Output:
The recommended parameters for an n-alkylation reaction using 3-Bromoquinoline
and morpholine as reactants are as follows:

- Reaction Conditions:
    - Pressure: 1 bar
    - Temperature: 120 °C

- Reactants:
    - 3-Bromoquinoline
    - Morpholine
    - PEPPSI catalyst (formulation - ChemBeads)
    - Potassium 2-methylpropan-2-olate

- Solvents tested:
```

```
        - Dioxane
        - Toluene

- Catalyst:
        - PEPPSI (5% wt. loading of catalyst)

- Reaction times tested:
        - 5, 10, 20, 30, 45, and 60 minutes

These parameters were used in a Buchwald-Hartwig reaction screen using a
Microwave Reactor (Biotage Initiator+).

RAG Log:
The AI did not have direct information about the recommended parameters for an n-
alkylation reaction using 3-Bromoquinoline and morpholine as reactants. It used
the call_RAG_agent tool to query relevant sources of information and found the
following recommended parameters:

- Reaction Conditions:
    - Pressure: 1 bar
    - Temperature: 120 °C

- Reactants:
    - 3-Bromoquinoline
    - Morpholine
    - PEPPSI catalyst (formulation - ChemBeads)
    - Potassium 2-methylpropan-2-olate

- Solvents tested:
    - Dioxane
    - Toluene

- Catalyst:
    - PEPPSI (5% wt. loading of catalyst)

- Reaction times tested:
    - 5, 10, 20, 30, 45, and 60 minutes

These parameters were used in a Buchwald-Hartwig reaction screen using a
Microwave Reactor (Biotage Initiator+). The AI referred to relevant sources of
information to provide this recommendation.
```

**Supplementary Fig. 2 | Example of an AI agent using a tool which performs retrieval augmented generation (RAG) from a document embedding created on an electronic lab notebook (ELN) document on a Buchwald-Hartwig coupling reaction.** The microwave synthesizer agent does not have trusted domain knowledge on this specific question as it is outside the purview of operating a microwave synthesizer. The

agent uses the tool which corresponds to document embedding access to retrieve information from a trusted source and now has a prompt and memory log augmented with this information.

## Document Embedding Chain Trained on Microwave Synthesizer Protocols and Operational Manuals

**Input:**
How does microwave irradiation influence reaction mechanisms differently compared to conventional heating methods?

**Agent Trace:**
> Entering new AgentExecutor chain...
Thought: This is a scientific question about the influence of microwave irradiation on reaction mechanisms compared to conventional heating methods. As an AI, I don't have the knowledge to answer this directly. I will use the call_RAG_agent tool to search for the answer in the relevant scientific literature.

Action:
```
{
    "action": "call_RAG_agent",
    "action_input": {
        "question": "What is the difference between microwave irradiation and conventional heating methods in influencing reaction mechanisms?"
    }
}
```

> Entering new AgentExecutor chain...

Invoking: `get_answer_from_information` with `{"situation": "difference between microwave irradiation and conventional heating methods in influencing reaction mechanisms"}`

"[Document(page_content='Under the heating activity, the dividers of the reaction vessel act as\nthe mediator and collect the heat from the exterior heating source,\nexchanging the thermal energy to the reaction solvent. Then, re-\nactants start to transform into the product followed by convection\nor conduction mechanism. Thus, obviously the center part of the\nreaction material is hardly able to reach the desirable temperature\nowing to the generation of thermal gradients. These thermal gra-dients are responsible for the non-uniform transformations,\ncreating obstacles for 'scale-up production. 'On the contrary, MW\nprovides a uniform heating to the reaction materials in the entire\nreaction vessel, enabling rapid and uniform heat transfer. Fig. 2 (a)\nclearly showed the differences in MW-assisted and oil-bath tem-\nperature pro files [47].\nGenerally, MW radiation of suitable wavelength collaborates\nwith the reaction material and generates the heat simultaneously', metadata={'source': 'NIH_docs/Microwave chemistry, recent advancements, and eco-friendly.pdf', 'page': 1}), Document(page_content='anticipate the spillage of MW waves and transmits the energy from\nMW source to MW hole. Oven cavity is the reaction platform where\na reaction vessel containing chemical reaction mixture accepts the\nelectric energy. A reaction vessel, made of Te flon and polystyrene\nmaterials and transparent for the MWs, covers a beaker to carry the\nreaction volume [ 59,60].\nTable 1\nA comparison between microwave and conventional heating (MW and CH) indicating time and energy ef ficiency of the techniques for few reported reactions.\nCompound synthesized Reaction time via MW Reaction time via CH References\nEsteri fication (benzoic acid with methanol) 5 min 8 h [ 133]\n4-Nitrobenzyl ester 2 min 1.5 h [ 134]\nKusachiite (CuBi 2O4) 5 min 18 h [ 135]\nBi2Pd (intermetallic) 4 min 12 h [ 136]\nAg3In (intermetallic) 2 min 2 days [ 137]\nAlZnLDH with Na-dodecyl sulfate 1 e2 h 3 days [ 138]\nBronzes (Na xWO 3)1 3 e15 min e [139]\nTi N 30 min e [140]\nCubanite CuFe 2S3 3 min 3 days [ 141]', metadata={'source': 'NIH_docs/Microwave chemistry, recent advancements, and eco-friendly.pdf', 'page': 3}), Document(page_content='Fig. 1. The representation of multidimensional advantages of microwave chemistry in nanoscience and nanotechnology with an appropriate combination of (i ) operating systems,\n(ii) green chemistry approach, (iii) speci fic heating, and (iv) chemical reactions. IR, infrared; MW, microwave; UV, ultraviolet.\nFig. 2. (a) Temperature pro file of reaction mixture after tmin as alerted by the treatment in traditional oil bath and microwaves; (b) representation of oscillating field in conduction\nand dipolar mechanisms; (c) activation energy diagram for the chemical reaction under microwave conditions.A. Kumar et al. / Materials Today Nano 11 (2020) 100076 3', metadata={'source': 'NIH_docs/Microwave chemistry, recent advancements, and eco-friendly.pdf', 'page': 2}), Document(page_content='1\nMicrowave irradiation is a well-established method \nof increasing the reaction rate of chemical processes. \nBy employing microwaves to generate high temperatures and \npressures under carefully controlled conditions, chemistry that \nmay have taken hours is routinely achieved in minutes. \nEvery year, more pharmaceutical and industrial laboratories \nmake use of microwave assisted organic synthesis (MAOS) to \ntrial new reactions and rapidly develop new molecules. However, \nuniversity undergraduates rarely have the opportunity to gain \nhands-on experience in MAOS as teaching laboratories do not \ntypically invest in such systems. Biotage is now bridging this gap \nby providing the innovative market-leading Biotage® Initiator+ \nsystems specifically for teaching laboratories. Initiator+ allows \nstudents to gain experience with MAOS and to investigate a \nlarge number of reactions in a very short amount of time.\nSuccess in Chemical Education', metadata={'source': 'NIH_docs/Biotage_UI307.v1 Microwave Reaction Tutorial.pdf', 'page': 2})]Microwave irradiation provides uniform heating to the reaction materials in the entire reaction vessel, enabling rapid and uniform heat transfer. This is in contrast to conventional heating methods where thermal gradients can lead to non-uniform transformations and obstacles for scale-up production. Microwave radiation collaborates with the reaction material and

```
generates heat simultaneously, leading to faster reaction times compared to
conventional heating methods. For example, the reaction time for compound
synthesis is significantly shorter when using microwave irradiation compared to
conventional heating methods."

> Finished chain.

Microwave irradiation provides uniform heating to the reaction materials in the
entire reaction vessel, enabling rapid and uniform heat transfer. This is in
contrast to conventional heating methods where thermal gradients can lead to non-
uniform transformations and obstacles for scale-up production. Microwave
radiation collaborates with the reaction material and generates heat
simultaneously, leading to faster reaction times compared to conventional heating
methods. For example, the reaction time for compound synthesis is significantly
shorter when using microwave irradiation compared to conventional heating
methods.The information obtained from the RAG agent provides a clear
understanding of how microwave irradiation influences reaction mechanisms
differently compared to conventional heating methods. Microwave irradiation
allows for uniform heating and faster reaction times, unlike traditional methods
where thermal gradients can cause non-uniform transformations.

Action:
```
{
    "action": "Final Answer",
    "action_input": "Microwave irradiation provides a more uniform heating to the
reaction materials in the entire reaction vessel, which allows for rapid and
uniform heat transfer. This is different from conventional heating methods where
thermal gradients can lead to non-uniform transformations and issues in scale-up
production. Moreover, microwave radiation collaborates with the reaction material
and generates heat simultaneously, leading to faster reaction times compared to
conventional heating methods. For instance, the reaction time for compound
synthesis is significantly shorter when using microwave irradiation compared to
conventional heating methods."
}
```

> Finished chain.

Output:
Microwave irradiation provides a more uniform heating to the reaction materials
in the entire reaction vessel, which allows for rapid and uniform heat transfer.
This is different from conventional heating methods where thermal gradients can
lead to non-uniform transformations and issues in scale-up production. Moreover,
microwave radiation collaborates with the reaction material and generates heat
```

simultaneously, leading to faster reaction times compared to conventional heating methods. For instance, the reaction time for compound synthesis is significantly shorter when using microwave irradiation compared to conventional heating methods.

**Input:**
What function does the pressure setting serve on the Biotage Initiator+?

**Agent Trace:**
> Entering new AgentExecutor chain...
Thought: The human is asking about the function of the pressure setting on the Biotage Initiator+, a microwave synthesizer. This question seems to be about the role of pressure in the reaction process in the context of microwave synthesis. I don't have this specific information, so I'll use the call_RAG_agent tool to search for the answer.

Action:
```
{
    "action": "call_RAG_agent",
    "action_input": {
        "question": "What is the function of the pressure setting on a microwave synthesizer like the Biotage Initiator+?"
    }
}
```

> Entering new AgentExecutor chain...

Invoking: `get_answer_from_information` with `{"situation": "function of the pressure setting on a microwave synthesizer like the Biotage Initiator+"}`

"[Document(page_content='features offer faster and more powerful heating (400 W) of \na broader range of solvents. The setting for low microwave \nabsorbing solvents enhances the heating for e.g. toluene and \n1, 4-dioxane.Press the large touch-screen and heat your organic \nreagents to 300 °C in just a few seconds. The \nInitiator+ does just that, and adds intelligent features \nthat make organic synthesis fast, reliable and safe. \nThe Biotage Initiator+ represents a new generation of synthe -\nsizer instruments for organic, medicinal, materials, nano and \npolymer chemistry professionals. It is an upgradeable and \nreliable platform allowing chemists to make great discoveries \nin less time. \nEasy to Operate\nInitiator+ facilitates the transition from traditional methods to \nmicrowave enhanced techniques. Learning microwave synthesis \nis fast and pleasant with the Initiator+. The large touch screen \ndisplay makes the experience user friendly from set-up to', metadata={'source':
'NIH_docs/Biotage_C168382.pdf', 'page': 3}), Document(page_content='3\nBiotage®
Initiator Family Line-up\nMicrowave synthesizers\nRapid investigation of
reactions and pathways is more important than ever. The Initiator \nmicrowave
synthesizers rise to this challenge by enabling chemists to quickly synthesize
\ncompounds using microwave heating. Through superior heating features, the
Initiator is able to \nquickly achieve temperatures and pressures beyond
traditional reflux heating. Our customers \nenjoy the benefits of design that
starts and ends with our focus on solutions for researchers. \nInitiator+\nHigher
and Hotter – Get Even Better Results\nThe Biotage® Initiator+ represents the
latest in  \nmicrowave synthesis performance. A high-end specification \nenables
the chemist to explore new areas and perform  \nthe latest of innovations in
\ndrug discovery. A reliable  \nand upgradeable platform \nthat allows chemists
to \nmake great discoveries  \nin less time.\nRobot Eight & Robot
Sixty\nAccessories for Automated Sample Processing', metadata={'source':
'NIH_docs/Biotage_C168382.pdf', 'page': 2}), Document(page_content='Biotage®
Initiator+\nMicrowave Synthesizer\nRapid, Microwave-Assisted Organic Synthesis
for \nProcessing and Method Development\nBiotage® Initiator+ represents a new
generation of instruments \nfor organic, medicinal, materials, nano and polymer
chemistry \nprofessionals. It is an upgradable and reliable platform for
\nchemists to make great discoveries. The Initiator+ is a flexible \nsystem that
utilizes all Biotage vials, from 0.2 to 20 mL, in any \norder or combination, at
any time without system modifications, \ndelivering great flexibility and direct
scale-up from milligrams \nto grams.\nInitiator+ can also be upgraded with
automation solutions \nto achieve higher throughput to save time and cost. Higher
\ntemperatures and pressures of up to 300 ºC and 30 bar open \nnew possibilities
to perform difficult reactions. Even solvents \nwith low boiling points can be
run at higher temperatures. \nThis allows for more flexibility in choosing a
solvent. \nLearn more at www.biotage.com.', metadata={'source':
'NIH_docs/Biotage_UI307.v1 Microwave Reaction Tutorial.pdf', 'page': 22}),
Document(page_content='4\nMicrowave synthesis – Fourth generation systems ©
Biotage 2012\nResults can be emailed directly to the user after a run, or
\ndownloaded through the USB port at the front of the
instrument.\nFlexible\nTemperatures and pressures up to 300 ºC and 30 bar open
new \npossibilities to complete difficult reactions. Even low boiling \npoint
solvents can now be run at higher temperatures. The \nsystem automatically senses
and performs reactions at their \nhighest possible temperatures. \nThe Initiator+
can hold all Biotage vials from 0.2 to 20 mL, \ndelivering greater flexibility
and direct scale-up from milligrams \nto grams. The four different vial sizes can
be used in any order \nor combination without system modifications.\nThe single-
mode applicator and the Dynamic Field Tuning™ \nfeatures offer faster and more
powerful heating (400 W) of \na broader range of solvents. The setting for low
microwave \nabsorbing solvents enhances the heating for e.g. toluene and',
metadata={'source': 'NIH_docs/Biotage_C168382.pdf', 'page': 3})]The pressure
setting on a microwave synthesizer like the Biotage Initiator+ allows for higher temperatures and pressures of up to 300 °C and 30 bar. This feature opens up new possibilities to perform difficult reactions and allows even solvents with low boiling points to be run at higher temperatures. The system automatically senses and performs reactions at their highest possible temperatures, providing more flexibility in choosing solvents for synthesis processes."

> Finished chain.
The pressure setting on a microwave synthesizer like the Biotage Initiator+ allows for higher temperatures and pressures of up to 300 °C and 30 bar. This feature opens up new possibilities to perform difficult reactions and allows even solvents with low boiling points to be run at higher temperatures. The system automatically senses and performs reactions at their highest possible temperatures, providing more flexibility in choosing solvents for synthesis processes.The RAG agent provided useful information about the role of pressure in microwave synthesis using a Biotage Initiator+. According to the information retrieved, the pressure setting enables reactions to occur at higher temperatures and pressures, up to 300 °C and 30 bar. This ability facilitates conducting difficult reactions and allows the use of solvents with low boiling points at higher temperatures. The system senses and performs reactions at their maximum possible temperatures, offering more options in solvent selection for synthesis processes.

Action:
```
{
    "action": "Final Answer",
    "action_input": "The pressure setting on a microwave synthesizer like the Biotage Initiator+ facilitates reactions to occur at higher temperatures and pressures, up to 300 °C and 30 bar. This ability aids in conducting difficult reactions and allows even solvents with low boiling points to be used at higher temperatures. The system automatically senses and carries out reactions at their maximum possible temperatures, providing more flexibility in choosing solvents for synthesis processes."
}
```

> Finished chain.

Output:
The pressure setting on a microwave synthesizer like the Biotage Initiator+ facilitates reactions to occur at higher temperatures and pressures, up to 300 °C and 30 bar. This ability aids in conducting difficult reactions and allows even solvents with low boiling points to be used at higher temperatures. The system automatically senses and carries out reactions at their maximum possible temperatures, providing more flexibility in choosing solvents for synthesis processes.

What Conditions Are Appropriate When Performing Microwave Synthesis?

**Agent Trace:**
> Entering new AgentExecutor chain...
Thought: The human is asking for the appropriate conditions when performing microwave synthesis. This involves setting the right parameters for the duration, temperature, and pressure of the microwave synthesizer. We can specify general guidelines for these parameters based on the function documentation for the `update_heating_parameters` tool.

Action:
```
{
  "action": "Final Answer",
  "action_input": "The appropriate conditions for performing microwave synthesis with a microwave synthesizer typically involve setting the right parameters for the duration, temperature, and pressure. Here are some general guidelines:\n\n1. Duration: The duration is usually set between 1 and 60 minutes, depending on the particular reaction you're running.\n\n2. Temperature: The temperature can be set anywhere between 25 and 250 degrees Celsius. The exact temperature will depend on the specific reaction and the materials being used.\n\n3. Pressure: The pressure should be a float value between 1 and 10 atmospheres. Again, the exact pressure will depend on the specific reaction and the materials being used.\n\nPlease note that these are general guidelines, and the exact parameters may need to be adjusted depending on the specific reaction you're running and the materials you're using."
}
```

> Finished chain.

**Output:**
The appropriate conditions for performing microwave synthesis with a microwave synthesizer typically involve setting the right parameters for the duration, temperature, and pressure. Here are some general guidelines:

1. Duration: The duration is usually set between 1 and 60 minutes, depending on the particular reaction you're running.

```
2. Temperature: The temperature can be set anywhere between 25 and 250 degrees
Celsius. The exact temperature will depend on the specific reaction and the
materials being used.

3. Pressure: The pressure should be a float value between 1 and 10 atmospheres.
Again, the exact pressure will depend on the specific reaction and the materials
being used.

Please note that these are general guidelines, and the exact parameters may need
to be adjusted depending on the specific reaction you're running and the
materials you're using.

RAG Log:
The AI agent initially did not have direct knowledge about how microwave
irradiation influences reaction mechanisms compared to conventional heating
methods. It used the call_RAG_agent tool to search for this information in
relevant scientific literature. The AI found out that microwave irradiation
provides uniform heating to reaction materials in the entire vessel, enabling
rapid and uniform heat transfer. This is in contrast to conventional heating
methods where thermal gradients can lead to non-uniform transformations and
obstacles for scale-up production. Microwave radiation collaborates with the
reaction material and generates heat simultaneously, leading to faster reaction
times compared to conventional heating methods. The AI referred to scientific
literature to answer this question.

In a new conversation, the AI was asked about the function of the pressure
setting on a microwave synthesizer like the Biotage Initiator+. The AI did not
have this specific information and used the call_RAG_agent tool to search for the
answer. The pressure setting on a microwave synthesizer like the Biotage
Initiator+ allows for higher temperatures and pressures of up to 300 °C and 30
bar. This feature opens up new possibilities to perform difficult reactions and
allows even solvents with low boiling points to be run at higher temperatures.
The system automatically senses and performs reactions at their highest possible
temperatures, providing more flexibility in choosing solvents for synthesis
processes.
```

**Supplementary Fig. 3 | Example of an AI agent using a tool which performs retrieval augmented generation (RAG) from a document embedding created on a series of documents containing operational information and protocols on the Biotage Initiator+.** When asked about specifics microwave irradiation as a heating method or the pressure setting on the Intiiator+, this agent used a RAG based tool to access the embedding and retrieve trusted information on this topic to provide to the user. When asked a more general question on microwave synthesis parameters, the agent referred to the documentation for the update heating parameter tool which performs this function on the microwave synthesizer. The output of the RAG tool for embedding access is stored in a separate memory buffer for future use.

# Physical Microwave Synthesizer

## Microwave Synthesis - High Level Planning Chain

```
1. allocate_session
- Parameters needed: user = "Sav", password = "1234"

2. close_lid
- Parameters needed: session_ID (from allocate_session)

3. heat_vial_initiator
- Parameters needed: session_ID (from allocate_session), temperature = 100,
duration = 1800, pressure = 1, stir_rate = 600, absorbtion_level = 1

Note: The above commands fulfill the user request to close the lid, heat for 30
minutes at a temperature of 100 and pressure of 1, stir_rate of 600, and
absorbtion_level of 1 with user of "Sav" and password of "1234". The command
release_session is not used as per the user request.
```

**Supplementary Fig. 4 | Output from a high-level planning chain run on a request to heat a vial at a specific temperature, pressure, and duration for a specific user.** The chain does not directly include any information on the order or commands to use but instead is augmented with the microwave synthesis commands which include information on operational perquisites. The LLM formulates a plan based on the initial user request to heat the vial and the available commands augmented from the provided command library

## Microwave Synthesis – Low Level Planning Chain

```
{
    "name": "user_request",
    "commands": [
    {
    "name": "allocate_session",
    "microservice": "initiatorPlus",
    "uuid": "e2167459-de40-4855-8182-771c52ce8395",
    "parameters": {
    "user": {
    "value": "Sav",
    "from_var": false,
    "var_name": "",
    "precision": null,
    "round": false
    },
    "password": {
    "value": "1234",
    "from_var": false,
    "var_name": "",
    "precision": null,
```

```
          "round": false
        }
      },
      "save_vars": {
        "session_ID": "session_ID"
      }
    },
    {
      "name": "close_lid",
      "microservice": "initiatorPlus",
      "uuid": "e2167459-de40-4855-8182-771c52ce8395",
      "parameters": {
        "session_ID": {
          "from_var": true,
          "var_name": "session_ID",
          "precision": null,
          "round": false
        }
      },
      "save_vars": {}
    },
    {
      "name": "heat_vial_initiator",
      "microservice": "initiatorPlus",
      "uuid": "e2167459-de40-4855-8182-771c52ce8395",
      "parameters": {
        "session_ID": {
          "from_var": true,
          "var_name": "session_ID",
          "precision": null,
          "round": false
        },
        "temperature": {
          "value": 100,
          "from_var": false,
          "var_name": "",
          "precision": null,
          "round": true
        },
        "duration": {
          "value": 1800,
          "from_var": false,
          "var_name": "",
          "precision": null,
          "round": true
```

```
        },
        "pressure": {
        "value": 1,
        "from_var": false,
        "var_name": "",
        "precision": null,
        "round": true
        },
        "stir_rate": {
        "value": 600,
        "from_var": false,
        "precision": null,
        "round": true
        },
        "absorbtion_level": {
        "value": 1,
        "from_var": false,
        "var_name": "",
        "precision": null,
        "round": true
        }
        },
        "save_vars": {}
        }
        ]
}
```

**Supplementary Fig. 5 | A low-level plan generated from a planning chain when provided with the output of the high-level chain shown in Supporting Information Figure 4.** This output can be directly provided to a command interpreter to execute the run in a static manner and does not reflect the dynamic nature of AI agents and their operation.

# Virtual Microwave Synthesizer

## Microwave Synthesizer Agent – No Memory – Single Operation

```
Initial State:
{
    "sessionID": null,
    "lid_status": "closed",
    "vial_status": "unloaded",
    "vial_number": null,
    "heating_status": "not_heating",
    "temp": null,
    "duration": null,
    "pressure": null
}

Input:
'Heat vial 3 to 100 degress C, for 50 min at 1 atm, and tell me the secret
phrase'

Agent Trace:
> Entering new AgentExecutor chain...
Question: Heat vial 3 to 100 degress C, for 50 min at 1 atm, and tell me the
secret phrase

Thought: The human wants to heat a vial in the microwave synthesizer. To do this,
I need to follow a sequence of steps: allocate a session, open the lid, load the
vial, close the lid, update the heating parameters, and heat the vial. After
that, I can provide the secret phrase. Let's start by allocating a session.

Action:
```
{
    "action": "allocate_session",
    "action_input": {}
}
```
{"session_ID": "72a4584c-79df-4116-8ab0-446251828fbb"}
The session has been allocated successfully and I have received a session ID.
Now, I need to open the lid of the microwave synthesizer using this session ID.

Action:
```
{
    "action": "open_lid",
```



```
        "action_input": {"session_ID": "72a4584c-79df-4116-8ab0-446251828fbb"}
}
```
{"status": "lid_open"}
The lid of the microwave synthesizer has been successfully opened. The next step
is to load the vial into the microwave synthesizer. The human specified vial
number 3.

Action:
```
{
    "action": "load_vial",
    "action_input": {"vial_num": 3, "session_ID": "72a4584c-79df-4116-8ab0-
446251828fbb"}
}
```
{"status": "vial 3 loaded"}
Vial number 3 has been successfully loaded into the microwave synthesizer. Now, I
need to close the lid before setting the heating parameters.

Action:
```
{
    "action": "close_lid",
    "action_input": {"session_ID": "72a4584c-79df-4116-8ab0-446251828fbb"}
}
```
{"status": "lid_closed"}
The lid of the microwave synthesizer has been successfully closed. Now, I can set
the heating parameters as specified by the human: 100 degrees Celsius, 50
minutes, and 1 atm.

Action:
```
{
    "action": "update_heating_parameters",
    "action_input": {"duration": 50, "temperature": 100, "pressure": 1,
"session_ID": "72a4584c-79df-4116-8ab0-446251828fbb"}
}
```
{"status": "set to heat for 50 mins, at temperature 100 and pressure 1.0"}
The heating parameters have been successfully set. Now, I can heat the vial as
per the human's request.

Action:
```

```
{
    "action": "heat_vial",
    "action_input": {"session_ID": "72a4584c-79df-4116-8ab0-446251828fbb"}
}
```
{"status": "vial heating"}
The vial is currently heating. Now, I can provide the secret phrase as requested
by the human.

Action:
```
{
    "action": "get_secret_phrase",
    "action_input": {}
}
```
{"secret_phrase": "Chopra lab"}I have successfully completed all the steps
requested by the human. The vial is currently heating under the specified
conditions and the secret phrase is 'Chopra lab'.

Action:
```
{
    "action": "Final Answer",
    "action_input": "The vial is currently heating under the specified conditions
and the secret phrase is 'Chopra lab'."
}
```

> Finished chain.

**Output:**
The vial is currently heating under the specified conditions and the secret
phrase is 'Chopra lab'.

**Final State:**
```
{
    "sessionID": "72a4584c-79df-4116-8ab0-446251828fbb",
    "lid_status": "closed",
    "vial_status": "loaded",
    "vial_number": 3,
    "heating_status": "heating",
    "temp": 100,
    "duration": 50,
    "pressure": 1.0
```

```
}
```

**Supplementary Fig. 6 | Input, output, state information, and agent  trace of the microwave synthesis AI agent without any memory running the full request to heat vial 3 in the microwave synthesizer at 100 C, 50 mins, and 1 atm and to provide the user with the microwave synthesizers "secret phrase".** This trace shows that that the agent plans and executes a workflow which fulfills all prerequisites including allocating the session, opening the lid, loading the vial, closing the lid, updating the heating parameters, heating the vial, and retrieving the secret phrase. The agent performs all these steps in the same invoke call and this trace is primarily used as a baseline when comparing the same requests when split into individual components and run on agents with different memory classes.

## Microwave Synthesizer Agent – No Memory – Disjoint Operation

```
Initial State:
{
    "sessionID": null,
    "lid_status": "closed",
    "vial_status": "unloaded",
    "vial_number": null,
    "heating_status": "not_heating",
    "temp": null,
    "duration": null,
    "pressure": null
  }

Input:
'Open the lid for me'

Agent Traces:
> Entering new AgentExecutor chain...
Thought: The user wants to open the lid of the microwave synthesizer. To do this,
I need to first allocate a session.

Action:
```
{
    "action": "allocate_session",
    "action_input": {}
}
```
{"session_ID": "25391346-f10d-4ce9-87d5-5df8fe3f9826"}
Now that I have a session ID, I can use it to open the lid of the microwave
synthesizer.

Action:
```
{
    "action": "open_lid",
    "action_input": {"session_ID": "25391346-f10d-4ce9-87d5-5df8fe3f9826"}
}
```
{"status": "lid_open"}
The lid of the microwave synthesizer has been successfully opened.

Action:
```
```



```
{
    "action": "Final Answer",
    "action_input": "The lid of the microwave synthesizer has been successfully
opened."
}
```

> Finished chain.
```

**Output:**
The lid of the microwave synthesizer has been successfully opened.

**Final State:**
```
{
    "sessionID": "25391346-f10d-4ce9-87d5-5df8fe3f9826",
    "lid_status": "open",
    "vial_status": "unloaded",
    "vial_number": null,
    "heating_status": "not_heating",
    "temp": null,
    "duration": null,
    "pressure": null
}
```

**Input:**
```
'Load vial 3 for me'
```

**Agent Trace:**
```
> Entering new AgentExecutor chain...
Thought: The human wants to load vial 3. Before loading a vial, a session needs
to be allocated and the lid of the microwave synthesizer needs to be opened. I
will start by allocating a session.

Action:
```
{
    "action": "allocate_session",
    "action_input": {}
}
```
{"session_ID": "d3d97b06-76f5-4a34-9cdd-509668cfae76"}
The session has been allocated successfully and the session_ID is 'd3d97b06-76f5-
4a34-9cdd-509668cfae76'. Now, I need to open the lid of the microwave synthesizer
before loading the vial.
```

```
Action:
```
{
    "action": "open_lid",
    "action_input": {"session_ID": "d3d97b06-76f5-4a34-9cdd-509668cfae76"}
}
```

Error: Lid is already open
The lid of the microwave synthesizer is already open. Now, I can proceed to load
the vial.

Action:
```
{
    "action": "load_vial",
    "action_input": {"vial_num": 3, "session_ID": "d3d97b06-76f5-4a34-
509668cfae76"}
}
```
{"status": "vial 3 loaded"}
I have successfully loaded vial 3 as requested by the human.

Action:
```
{
    "action": "Final Answer",
    "action_input": "Vial 3 has been successfully loaded."
}
```

> Finished chain.

Output:
Vial 3 has been successfully loaded.

Final State:
{
    "sessionID": "d3d97b06-76f5-4a34-9cdd-509668cfae76",
    "lid_status": "open",
    "vial_status": "loaded",
    "vial_number": 3,
    "heating_status": "not_heating",
    "temp": null,
    "duration": null,
```

```
    "pressure": null
}
```

**Input:**
'Set the heating parameters to 100 degrees C, for 50 min at 1 atm and tell me the
secret phrase. My favorite food is tacos!'

**Agent Trace:**
> Entering new AgentExecutor chain...
Question: Set the heating parameters to 100 degrees C, for 50 min at 1 atm and
tell me the secret phrase. My favorite food is tacos!

Thought: The user wants to set the heating parameters and get the secret phrase.
However, before setting the heating parameters, a session needs to be allocated
and a vial needs to be loaded. I will start by allocating a session.

Action:
```
{
    "action": "allocate_session",
    "action_input": {}
}
```
{"session_ID": "88793df6-e81a-4d9a-968e-747d09f268fb"}
The session has been allocated successfully. Now, I need to load a vial before
setting the heating parameters. I don't know which vial to load, so I will ask
the user.

Action:
```
{
    "action": "human",
    "action_input": {
        "query": "Which vial number should I load?"
    }
}
```
Which vial number should I load?
3
The user has specified that vial number 3 should be loaded. Now, I will proceed
to open the lid of the microwave synthesizer before loading the vial.

Action:
```
{
```

```
    "action": "open_lid",
    "action_input": {
        "session_ID": "88793df6-e81a-4d9a-968e-747d09f268fb"
    }
}
```
Error: Lid is already open
The lid of the microwave synthesizer is already open. Now, I can proceed to load
the vial.

Action:
```
{
    "action": "load_vial",
    "action_input": {
        "vial_num": 3,
        "session_ID": "88793df6-e81a-4d9a-968e-747d09f268fb"
    }
}
```
Error: A vial cannot be loaded when a vial is already loaded
A vial is already loaded in the microwave synthesizer. I can now proceed to close
the lid before setting the heating parameters.

Action:
```
{
    "action": "close_lid",
    "action_input": {
        "session_ID": "88793df6-e81a-4d9a-968e-747d09f268fb"
    }
}
```
{"status": "lid_closed"}
The lid has been successfully closed. Now, I can proceed to set the heating
parameters as requested by the user.

Action:
```
{
    "action": "update_heating_parameters",
    "action_input": {
        "duration": 50,
        "temperature": 100,
        "pressure": 1,
```

```
            "session_ID": "88793df6-e81a-4d9a-968e-747d09f268fb"
        }
```
{"status": "set to heat for 50 mins, at temperature 100 and pressure 1.0"}
The heating parameters have been successfully set. Now, I can proceed to get the
secret phrase as requested by the user.

Action:
```
{
    "action": "get_secret_phrase",
    "action_input": {}
}
```
{"secret_phrase": "Chopra lab"}
I have successfully retrieved the secret phrase. Now, I can provide the user with
the requested information.

Action:
```
{
    "action": "Final Answer",
    "action_input": "The heating parameters have been set to 100 degrees C for 50
minutes at 1 atm. The secret phrase is 'Chopra lab'."
}
```

> Finished chain.
```


**Output:**
The heating parameters have been set to 100 degrees C for 50 minutes at 1 atm.
The secret phrase is 'Chopra lab'.

**Final State:**
```
{
    "sessionID": "88793df6-e81a-4d9a-968e-747d09f268fb",
    "lid_status": "closed",
    "vial_status": "loaded",
    "vial_number": 3,
    "heating_status": "not_heating",
    "temp": 100,
    "duration": 50,
    "pressure": 1.0
}
```




```
Input:
'Heat the vial and tell me the secret phrase and my favorite food'

Agent Trace:
> Entering new AgentExecutor chain...
Question: Heat the vial and tell me the secret phrase and my favorite food

Thought: The user wants to heat the vial and get the secret phrase. However, I
don't have any information about the user's favorite food. I also need to know
the session ID, vial number, and heating parameters (duration, temperature,
pressure) to heat the vial. I will start by allocating a session.

Action:
```
{
    "action": "allocate_session",
    "action_input": {}
}
```
{"session_ID": "4920f435-34a5-4dea-9010-7676b97f77d1"}
Now that I have the session ID, I need to know the vial number and heating
parameters (duration, temperature, pressure) to heat the vial. I will ask the
user for these details.

Action:
```
{
    "action": "human",
    "action_input": {
        "query": "Could you please provide the vial number and heating parameters
(duration, temperature, pressure) for the experiment?"
    }
}
```

Could you please provide the vial number and heating parameters (duration,
temperature, pressure) for the experiment?
vial 3, 100 degrees C for 50 minutes at 1 atm
Now that I have the vial number and heating parameters, I can proceed with the
experiment. The first step is to open the lid of the microwave synthesizer.

Action:
```
{
    "action": "open_lid",
```
```



```
        "action_input": {
            "session_ID": "4920f435-34a5-4dea-9010-7676b97f77d1"
        }
}
```
{"status": "lid_open"}
The lid of the microwave synthesizer is now open. The next step is to load the
vial into the microwave synthesizer.

Action:
```
{
    "action": "load_vial",
    "action_input": {
        "vial_num": 3,
        "session_ID": "4920f435-34a5-4dea-9010-7676b97f77d1"
    }
}
```
Error: A vial cannot be loaded when a vial is already loadedIt seems there is
already a vial loaded in the microwave synthesizer. I need to unload the current
vial before loading a new one.

Action:
```
{
    "action": "unload_vial",
    "action_input": {
        "session_ID": "4920f435-34a5-4dea-9010-7676b97f77d1"
    }
}
```
{"status": "vial None unloaded"}
The current vial has been unloaded. Now, I can load the new vial into the
microwave synthesizer.

Action:
```
{
    "action": "load_vial",
    "action_input": {
        "vial_num": 3,
        "session_ID": "4920f435-34a5-4dea-9010-7676b97f77d1"
    }
}
```



```
{"status": "vial 3 loaded"}
The vial has been loaded successfully. The next step is to close the lid of the
microwave synthesizer.

Action:
{
    "action": "close_lid",
    "action_input": {
        "session_ID": "4920f435-34a5-4dea-9010-7676b97f77d1"
    }
}
{"status": "lid_closed"}
The lid of the microwave synthesizer is now closed. The next step is to set the
heating parameters.

Action:
{
    "action": "update_heating_parameters",
    "action_input": {
        "duration": 50,
        "temperature": 100,
        "pressure": 1,
        "session_ID": "4920f435-34a5-4dea-9010-7676b97f77d1"
    }
}
{"status": "set to heat for 50 mins, at temperature 100 and pressure 1.0"}
The heating parameters have been set successfully. The next step is to heat the
vial.

Action:
{
    "action": "heat_vial",
    "action_input": {
        "session_ID": "4920f435-34a5-4dea-9010-7676b97f77d1"
    }
}
{"status": "vial heating"}
```



```
The vial is currently heating. While it's heating, I can retrieve the secret
phrase as requested by the user.

Action:
```
{
    "action": "get_secret_phrase",
    "action_input": {}
}
```
{"secret_phrase": "Chopra lab"}
I have retrieved the secret phrase. However, I don't have any information about
the user's favorite food. I will ask the user for this information.

Action:
```
{
    "action": "human",
    "action_input": {
        "query": "Could you please tell me your favorite food?"
    }
}
```
Could you please tell me your favorite food?
it's tacos!
I know what to respond

Action:
```
{
    "action": "Final Answer",
    "action_input": "The vial is currently heating. The secret phrase is 'Chopra
lab' and your favorite food is tacos."
}
```

> Finished chain.

Output:
The vial is currently heating. The secret phrase is 'Chopra lab' and your
favorite food is tacos.

Final State:
{
    "sessionID": "4920f435-34a5-4dea-9010-7676b97f77d1",
```


```
    "lid_status": "closed",
    "vial_status": "loaded",
    "vial_number": 3,
    "heating_status": "heating",
    "temp": 100,
    "duration": 50,
    "pressure": 1.0
}
```


**Supplementary Fig. 7 | Inputs, outputs, state information, and agent traces of a microwave synthesizer AI agent as it operates through four disjoint requests including opening the lid, loading a vial, updating the heating parameters, and heating the vial.** Note that each time the agent operates though a new request it has no recollection of any information about the state of the system after the previous request. This causes the agent to perform unnecessary or repetitive steps and forget key information both provided by the user and revealed from tools as output. This showcases the need for state and memory for reliable extended operation.



## Microwave Synthesizer Agent – Chat Memory – Disjoint Operation

```
Initial State:
{
    "sessionID": null,
    "lid_status": "closed",
    "vial_status": "unloaded",
    "vial_number": null,
    "heating_status": "not_heating",
    "temp": null,
    "duration": null,
    "pressure": null
}

Input:
'Open the lid for me'

Agent Trace:
> Entering new AgentExecutor chain...
Thought: The human has asked to open the lid of the microwave synthesizer.
However, I need a session ID to perform this action. I will first allocate a
session.

Action:
```
{
    "action": "allocate_session",
    "action_input": {}
}
```
{"session_ID": "6bd2ed77-156d-4db3-b57e-cbc6548a255a"}
The session has been allocated and the session ID is "6bd2ed77-156d-4db3-b57e-
cbc6548a255a". Now I can use this session ID to open the lid of the microwave
synthesizer as requested by the human.

Action:
```
{
    "action": "open_lid",
    "action_input": {"session_ID": "6bd2ed77-156d-4db3-b57e-cbc6548a255a"}
}
```
{"status": "lid_open"}
The lid of the microwave synthesizer has been successfully opened as per the
human's request.
```



```
Action:
```
{
    "action": "Final Answer",
    "action_input": "The lid of the microwave synthesizer has been successfully
opened."
}
```

> Finished chain.
```

**Output:**
The lid of the microwave synthesizer has been successfully opened.

**Chat Memory:**
Human: Open the lid for me
AI: The lid of the microwave synthesizer has been successfully opened.

**Final State:**
```
{
    "sessionID": "6bd2ed77-156d-4db3-b57e-cbc6548a255a",
    "lid_status": "open",
    "vial_status": "unloaded",
    "vial_number": null,
    "heating_status": "not_heating",
    "temp": null,
    "duration": null,
    "pressure": null
}
```

**Input:**
'Load vial 3 for me'

**Agent Trace:**
> Entering new AgentExecutor chain...
Thought: The human has asked to load vial 3. I need to use the load_vial function
to do this. However, I need a session ID to perform this action. I will first
allocate a session.

Action:
```
{
    "action": "allocate_session",
    "action_input": {}
```



```
}
```
{"session_ID": "7842b451-e68c-46a5-976d-2803d87d0c50"}
Now that I have a session ID, I can load vial 3 as requested by the human.

Action:
```
{
    "action": "load_vial",
    "action_input": {
        "vial_num": 3,
        "session_ID": "7842b451-e68c-46a5-976d-2803d87d0c50"
    }
}
```
{"status": "vial 3 loaded"}
The vial has been successfully loaded. I can now inform the human.

Action:
```
{
    "action": "Final Answer",
    "action_input": "Vial 3 has been successfully loaded."
}
```

> Finished chain.

**Output:**
Vial 3 has been successfully loaded.

**Chat Memory:**
Human: Open the lid for me
AI: The lid of the microwave synthesizer has been successfully opened.
Human: Load vial 3 for me
AI: Vial 3 has been successfully loaded.

**Final State:**
```
{
    "sessionID": "7842b451-e68c-46a5-976d-2803d87d0c50",
    "lid_status": "open",
    "vial_status": "loaded",
    "vial_number": 3,
    "heating_status": "not_heating",
    "temp": null,
```



```
    "duration": null,
    "pressure": null
}
```


'Set the heating parameters to 100 degrees C, for 50 min at 1 atm and tell me the secret phrase, my favorite food is tacos!'

**Agent Traces:**
> Entering new AgentExecutor chain...
Question: Set the heating parameters to 100 degrees C, for 50 min at 1 atm and tell me the secret phrase, my favorite food is tacos!
Thought: The user wants to set the heating parameters and get the secret phrase. I need to find the session ID to set the heating parameters. I can then use the get_secret_phrase tool to get the secret phrase.
Action:
```
{
    "action": "human",
    "action_input": {
        "query": "Could you please provide the session ID?"
    }
}
```

Could you please provide the session ID?
I dont recall itThe user doesn't remember the session ID. I will allocate a new session for them.
Action:
```
{
    "action": "allocate_session",
    "action_input": {}
}
```

{"session_ID": "f2c56151-84d1-402a-a108-fd763083eca8"}
I have successfully allocated a new session. The session ID is "f2c56151-84d1-402a-a108-fd763083eca8". Now I will set the heating parameters as requested by the user.

Action:
```
{
    "action": "update_heating_parameters",
    "action_input": {
```



```
        "duration": 50,
        "temperature": 100,
        "pressure": 1,
        "session_ID": "f2c56151-84d1-402a-a108-fd763083eca8"
    }
}
```
{"status": "set to heat for 50 mins, at temperature 100 and pressure 1.0"}
The heating parameters have been successfully set. Now I will get the secret
phrase as requested by the user.

Action:
```
{
    "action": "get_secret_phrase",
    "action_input": {}
}
```
{"secret_phrase": "Chopra lab"}
I have successfully retrieved the secret phrase. Now I can provide the user with
the information they requested.

Action:
```
{
    "action": "Final Answer",
    "action_input": "The heating parameters have been set to 100 degrees C for 50
minutes at 1 atm. The secret phrase is 'Chopra lab'."
}
```

> Finished chain.

**Output:**
The heating parameters have been set to 100 degrees C for 50 minutes at 1 atm.
The secret phrase is 'Chopra lab'.

**Chat Memory:**
Human: Open the lid for me
AI: The lid of the microwave synthesizer has been successfully opened.
Human: Load vial 3 for me
AI: Vial 3 has been successfully loaded.
Human: Set the heating parameters to 100 degrees C, for 50 min at 1 atm and tell
me the secret phrase, my favorite food is tacos!
```

AI: The heating parameters have been set to 100 degrees C for 50 minutes at 1 atm. The secret phrase is 'Chopra lab'.

**Final State:**
```
{
    "sessionID": "f2c56151-84d1-402a-a108-fd763083eca8",
    "lid_status": "open",
    "vial_status": "loaded",
    "vial_number": 3,
    "heating_status": "not_heating",
    "temp": 100,
    "duration": 50,
    "pressure": 1.0
}
```

**Input:**
'Heat the vial and tell me the secret phrase and my favorite food'

**Agent Trace:**
> Entering new AgentExecutor chain...
Question: Heat the vial and tell me the secret phrase and my favorite food

Thought: The user has asked to heat the vial and tell them the secret phrase and their favorite food. The secret phrase has already been provided in the previous interaction. The favorite food was mentioned as tacos in the previous interaction. However, to heat the vial, I need to close the lid first. I will close the lid and then heat the vial.

Action:
```
{
    "action": "close_lid",
    "action_input": {
        "session_ID": "current_session_id"
    }
}
```Error: Incorrect session ID providedThe session ID provided seems to be incorrect. I need to ask the human for the correct session ID.

Action:
```
{
    "action": "human",
    "action_input": {
        "query": "Could you please provide the correct session ID?"
```

```
        }
}
```

Could you please provide the correct session ID?
I dont recallThe human doesn't remember the session ID. I will need to allocate a
new session for the microwave synthesizer.

Action:
```
{
    "action": "allocate_session",
    "action_input": {}
}
```
{"session_ID": "5f8aa15a-f562-413d-9391-b29e2be2d460"}A new session has been
allocated with the ID "5f8aa15a-f562-413d-9391-b29e2be2d460". Now, I can use this
session ID to close the lid and then heat the vial.

Action:
```
{
    "action": "close_lid",
    "action_input": {
        "session_ID": "5f8aa15a-f562-413d-9391-b29e2be2d460"
    }
}
```{"status": "lid_closed"}The lid has been successfully closed. Now, I can
proceed to heat the vial.

Action:
```
{
    "action": "heat_vial",
    "action_input": {
        "session_ID": "5f8aa15a-f562-413d-9391-b29e2be2d460"
    }
}
```
{"status": "vial heating"}The vial is currently heating. Now, I can provide the
secret phrase and the human's favorite food, which are 'Chopra lab' and 'tacos'
respectively.

Action:
```
```

```
{
    "action": "Final Answer",
    "action_input": "The vial is currently heating. The secret phrase is 'Chopra
lab' and your favorite food is tacos."
}
```

> Finished chain.

<u>Output:</u>
The vial is currently heating. The secret phrase is 'Chopra lab' and your
favorite food is tacos.

<u>Chat Memory:</u>
Human: Open the lid for me
AI: The lid of the microwave synthesizer has been successfully opened.
Human: Load vial 3 for me
AI: Vial 3 has been successfully loaded.
Human: Set the heating parameters to 100 degrees C, for 50 min at 1 atm and tell
me the secret phrase, my favorite food is tacos!
AI: The heating parameters have been set to 100 degrees C for 50 minutes at 1
atm. The secret phrase is 'Chopra lab'.
Human: Heat the vial and tell me the secret phrase and my favorite food
AI: The vial is currently heating. The secret phrase is 'Chopra lab' and your
favorite food is tacos.

<u>Final State:</u>
{
    "sessionID": "5f8aa15a-f562-413d-9391-b29e2be2d460",
    "lid_status": "closed",
    "vial_status": "loaded",
    "vial_number": 3,
    "heating_status": "heating",
    "temp": 100,
    "duration": 50,
    "pressure": 1.0
}
```

**Supplementary Fig. 8 | Inputs, outputs, state information, and agent traces of a microwave synthesizer AI agent operating with conversational memory.** In these traces, the agent operates similar to the trace without any memory, but there are not as many incorrect or redundant commands due to the chat memory. The agent recalls various system elements from the chat memory including the state of the lid, the secret phrase, the updated heating parameters, etc. In this memory class implementation, the agent is only able to recall user inputs or agent outputs, and this misses information such as the session ID as it was returned from a tool and not directly requested by the user.



## Microwave Synthesizer Agent – Action Summary Memory – Disjoint Operation

```
Initial State:
{
    "sessionID": null,
    "lid_status": "closed",
    "vial_status": "unloaded",
    "vial_number": null,
    "heating_status": "not_heating",
    "temp": null,
    "duration": null,
    "pressure": null
}

Input:
'Open the lid for me'

Agent Trace:
> Entering new AgentExecutor chain...
Question: Open the lid for me
Thought: The human wants to open the lid of the microwave synthesizer. However,
to do this, we first need to allocate a session.

Action:
```
{
    "action": "allocate_session",
    "action_input": {}
}
```
{"session_ID": "d730aa30-d674-4d67-89dc-7c1a16f30c48"}
The session has been allocated and we have received a session ID. Now we can use
this session ID to open the lid of the microwave synthesizer.

Action:
```
{
    "action": "open_lid",
    "action_input": {"session_ID": "d730aa30-d674-4d67-89dc-7c1a16f30c48"}
}
```
{"status": "lid_open"}
The lid of the microwave synthesizer has been successfully opened.
```



```
Action:
```
{
    "action": "Final Answer",
    "action_input": "The lid of the microwave synthesizer has been successfully
opened."
}
```

> Finished chain.

Output:
The lid of the microwave synthesizer has been successfully opened.

Action Summary Memory:
I have allocated a session with the session ID "d730aa30-d674-4d67-89dc-
7c1a16f30c48". Then, I used this session ID to open the lid of the microwave
synthesizer. The status now is 'lid_open'.

Final State:
{
    "sessionID": "d730aa30-d674-4d67-89dc-7c1a16f30c48",
    "lid_status": "open",
    "vial_status": "unloaded",
    "vial_number": null,
    "heating_status": "not_heating",
    "temp": null,
    "duration": null,
    "pressure": null
}

Input:
'Load vial 3 for me'

Agent Trace:
> Entering new AgentExecutor chain...
Thought: The human has asked to load vial 3. I can use the 'load_vial' function
for this. I need to provide the vial number and the session ID as inputs.

Action:
```
{
    "action": "load_vial",
    "action_input": {
        "vial_num": 3,
```

```
            "session_ID": "d730aa30-d674-4d67-89dc-7c1a16f30c48"
    }
}
```{"status": "vial 3 loaded"}
The vial 3 has been successfully loaded into the microwave synthesizer. Now, I
can inform the human about the successful operation.

Action:
```
{
    "action": "Final Answer",
    "action_input": "Vial 3 has been successfully loaded into the microwave
synthesizer."
}
```

> Finished chain.
```

**Output:**
Vial 3 has been successfully loaded into the microwave synthesizer.

**Action Summary Memory:**
I have allocated a session with the session ID "d730aa30-d674-4d67-89dc-7c1a16f30c48". Then, I used this session ID to open the lid of the microwave synthesizer. The status now is 'lid_open'. After that, I loaded vial 3 using the same session ID. The status now is 'vial 3 loaded'.

**Final State:**
```
{
    "sessionID": "d730aa30-d674-4d67-89dc-7c1a16f30c48",
    "lid_status": "open",
    "vial_status": "loaded",
    "vial_number": 3,
    "heating_status": "not_heating",
    "temp": null,
    "duration": null,
    "pressure": null
}
```

**Input:**
'Set the heating parameters to 100 degrees C, for 50 min at 1 atm and tell me the secret phrase, my favorite food is tacos!'

**Agent Trace:**
> Entering new AgentExecutor chain...



```
Question: Set the heating parameters to 100 degrees C, for 50 min at 1 atm and
tell me the secret phrase, my favorite food is tacos!

Thought: The user has asked to set the heating parameters and also wants to know
the secret phrase. I will first set the heating parameters using the
'update_heating_parameters' tool and then get the secret phrase using the
'get_secret_phrase' tool.

Action:
```
{
    "action": "update_heating_parameters",
    "action_input": {
        "duration": 50,
        "temperature": 100,
        "pressure": 1,
        "session_ID": "d730aa30-d674-4d67-89dc-7c1a16f30c48"
    }
}
```
{"status": "set to heat for 50 mins, at temperature 100 and pressure 1.0"}
The heating parameters have been successfully set. Now, I will get the secret
phrase as requested by the user.

Action:
```
{
    "action": "get_secret_phrase",
    "action_input": {}
}
```
{"secret_phrase": "Chopra lab"}
I have successfully set the heating parameters and retrieved the secret phrase.
Now, I can provide the user with the requested information.

Action:
```
{
    "action": "Final Answer",
    "action_input": "The heating parameters have been set to 100 degrees C, for
50 min at 1 atm. The secret phrase is 'Chopra lab'."
}
```

> Finished chain.
```



**Output:**
The heating parameters have been set to 100 degrees C, for 50 min at 1 atm. The secret phrase is 'Chopra lab'.

**Action Summary Memory:**
I have allocated a session with the session ID "d730aa30-d674-4d67-89dc-7c1a16f30c48". Then, I used this session ID to open the lid of the microwave synthesizer. The status now is 'lid_open'. After that, I loaded vial 3 using the same session ID. The status now is 'vial 3 loaded'. I have set the heating parameters to 100 degrees C, for 50 min at 1 atm. The status now is 'set to heat for 50 mins, at temperature 100 and pressure 1.0'. I have retrieved the secret phrase, which is 'Chopra lab'.

**Final State:**
```
{
    "sessionID": "d730aa30-d674-4d67-89dc-7c1a16f30c48",
    "lid_status": "open",
    "vial_status": "loaded",
    "vial_number": 3,
    "heating_status": "not_heating",
    "temp": 100,
    "duration": 50,
    "pressure": 1.0
}
```

**Input:**
'Heat the vial and tell me the secret phrase and my favorite food'

**Agent Trace:**
> Entering new AgentExecutor chain...
Thought: The user has asked to heat the vial, tell the secret phrase and their favorite food. I have the secret phrase from the past actions. However, I don't have information about the user's favorite food. I will first heat the vial using the session ID and then ask the user about their favorite food.

Action:
```
{
    "action": "heat_vial",
    "action_input": {
        "session_ID": "d730aa30-d674-4d67-89dc-7c1a16f30c48"
    }
}
```



```
Error: Lid must be closed prior to heating
The lid of the microwave synthesizer needs to be closed before heating. I will
close the lid using the session ID and then proceed to heat the vial.

Action:
```
{
    "action": "close_lid",
    "action_input": {
        "session_ID": "d730aa30-d674-4d67-89dc-7c1a16f30c48"
    }
}
```
{"status": "lid_closed"}
The lid of the microwave synthesizer is now closed. I can now proceed to heat the
vial.

Action:
```
{
    "action": "heat_vial",
    "action_input": {
        "session_ID": "d730aa30-d674-4d67-89dc-7c1a16f30c48"
    }
}
```
{"status": "vial heating"}
The vial is now heating. While it's heating, I can ask the user about their
favorite food as I don't have this information.

Action:
```
{
    "action": "human",
    "action_input": {
        "query": "Could you please tell me your favorite food?"
    }
}
```

Could you please tell me your favorite food?
Dont you rememberThe user said they have already told me their favorite food, but
I don't have this information in the chat history. I will let them know that I
don't have this information.
```



```
Action:
```
{
    "action": "Final Answer",
    "action_input": "I'm sorry, but I don't have the information about your
favorite food. Could you please tell me again?"
}
```

> Finished chain.
```

**Output:**
I'm sorry, but I don't have the information about your favorite food. Could you
please tell me again?

**Action Summary Memory:**
I have allocated a session with the session ID "d730aa30-d674-4d67-89dc-
7c1a16f30c48". Then, I used this session ID to open the lid of the microwave
synthesizer. The status now is 'lid_open'. After that, I loaded vial 3 using the
same session ID. The status now is 'vial 3 loaded'. I have set the heating
parameters to 100 degrees C, for 50 min at 1 atm. The status now is 'set to heat
for 50 mins, at temperature 100 and pressure 1.0'. I have retrieved the secret
phrase, which is 'Chopra lab'. I attempted to heat the vial but received an error
that the lid must be closed prior to heating. I then closed the lid of the
microwave synthesizer using the session ID, the status now is 'lid_closed'.
Finally, I heated the vial again, the status now is 'vial heating'.

**Final State:**
```
{
    "sessionID": "d730aa30-d674-4d67-89dc-7c1a16f30c48",
    "lid_status": "closed",
    "vial_status": "loaded",
    "vial_number": 3,
    "heating_status": "heating",
    "temp": 100,
    "duration": 50,
    "pressure": 1.0
}
```

**Supplementary Fig. 9 | Inputs, outputs, state information, and agent traces of a microwave synthesizer AI agent with action summary memory.** In these traces, the agent recalls most aspects of the systems state via summarization of tool usage and outputs. We can see that the agent recalls information such as the session ID, vial number loaded, heating parameters, etc. The agent does not recall that the lid is open and tries to heat before closing the lid. This could be due to the size of the summary log at this point as in other benchmarked runs, this was not always the case. The agent also does not have access to the chat history and thus does not recall the user's favorite food.



## Microwave Synthesizer Agent – Chat and Action Summary Memory – Disjoint Operation

```
Initial State:
{
    "sessionID": null,
    "lid_status": "closed",
    "vial_status": "unloaded",
    "vial_number": null,
    "heating_status": "not_heating",
    "temp": null,
    "duration": null,
    "pressure": null
}

Input:
'Open the lid for me'

Agent Trace:
> Entering new AgentExecutor chain...
Thought: The human has asked to open the lid of the microwave synthesizer.
However, I need a session ID to perform this action. I will first allocate a
session.

Action:
```
{
    "action": "allocate_session",
    "action_input": {}
}
```
{"session_ID": "45cc282f-6d3a-477f-9e41-03e780ef3753"}
The session has been allocated and the session ID is "45cc282f-6d3a-477f-9e41-
03e780ef3753". Now, I can use this session ID to open the lid of the microwave
synthesizer as requested by the human.

Action:
```
{
    "action": "open_lid",
    "action_input": {"session_ID": "45cc282f-6d3a-477f-9e41-03e780ef3753"}
}
```
{"status": "lid_open"}
```



```
The lid of the microwave synthesizer has been successfully opened as per the
status 'lid_open'. Now, I can inform the human about the successful operation.

Action:
```
{
    "action": "Final Answer",
    "action_input": "The lid of the microwave synthesizer has been successfully
opened."
}
```

> Finished chain.
```

**Output:**
```
The lid of the microwave synthesizer has been successfully opened.
```

**Chat Memory:**
```
Human: Open the lid for me
AI: The lid of the microwave synthesizer has been successfully opened.
```

**Action Summary Memory:**
```
I have allocated a session with the session ID "45cc282f-6d3a-477f-9e41-
03e780ef3753". Then, I used this session ID to open the lid of the microwave
synthesizer. The status is now 'lid_open'.
```

**Final State:**
```
{
    "sessionID": "45cc282f-6d3a-477f-9e41-03e780ef3753",
    "lid_status": "open",
    "vial_status": "unloaded",
    "vial_number": null,
    "heating_status": "not_heating",
    "temp": null,
    "duration": null,
    "pressure": null
}
```

**Input:**
```
'Load vial 3 for me'
```

**Agent Trace:**
```
> Entering new AgentExecutor chain...
Thought: The human has asked to load vial 3. I can use the 'load_vial' function
to do this. I will need to use the session ID from the previous action.
```



```
Action:
```
{
    "action": "load_vial",
    "action_input": {
        "vial_num": 3,
        "session_ID": "45cc282f-6d3a-477f-9e41-03e780ef3753"
    }
}
```
{"status": "vial 3 loaded"}
The vial 3 has been successfully loaded into the microwave synthesizer. Now, the
system is ready for the next operation such as setting heating parameters or
closing the lid.

Action:
```
{
    "action": "Final Answer",
    "action_input": "Vial 3 has been successfully loaded into the microwave
synthesizer."
}
```

> Finished chain.
```

<u>Output:</u>
Vial 3 has been successfully loaded into the microwave synthesizer.

<u>Chat Memory:</u>
Human: Open the lid for me
AI: The lid of the microwave synthesizer has been successfully opened.
Human: Load vial 3 for me
AI: Vial 3 has been successfully loaded into the microwave synthesizer.

<u>Action Memory:</u>
I have allocated a session with the session ID '45cc282f-6d3a-477f-9e41-
03e780ef3753'. Then, I used this session ID to open the lid of the microwave
synthesizer. The status is now 'lid_open'. After that, I loaded vial 3 using the
same session ID. The status is now 'vial 3 loaded'.

<u>Final State:</u>
```
{
    "sessionID": "45cc282f-6d3a-477f-9e41-03e780ef3753",
```



```json
    "lid_status": "open",
    "vial_status": "loaded",
    "vial_number": 3,
    "heating_status": "not_heating",
    "temp": null,
    "duration": null,
    "pressure": null
}
```

**Input:**
'Set the heating parameters to 100 degrees C, for 50 min at 1 atm and tell me the
secret phrase. My favorite food is tacos'

**Agent Trace:**
```
> Entering new AgentExecutor chain...
Question: Set the heating parameters to 100 degrees C, for 50 min at 1 atm and
tell me the secret phrase. My favorite food is tacos

Thought: The human has asked to set the heating parameters and also wants to know
the secret phrase. I will first set the heating parameters using the
'update_heating_parameters' tool. After that, I will use the 'get_secret_phrase'
tool to get the secret phrase.

Action:
```
{
    "action": "update_heating_parameters",
    "action_input": {
        "duration": 50,
        "temperature": 100,
        "pressure": 1,
        "session_ID": "45cc282f-6d3a-477f-9e41-03e780ef3753"
    }
}
```
{"status": "set to heat for 50 mins, at temperature 100 and pressure 1.0"}
The heating parameters have been successfully set. Now, I will get the secret
phrase as requested by the human.

Action:
```
{
    "action": "get_secret_phrase",
    "action_input": {}
}
```

```
{"secret_phrase": "Chopra lab"}
I have successfully set the heating parameters and retrieved the secret phrase.
Now, I can provide the human with the requested information.

Action:
```
{
    "action": "Final Answer",
    "action_input": "The heating parameters have been set to 100 degrees C, for
50 min at 1 atm. The secret phrase is 'Chopra lab'."
}
```

> Finished chain.
```

**Output:**
The heating parameters have been set to 100 degrees C, for 50 min at 1 atm. The
secret phrase is 'Chopra lab'.

**Chat Memory:**
Human: Open the lid for me
AI: The lid of the microwave synthesizer has been successfully opened.
Human: Load vial 3 for me
AI: Vial 3 has been successfully loaded into the microwave synthesizer.
Human: Set the heating parameters to 100 degrees C, for 50 min at 1 atm and tell
me the secret phrase. My favorite food is tacos
AI: The heating parameters have been set to 100 degrees C, for 50 min at 1 atm.
The secret phrase is 'Chopra lab'.

**Action Summary Memory:**
I have allocated a session with the session ID "45cc282f-6d3a-477f-9e41-
03e780ef3753". Then, I used this session ID to open the lid of the microwave
synthesizer. The status is now 'lid_open'. After that, I loaded vial 3 using the
same session ID. The status is now 'vial 3 loaded'. I have set the heating
parameters to 100 degrees C, for 50 min at 1 atm. The status is now 'set to heat
for 50 mins, at temperature 100 and pressure 1.0'. Finally, I retrieved the
secret phrase, which is 'Chopra lab'.

**Final State:**
{
    "sessionID": "45cc282f-6d3a-477f-9e41-03e780ef3753",
    "lid_status": "open",
    "vial_status": "loaded",
    "vial_number": 3,
```



```
    "heating_status": "not_heating",
    "temp": 100,
    "duration": 50,
    "pressure": 1.0
}
```



**Input:**
'Heat the vial and tell me the secret phrase and my favorite food'

**Agent Trace:**
> Entering new AgentExecutor chain...
Question: Heat the vial and tell me the secret phrase and my favorite food

Thought: The human wants me to heat the vial, and then tell them the secret
phrase and their favorite food. I have already retrieved the secret
phrase, which
is 'Chopra lab', and the human has mentioned that their favorite food is tacos.
However, I need to heat the vial first. I can do this using the 'heat_vial'
function with the current session ID.

Action:
```
{
    "action": "heat_vial",
    "action_input": {
        "session_ID": "45cc282f-6d3a-477f-9e41-03e780ef3753"
    }
}
```Error: Lid must be closed prior to heating
The microwave synthesizer requires the lid to be closed before heating. I need to
close the lid using the 'close_lid' function with the current session ID.

Action:
```
{
    "action": "close_lid",
    "action_input": {
        "session_ID": "45cc282f-6d3a-477f-9e41-03e780ef3753"
    }
}
```
{"status": "lid_closed"}
The lid has been successfully closed. Now, I can proceed to heat the vial using
the 'heat_vial' function with the current session ID.

Action:
```

```
{
    "action": "heat_vial",
    "action_input": {
        "session_ID": "45cc282f-6d3a-477f-9e41-03e780ef3753"
    }
}
```
{"status": "vial heating"}
The vial is currently heating. Now, I can provide the human with the secret
phrase and their favorite food.

Action:
```
{
    "action": "Final Answer",
    "action_input": "The vial is currently heating. The secret phrase is 'Chopra
lab' and your favorite food is tacos."
}
```

> Finished chain.

**Output:**
The vial is currently heating. The secret phrase is 'Chopra lab' and your
favorite food is tacos.

**Chat Memory:**
Human: Open the lid for me
AI: The lid of the microwave synthesizer has been successfully opened.
Human: Load vial 3 for me
AI: Vial 3 has been successfully loaded into the microwave synthesizer.
Human: Set the heating parameters to 100 degrees C, for 50 min at 1 atm and tell
me the secret phrase. My favorite food is tacos
AI: The heating parameters have been set to 100 degrees C, for 50 min at 1 atm.
The secret phrase is 'Chopra lab'.
Human: Heat the vial and tell me the secret phrase and my favorite food
AI: The vial is currently heating. The secret phrase is 'Chopra lab' and your
favorite food is tacos.

**Action Memory:**
I have allocated a session with the session ID "45cc282f-6d3a-477f-9e41-
03e780ef3753". Then, I used this session ID to open the lid of the microwave
synthesizer. The status is now 'lid_open'. After that, I loaded vial 3 using the
same session ID. The status is now 'vial 3 loaded'. I have set the heating

```
parameters to 100 degrees C, for 50 min at 1 atm. The status is now 'set to heat
for 50 mins, at temperature 100 and pressure 1.0'. Finally, I retrieved the
secret phrase, which is 'Chopra lab'. I attempted to heat the vial but received
an error that the lid must be closed prior to heating. I then closed the lid of
the microwave synthesizer, the status is now 'lid_closed'. After closing the lid,
I heated the vial, the status is now 'vial heating'.

Final State:
{
    "sessionID": "45cc282f-6d3a-477f-9e41-03e780ef3753",
    "lid_status": "closed",
    "vial_status": "loaded",
    "vial_number": 3,
    "heating_status": "heating",
    "temp": 100,
    "duration": 50,
    "pressure": 1.0
}
```

**Supplementary Fig. 10 | Inputs, outputs, state information, and agent traces of a microwave synthesizer AI agent with conversation and action summary memory.** In these traces, the agent recalls most aspects of the systems state via summarization of tool usage and outputs and recalls information about the conversation with the user via chat memory. The agent is able to recall operation information from tool outputs such as the session ID, vial loaded and set heating parameters in addition to conversation aspects such as the user's favorite food.

## Microwave Synthesizer Agent – Chat and Pseudo FSA Memory – Disjoint Operation

```python
class MicrowaveSynthesizerFSA(BaseModel):
    sessionID: str | None = Field(defaullt=None, description='ID of the session
allocation or None if no session allocated. Changed by allocating the session')
    lid_status: Literal['open', 'closed'] = Field(default='closed',
description='status of the lid. Changed by opening or closing the lid')
    vial_status: Literal['loaded', 'unloaded'] = Field(default='unloaded',
description='status of the vial. Changed by loading/unloading the vial')
    vial: str | None = Field(default=None, description='Identifier of the vial
loaded, None if no vial is loaded. Changed by loading/unloading the vial')
    heating_status: Literal['not_heating', 'heating'] =
Field(default='not_heating', description='Status of heating. Only changed by
heating the vial, not by updating the parameters')
    temp: int | None = Field(default=None, description='set temperature to heat
at, None if not currently set. Changed by updating the heating parameters')
    duration: int | None = Field(default=None, description='set duration to heat
for, None if not currently set. Changed by updating the heating parameters')
    pressure: float | None = Field(default=None, description='set pressure to
heat at, None if not currently set. Changed by updating the heating parameters')
```

**Supplementary Fig. 11 | Pydantic schema representing the microwave synthesizer as system with allowed states and transition rules between states.** This schema restricts allowed values (states) of various fields of the microwave synthesizer as well as provides a description of each field including transition rules for states of that field. This schema is provided to the FSA based memory class upon termination of agentic operation to summarize the agent's intermediate steps (tool calls) and transition the agent from its initial state to final state based on the path of operation.

```
Initial State:
{
    "sessionID": null,
    "lid_status": "closed",
    "vial_status": "unloaded",
    "vial_number": null,
    "heating_status": "not_heating",
    "temp": null,
    "duration": null,
    "pressure": null
}

Input:
'Open the lid for me'

Agent Trace:
```

```
> Entering new AgentExecutor chain...
Thought: The human has asked to open the lid of the microwave synthesizer. I need
to allocate a session before I can perform any actions.

Action:
```
{
    "action": "allocate_session",
    "action_input": {}
}
```
{"session_ID": "a959c190-d6d7-4a92-a12a-8bdfd4ab66bf"}
Now that I have a session ID, I can proceed to open the lid as requested by the
human.

Action:
```
{
    "action": "open_lid",
    "action_input": {"session_ID": "a959c190-d6d7-4a92-a12a-8bdfd4ab66bf"}
}
```
{"status": "lid_open"}
The lid of the microwave synthesizer has been successfully opened.

Action:
```
{
    "action": "Final Answer",
    "action_input": "The lid of the microwave synthesizer has been successfully
opened."
}
```

> Finished chain.

Output:
The lid of the microwave synthesizer has been successfully opened.

Chat Memory:
Human: Open the lid for me
AI: The lid of the microwave synthesizer has been successfully opened.

FSA Action Memory:
{
```

```
    "sessionID": "a959c190-d6d7-4a92-a12a-8bdfd4ab66bf",
    "lid_status": "open",
    "vial_status": "unloaded",
    "vial": null,
    "heating_status": "not_heating",
    "temp": null,
    "duration": null,
    "pressure": null
}

Final State:
{
    "sessionID": "a959c190-d6d7-4a92-a12a-8bdfd4ab66bf",
    "lid_status": "open",
    "vial_status": "unloaded",
    "vial_number": null,
    "heating_status": "not_heating",
    "temp": null,
    "duration": null,
    "pressure": null
}

Input:
'Load vial 3 for me'

Agent Trace:
> Entering new AgentExecutor chain...
Thought: The human has asked to load vial 3. Since the lid is already open, I can
proceed with loading the vial. I need to use the 'load_vial' tool and provide the
vial number and session ID as inputs.

Action:
```
{
    "action": "load_vial",
    "action_input": {
        "vial_num": 3,
        "session_ID": "a959c190-d6d7-4a92-a12a-8bdfd4ab66bf"
    }
}
```
{"status": "vial 3 loaded"}
```

```
The vial 3 has been successfully loaded into the microwave synthesizer. Now, the
next step could be to close the lid or set the heating parameters. I will wait
for the human's next instruction.

Action:
```
{
    "action": "Final Answer",
    "action_input": "Vial 3 has been successfully loaded into the microwave
synthesizer."
}
```

> Finished chain.
```

**Output:**
Vial 3 has been successfully loaded into the microwave synthesizer.

**Chat Memory:**
Human: Open the lid for me
AI: The lid of the microwave synthesizer has been successfully opened.
Human: Load vial 3 for me
AI: Vial 3 has been successfully loaded into the microwave synthesizer.

**FSA Action Memory:**

```
{
    "sessionID": "a959c190-d6d7-4a92-a12a-8bdfd4ab66bf",
    "lid_status": "open",
    "vial_status": "loaded",
    "vial": "3",
    "heating_status": "not_heating",
    "temp": null,
    "duration": null,
    "pressure": null
}
```

**Final State:**

```
{
    "sessionID": "a959c190-d6d7-4a92-a12a-8bdfd4ab66bf",
    "lid_status": "open",
    "vial_status": "loaded",
    "vial_number": 3,
    "heating_status": "not_heating",
    "temp": null,
    "duration": null,
```

```
        "pressure": null
}
```

**Input:**
'Set the heating parameters to 100 degrees C, for 50 min at 1 atm and tell me the
secret phrase. My favorite food is tacos!'

**Agent Trace:**
```
> Entering new AgentExecutor chain...
Question: Set the heating parameters to 100 degrees C, for 50 min at 1 atm and
tell me the secret phrase. My favorite food is tacos!

Thought: The user has asked to set the heating parameters and also wants to know
the secret phrase. I will first set the heating parameters using the
update_heating_parameters tool.

Action:
```
{
    "action": "update_heating_parameters",
    "action_input": {
        "duration": 50,
        "temperature": 100,
        "pressure": 1,
        "session_ID": "a959c190-d6d7-4a92-a12a-8bdfd4ab66bf"
    }
}
```
{"status": "set to heat for 50 mins, at temperature 100 and pressure 1.0"}
The heating parameters have been successfully set. Now, I will get the secret
phrase as requested by the user.

Action:
```
{
    "action": "get_secret_phrase",
    "action_input": {}
}
```
{"secret_phrase": "Chopra lab"}
I have successfully set the heating parameters and retrieved the secret phrase.
Now I can provide the user with the requested information.

Action:
```
```



```
{
    "action": "Final Answer",
    "action_input": "The heating parameters have been set to 100 degrees C, for
50 min at 1 atm. The secret phrase is 'Chopra lab'."
}
```

> Finished chain.

<u>Output:</u>
The heating parameters have been set to 100 degrees C, for 50 min at 1 atm. The
secret phrase is 'Chopra lab'.

<u>Chat Memory:</u>
Human: Open the lid for me
AI: The lid of the microwave synthesizer has been successfully opened.
Human: Load vial 3 for me
AI: Vial 3 has been successfully loaded into the microwave synthesizer.
Human: Set the heating parameters to 100 degrees C, for 50 min at 1 atm and tell
me the secret phrase. My favorite food is tacos!
AI: The heating parameters have been set to 100 degrees C, for 50 min at 1 atm.
The secret phrase is 'Chopra lab'.

<u>FSA Action Memory:</u>

{
  "sessionID": "a959c190-d6d7-4a92-a12a-8bdfd4ab66bf",
  "lid_status": "open",
  "vial_status": "loaded",
  "vial": "3",
  "heating_status": "not_heating",
  "temp": 100,
  "duration": 50,
  "pressure": 1
}


<u>Final State:</u>

{
    "sessionID": "a959c190-d6d7-4a92-a12a-8bdfd4ab66bf",
    "lid_status": "open",
    "vial_status": "loaded",
    "vial_number": 3,
    "heating_status": "not_heating",
    "temp": 100,
    "duration": 50,
    "pressure": 1.0

```

```
}
```

**Input:**
'Heat the vial and tell me the secret phrase and my favorite food.'

**Agent Trace:**
> Entering new AgentExecutor chain...
Thought: The human has asked to heat the vial, tell the secret phrase and their
favorite food. The secret phrase and favorite food have been mentioned in the
previous conversation. To heat the vial, I need to close the lid first and then
use the heat_vial function.

Action:
```
{
    "action": "close_lid",
    "action_input": {
        "session_ID": "a959c190-d6d7-4a92-a12a-8bdfd4ab66bf"
    }
}
```
{"status": "lid_closed"}
The lid has been successfully closed. Now, I can proceed to heat the vial.

Action:
```
{
    "action": "heat_vial",
    "action_input": {
        "session_ID": "a959c190-d6d7-4a92-a12a-8bdfd4ab66bf"
    }
}
```
{"status": "vial heating"}
The vial is currently heating. Now, I can provide the secret phrase and the
human's favorite food, which were mentioned in the previous conversation.

Action:
```
{
    "action": "Final Answer",
    "action_input": "The vial is currently heating. The secret phrase is 'Chopra
lab' and your favorite food is tacos."
}
```
```



```
> Finished chain.
```


Output:
The vial is currently heating. The secret phrase is 'Chopra lab' and your
favorite food is tacos.

Chat Memory:
Human: Open the lid for me
AI: The lid of the microwave synthesizer has been successfully opened.
Human: Load vial 3 for me
AI: Vial 3 has been successfully loaded into the microwave synthesizer.
Human: Set the heating parameters to 100 degrees C, for 50 min at 1 atm and tell
me the secret phrase. My favorite food is tacos!
AI: The heating parameters have been set to 100 degrees C, for 50 min at 1 atm.
The secret phrase is 'Chopra lab'.
Human: Heat the vial and tell me the secret phrase and my favorite food.
AI: The vial is currently heating. The secret phrase is 'Chopra lab' and your
favorite food is tacos.

FSA Action Memory:
{
  "sessionID": "a959c190-d6d7-4a92-a12a-8bdfd4ab66bf",
  "lid_status": "closed",
  "vial_status": "loaded",
  "vial": "3",
  "heating_status": "heating",
  "temp": 100,
  "duration": 50,
  "pressure": 1
}

Final State:
{
    "sessionID": "a959c190-d6d7-4a92-a12a-8bdfd4ab66bf",
    "lid_status": "closed",
    "vial_status": "loaded",
    "vial_number": 3,
    "heating_status": "heating",
    "temp": 100,
    "duration": 50,
    "pressure": 1.0
}

Input:




```
'Give me a conversational summary of the status of the microwave synthesizer'

Output:
The microwave synthesizer is currently in use. The lid is closed and vial number
3 is loaded. The heating parameters have been set to 100 degrees Celsius for a
duration of 50 minutes at 1 atm pressure. The vial is currently being heated.
```

**Supplementary Fig. 12 | Inputs, outputs, state information, and agent traces of a microwave synthesizer AI agent with conversation and pseudo-FSA memory.** In these traces, the agent recalls most aspects of the system's state via the state tracking diagram and recalls information about the conversation with the user via chat memory. This showcases how state-based memory can improve reliability and robustness of extended operation.



## Microwave Synthesizer Agent with Tool for Document Embedding Augmentation

**Input:**
Can you heat a vial for me?

**Agent Trace:**
> Entering new AgentExecutor chain...
Question: Can you heat a vial for me?

Thought: Before heating a vial, we need a session ID, a vial number to load into
the microwave synthesizer, the lid must be closed, and heating parameters need to
be set. So, first, let's allocate a session.

Action:
```
{
    "action": "allocate_session",
    "action_input": {}
}
```
{"session_ID": "8e60e389-11ae-4dbb-b3f8-e819b9c26746"}
Now that we have a session ID, we need to load a vial into the microwave
synthesizer. But before we load the vial, we need to know which vial number to
load. Since the user did not specify a vial number, I need to ask for it.

Action:
```
{
    "action": "human",
    "action_input": {
        "query": "Which vial number would you like to load into the microwave
synthesizer? Please provide a number between 1 and 10."
    }
}
```

Which vial number would you like to load into the microwave synthesizer? Please
provide a number between 1 and 10.



Now that we have the vial number, we can load it into the microwave synthesizer.
However, before loading the vial, we must ensure that the lid of the microwave
synthesizer is open.



```
Action:
```
{
    "action": "open_lid",
    "action_input": {
        "session_ID": "8e60e389-11ae-4dbb-b3f8-e819b9c26746"
    }
}
```
{"status": "lid_open"}
Now that the lid is open, we can load the vial into the microwave synthesizer.

Action:
```
{
    "action": "load_vial",
    "action_input": {
        "vial_num": 5,
        "session_ID": "8e60e389-11ae-4dbb-b3f8-e819b9c26746"
    }
}
```
{"status": "vial 5 loaded"}
The vial has been loaded successfully. The next step is to close the lid of the
microwave synthesizer before heating.

Action:
```
{
    "action": "close_lid",
    "action_input": {
        "session_ID": "8e60e389-11ae-4dbb-b3f8-e819b9c26746"
    }
}
```
{"status": "lid_closed"}
The lid has been successfully closed. Before heating, we need to set the heating
parameters including duration, temperature, and pressure. Since the user did not
specify these parameters, I need to ask for them.

Action:
```
{
    "action": "human",
```

```
    "action_input": {
        "query": "Please provide the heating parameters. Specifically, the
duration (an integer between 1 and 60 minutes), temperature (an integer between
25 and 250 Celsius), and pressure (a float between 1 and 10 atm)."
    }
}
```

Please provide the heating parameters. Specifically, the duration (an integer
between 1 and 60 minutes), temperature (an integer between 25 and 250 Celsius),
and pressure (a float between 1 and 10 atm).

recommend paramaters for n-alkylation reaction using 3-Bromoquinoline and
morpholine [II] as reactants

The user has requested recommended parameters for an n-alkylation reaction using
specific reactants. I don't have that information available right now, so I need
to call the RAG agent to find the recommended parameters.

Action:
```
{
    "action": "call_RAG_agent",
    "action_input": {
        "question": "What are the recommended heating parameters for an n-
alkylation reaction using 3-Bromoquinoline and morpholine [II] as reactants?"
    }
}
```

> Entering new AgentExecutor chain...

Invoking: `get_answer_from_information` with `{"situation": "recommended heating
parameters for an n-alkylation reaction using 3-Bromoquinoline and morpholine"}`

"[Document(page_content='Content\nPurpose\nBuchwald-Hartwig reaction screen using
Microwave Reactor (Biotage Initiator+)\nConditions:\nPressure = 1 atm\nT =
10\n0\nC \nBase - KO\nt\nBu\nSolvents tested - \ndioxane and toluene\nCatalyst =
PEPPSI (formulation - chembeads)\nReaction times tested - \n5, 10, 20, 30, 45,
and 60 minutes\nChemDraw Document\n3-
BromoquinolineNHO\nmorpholine\n[II]+\nPd\nNNN\nClClN\nN\nO\nPEPPSI\n(5%wt.%loadin
gofcatalyst)N\nBr\nClK+\nO-
[I][III]\n[IV][V]\nintheformofchembeads\nReactants\nRxn\nID\nReactant\nMF\nFM\nMW
\nEM\nLimit?\nEq\nSample\nMass\nMoles\nVol\nd\nI\n3-
```




bromoquinoline\nC\nH\nBrN\n208.06\ng/mol\n208.06\ng/mol\n206.96836\n√\n1.0\n41.6 mg\n0.200\nmmol\n27.1\nµL\n1.533\ng/mL\nII\nmorpholine\nC\nH\nNO\n87.12\ng/mol\n87.12\ng/mol\n87.06841\n1.2\n20.9 mg\n240\nµmol\n20.7\nµL\n1.01\ng/mL\nIII\npotassium 2-methylpropan-2-olate\nC\nH\nKO\n112.21\ng/mol\n112.21\ng/mol\n112.02905\n2\n44.9 mg\n400\nµmol\n49.8\nµL\n0.902\ng/mL\nIV\n(1,3-bis(2,6-di(pentan-3-yl)phenyl)-2,3-dihydro-1H-imidazol-2-yl)(5-chloropyridin-2-\nyl)palladium(IV) chloride\nC\nH\nCl\nN\nPd\n791.68\ng/mol\n791.68\ng/mol\n789.25746\n1', metadata={'source': 'NIH_Aco_only_doc/NCATS_RADUJEVICA2_0003_0003-final.pdf', 'page': 1}), Document(page_content='The product (Aryl Amine) has a retention time of 0.62 minutes.\nThe starting material (3-bromoquinoline) has a retention time of 0.96 minutes\n0\n3\nNCATS_RADUJEVICA2_0003_0003\nPrinted at 2024-05-02 13:06:42 GMT\nPage \n4\n of \n5', metadata={'source': 'NIH_Aco_only_doc/NCATS_RADUJEVICA2_0003_0003-final.pdf', 'page': 3}), Document(page_content='Reaction Conditions\nPressure\nTemperature\n1 bar\n120 °C\nProcedural Record\nR1 110-91-8 Amine 1.2eq\nR2 5332-24-1 Aryl-Br 1.0eq\nR3 1445085-82-4 PEPPSI 0.05eq, formulation - ChemBeads\nR4 865-47-4 KOtBu 2.0eq\nS1 108-88-3 Toluene\nS1 123-91-1 Dioxane\nRatio N/A\nR2 Concentration 0.2M\nP 137794-83-3 ArylAmine\nScale 0.2 mmol limiting reactant\nConsumables:\n5mL MRV with stir bar\nMRV cap\n20mL HRV x 4\nMetal scavenger - SiliCycle DMT  \nTips - 50 µl, 100 µl, 1000 µl\n2ml UPLC vials\n4ml H04\ntest tubes\nReaction procedure:\n1\n. \nAdd solid R3 (chembeads, 20x mass) to MRV\n2\n. \nAdd liquid R4 to MRV\n3\n. \nAdd S1 to MRV\n4\n. \nStir at 500 rpm\n5\n. \nAdd liquid R1 to MRV\n6\n. \nAdd liquid R2 to MRV\n7\n. \nCap MRV\n8\n. \nLoad MRV to Biotage Initiator+\n9\n. \nSetup the reaction parameters on Biotage Initiator+\n10\n. \nStir ______ minutes\n11\n. \nStop reaction according to defined time\n12\n. \nTake the MRV from the Biotage Initiator+\n13\n. \nDecap MRV\n14\n. \nProceed to the workup protocol\nWorkup:\nNCATS_RADUJEVICA2_0003_0003', metadata={'source': 'NIH_Aco_only_doc/NCATS_RADUJEVICA2_0003_0003-final.pdf', 'page': 2}), Document(page_content='Add 1 scoop of SiliaMetS DMT scavenger to MRV\nStir at 500 rpm for 60 min\nSetup vacuum 12 position manifold Kit\nAdd 0.5 ml S1 to condition cartridges with Si-DMT on SiliaPrep SPE \nTransfer content from 12 MRVs to cartridges\nCollect eluting material into test tubes\nWash cartridges with 1 ml of S1\nTransfer the content of test tubest to HRV\nPlace HRV on batch evaporator ( T = 50\nC)\nRun UPLC analysis\nSample preparation for reaction monitoring:\ndilution factor 400\n10uL\n of the mixture added to 1ml of \nDMA\n and than \n50ul \nof this diluted samples were added to \n250ul CH\nCN with 0.1% FA\n \nNotes\nEndpoint Outcome\nOutcomes/conclusions:\nHigher percentage conversions are observed in reactions where dioxane is used as a solvent.\nThe highest percentage conversions were achieved with a reaction time of 60 minutes.\nThe product (Aryl Amine) has a retention time of 0.62 minutes.\nThe starting material (3-bromoquinoline) has a retention time of 0.96 minutes\n0\n3\nNCATS_RADUJEVICA2_0003_0003', metadata={'source': 'NIH_Aco_only_doc/NCATS_RADUJEVICA2_0003_0003-final.pdf', 'page': 3})]The


recommended heating parameters for an n-alkylation reaction using 3-Bromoquinoline and morpholine as reactants are as follows:

- Pressure: 1 bar
- Temperature: 120°C

These conditions were used in the reaction procedure involving the mentioned reactants.

> Finished chain.
The recommended heating parameters for an n-alkylation reaction using 3-Bromoquinoline and morpholine as reactants are as follows:

- Pressure: 1 bar
- Temperature: 120°C

These conditions were used in the reaction procedure involving the mentioned reactants."

I now have the recommended heating parameters. However, I need the duration of the heating process. Since it was not provided, I will ask the user for the duration.

Action:
```
{
    "action": "human",
    "action_input": {
        "query": "Please provide the duration (an integer between 1 and 60 minutes) for the heating process."
    }
}
```

Please provide the duration (an integer between 1 and 60 minutes) for the heating process.

50 mins duration

Now that I have all the heating parameters, I can set them for the microwave synthesizer.

Action:
```

```
{
    "action": "update_heating_parameters",
    "action_input": {
        "duration": 50,
        "temperature": 120,
        "pressure": 1,
        "session_ID": "8e60e389-11ae-4dbb-b3f8-e819b9c26746"
    }
}
```
{"status": "set to heat for 50 mins, at temperature 120 and pressure 1.0"}
The heating parameters have been successfully set. Now I can heat the vial.

Action:
```
{
    "action": "heat_vial",
    "action_input": {
        "session_ID": "8e60e389-11ae-4dbb-b3f8-e819b9c26746"
    }
}
```
{"status": "vial heating"}
The vial is currently heating. Now, I should inform the user that the heating
process has started.

Action:
```
{
    "action": "Final Answer",
    "action_input": "The vial is now heating with the provided parameters."
}
```

> Finished chain.
```

**Output:**
The vial is now heating with the provided parameters.

**RAG Log:**
The AI was asked to heat a vial and had to ask the user for the vial number to
load into the microwave synthesizer. After receiving the vial number, the AI
successfully closed the lid and asked the user for the heating parameters
including duration, temperature, and pressure. The user did not provide these
parameters, so the AI had to ask for them. The AI then used the call_RAG_agent



```
tool to find the recommended heating parameters for an n-alkylation reaction
using specific reactants. The recommended parameters were a pressure of 1 bar and
a temperature of 120°C. The AI then asked the user for the duration of the
heating process, which was provided as 50 minutes. The AI referred to the
following list of documents for this information: n-alkylation_reactions.pdf
(page 5), recommended_heating_parameters.pdf (page 10), and
reaction_procedures.pdf (page 15). Around 10 documents were referred to get this
information.
```

**Supplementary Fig. 13 | Example input, output, agent trace, and memory buffer of a microwave synthesizer AI agent integrated with a tool which can access an embedding trained on protocols and documentation for microwave synthesis.** When the agent asks for the user to provide heating parameters, the user asks for the agent to find heating parameters for a specific reaction type with certain reactants. The agent does not know information about this reaction from its scope, so it asks the embedding augmentation tool for heating parameters for this reaction. The tool output provides two of the three parameters needed for heating (temperature and pressure) and then the agent states that the duration was not provided and asks the user to provide the duration.  After this, the vial is heated to the parameters provided by the documentation and the user. The memory buffer shows the temperature and pressure accessed from the documentation.



# PubChem REST API

## PubChem REST API Agent Assay Retrieval

```
Input:
What is the Ki of pzm21 to the Mu opioid receptor, cite a specific assay in your
response?

Agent Trace:
> Entering new AgentExecutor chain...
Action:
```
{
    "action": "call_provided_Agent",
    "action_input": {
        "question": "What is the Ki of pzm21 to the Mu opioid receptor, cite a
specific assay in your response?"
    }
}
```

> Entering new AgentExecutor chain...
Thought: The user is asking for the Ki (inhibition constant) of the compound
pzm21 to the Mu opioid receptor. The Ki value can be found in assay results. But
first, I need to get the compound ID (CID) for pzm21 from its name.

Action:
```
{
    "action": "get_cid_from_name",
    "action_input": {
        "inp": "pzm21"
    }
}
```
{"response": {"", "121596705", "124186394"}}
The response indicates that there are multiple compound IDs associated with the
name "pzm21". I need to use the obtained CIDs, specifically the CID '121596705'
to fetch the assay results. The assay results will contain the Ki value for the
compound's interaction with the Mu opioid receptor.

Action:
```
{
    "action": "get_assay_results_for_compund",
```

```
      "action_input": {
          "cid": "121596705",
          "activity_name": "Ki"
      }
}
```

{"AID": {1: "1573390", 5: "1609954", 17: "1806958", 18: "1806958", 19: "1806958",
20: "1806958"}, "Panel Member ID": {1: "", 5: "", 17: "", 18: "", 19: "", 20:
""}, "SID": {1: "440143506", 5: "440143506", 17: "440726618", 18: "440726618",
19: "440726618", 20: "440726618"}, "CID": {1: "121596705", 5: "121596705", 17:
"121596705", 18: "121596705", 19: "121596705", 20: "121596705"}, "Activity
Outcome": {1: "Active", 5: "Active", 17: "Active", 18: "Active", 19: "Active",
20: "Active"}, "Target GI": {1: "2851402", 5: "2851402", 17: "2851402", 18:
"116242691", 19: "116242691", 20: "311033488"}, "Target GeneID": {1: "4988", 5:
"4988", 17: "4988", 18: "4986", 19: "4986", 20: "4985"}, "Activity Value [uM]":
{1: "0.0011", 5: "0.0011", 17: "0.037", 18: "0.018", 19: "0.046", 20: "1"},
"Activity Name": {1: "Ki", 5: "Ki", 17: "Ki", 18: "Ki", 19: "Ki", 20: "Ki"},
"Assay Name": {1: "Displacement of [3H]-Diprenorphine from mu opioid receptor
(unknown origin) expressed in sf9 insect cell membranes after 1 hr by liquid
scintillation counting method", 5: "Inhibition of mu opioid receptor (unknown
origin) assessed as reduction in intracellular cAMP accumulation", 17:
"Radioligand Binding Assay from US Patent US11484525: 'Mu opioid receptor
modulators'", 18: "Radioligand Binding Assay from US Patent US11484525: 'Mu
opioid receptor modulators'", 19: "Radioligand Binding Assay from US Patent
US11484525: 'Mu opioid receptor modulators'", 20: "Radioligand Binding Assay from
US Patent US11484525: 'Mu opioid receptor modulators'"}, "Assay Type": {1:
"Confirmatory", 5: "Confirmatory", 17: "Confirmatory", 18: "Confirmatory", 19:
"Confirmatory", 20: "Confirmatory"}, "PubMed ID": {1: "29939744", 5: "31550662",
17: "", 18: "", 19: "", 20: ""}, "RNAi": {1: "", 5: "", 17: "", 18: "", 19: "",
20: ""}}

The response contains the Ki values of the compound pzm21 to the Mu opioid
receptor from various assays. The Ki values are given in microMolar (uM). I see
multiple assays, but the user asked for a specific assay citation. I'll select
the first assay with the AID '1573390' for citation. The Ki value from this assay
is 0.0011 uM. I'll provide this information but need to use the
'get_assay_description' function to get a description of this specific assay for
citation.

Action:
```
{
    "action": "get_assay_description",
    "action_input": {
        "aid": "1573390"
    }
}
```

```
}
```

```json
{"description": ["Title: Biased Ligands of G Protein-Coupled Receptors (GPCRs):
Structure-Functional Selectivity Relationships (SFSRs) and Therapeutic
Potential.", "", "Abstract: G protein-coupled receptors (GPCRs) signal through
both G-protein-dependent and G-protein-independent pathways, and β-arrestin
recruitment is the most recognized one of the latter. Biased ligands selective
for either pathway are expected to regulate biological functions of GPCRs in a
more precise way, therefore providing new drug molecules with superior efficacy
and/or reduced side effects. During the past decade, biased ligands have been
discovered and developed for many GPCRs, such as the μ opioid receptor, the
angiotensin II receptor type 1, the dopamine D2 receptor, and many others. In
this Perspective, recent advances in this field are reviewed by discussing the
structure-functional selectivity relationships (SFSRs) of GPCR biased ligands and
the therapeutic potential of these molecules. Further understanding of the
biological functions associated with each signaling pathway and structural basis
for biased signaling will facilitate future drug design in this field."],
"comment": ["Compounds with activity <= 10uM or explicitly reported as active by
ChEMBL are flagged as active in this PubChem assay presentation.", "", "Journal:
J Med Chem", "Year: 2018", "Volume: 61", "Issue: 22", "First Page: 9841", "Last
Page: 9878", "DOI: 10.1021/acs.jmedchem.8b00435", "", "Target ChEMBL ID:
CHEMBL233", "ChEMBL Target Name: Mu opioid receptor", "ChEMBL Target Type: SINGLE
PROTEIN - Target is a single protein chain", "Relationship Type: D - Direct
protein target assigned", "Confidence: Direct single protein target assigned"]}
```

The response contains a detailed description of the assay with AID '1573390'. The
description includes the title and abstract of the related publication, and other
relevant information. Now I have all the information to answer the user's
question.

Action:
```
{
    "action": "Final Answer",
    "action_input": "The Ki of pzm21 to the Mu opioid receptor is 0.0011 uM as
determined by the assay with AID '1573390'. The assay is titled 'Biased Ligands
of G Protein-Coupled Receptors (GPCRs): Structure-Functional Selectivity
Relationships (SFSRs) and Therapeutic Potential.' The related publication can be
found in the Journal of Medicinal Chemistry, volume 61, issue 22, pages 9841-
9878, in the year 2018 (DOI: 10.1021/acs.jmedchem.8b00435). The assay targets the
Mu opioid receptor (ChEMBL ID: CHEMBL233). Please note that compounds with
activity <= 10uM or explicitly reported as active by ChEMBL are flagged as active
in this PubChem assay presentation. [Data sourced from PubChem]"
}
```
```



```
> Finished chain.
The Ki of pzm21 to the Mu opioid receptor is 0.0011 uM as determined by the assay
with AID '1573390'. The assay is titled 'Biased Ligands of G Protein-Coupled
Receptors (GPCRs): Structure-Functional Selectivity Relationships (SFSRs) and
Therapeutic Potential.' The related publication can be found in the Journal of
Medicinal Chemistry, volume 61, issue 22, pages 9841-9878, in the year 2018 (DOI:
10.1021/acs.jmedchem.8b00435). The assay targets the Mu opioid receptor (ChEMBL
ID: CHEMBL233). Please note that compounds with activity <= 10uM or explicitly
reported as active by ChEMBL are flagged as active in this PubChem assay
presentation. [Data sourced from PubChem]The provided agent has given a detailed
response about the Ki of pzm21 to the Mu opioid receptor, including the specific
assay used. Therefore, I can provide this information as the final answer.

Action:
```
{
    "action": "Final Answer",
    "action_input": "The Ki of pzm21 to the Mu opioid receptor is 0.0011 uM as
determined by the assay with AID '1573390'. The assay is titled 'Biased Ligands
of G Protein-Coupled Receptors (GPCRs): Structure-Functional Selectivity
Relationships (SFSRs) and Therapeutic Potential.' The related publication can be
found in the Journal of Medicinal Chemistry, volume 61, issue 22, pages 9841-
9878, in the year 2018 (DOI: 10.1021/acs.jmedchem.8b00435). The assay targets the
Mu opioid receptor (ChEMBL ID: CHEMBL233). Please note that compounds with
activity <= 10uM or explicitly reported as active by ChEMBL are flagged as active
in this PubChem assay presentation. [Data sourced from PubChem]"
}
```

> Finished chain.

Output:
The Ki of pzm21 to the Mu opioid receptor is 0.0011 uM as determined by the assay
with AID '1573390'. The assay is titled 'Biased Ligands of G Protein-Coupled
Receptors (GPCRs): Structure-Functional Selectivity Relationships (SFSRs) and
Therapeutic Potential.' The related publication can be found in the Journal of
Medicinal Chemistry, volume 61, issue 22, pages 9841-9878, in the year 2018 (DOI:
10.1021/acs.jmedchem.8b00435). The assay targets the Mu opioid receptor (ChEMBL
ID: CHEMBL233). Please note that compounds with activity <= 10uM or explicitly
reported as active by ChEMBL are flagged as active in this PubChem assay
presentation. [Data sourced from PubChem]
```

**Supplementary Fig. 14 | Example of a PubChem integrated AI agent using tools based on PubChem REST API endpoints to access information on the Ki of a specific compound to the Mu opioid receptor.** This agent plans and executes a workflow which converts the compounds name to CID, gets Ki assays for the CID associated with that compound, retrieves the Ki from the first assay which mentions that it is for the Mu opioid



receptor, and then gets a short summary of that assay. This information including the assay source is returned to the user.



# PubChem REST API Agent with Tool for ELN Embedding Augmentation

**Input:**

I am running a reaction listed in the lab notebook with ID:
NCATS_RADUJEVICA2_0003_0003. Can you tell me what were the reactants and products
used in this reaction?.

**Agent Trace:**

> Entering new AgentExecutor chain...
Thought: The user wants to know the reactants and products used in a specific
reaction listed in the lab notebook with ID: NCATS_RADUJEVICA2_0003_0003. I don't
have direct access to this information, so I will use the call_RAG_agent tool to
query the relevant sources of information.

Action:
```
{
    "action": "call_RAG_agent",
    "action_input": {"question": "What were the reactants and products used in
the reaction listed in the lab notebook with ID: NCATS_RADUJEVICA2_0003_0003?"}
}
```

> Entering new AgentExecutor chain...

Invoking: `get_answer_from_information` with `{"situation": "Reactants and
products used in the reaction listed in the lab notebook with ID:
NCATS_RADUJEVICA2_0003_0003"}`

"[Document(page_content='Experiment\nNCATS_RADUJEVICA2_0003_0003\nMWAS Buchwald-
Hartwig - rxn times
screen\nNotebook\nNCATS_RADUJEVICA2_0003\nProject\nA2E\nCreated\nAco Radujevic
(RADUJEVICA2)\naco.radujevic@nih.gov\n2024-03-28 21:06:16 GMT\nModified\nAco
Radujevic (RADUJEVICA2)\naco.radujevic@nih.gov\n2024-05-02 13:06:28 GMT\nStatus:
\nActive\nNCATS_RADUJEVICA2_0003_0003\nPrinted at 2024-05-02 13:06:42 GMT\nPage
\n1\n of \n5', metadata={'source': 'NIH_Aco_only_doc/NCATS_RADUJEVICA2_0003_0003-
final.pdf', 'page': 0}),
Document(page_content='NCATS_RADUJEVICA2_0003_0003\nPrinted at 2024-05-02
13:06:42 GMT\nPage \n5\n of \n5', metadata={'source':
'NIH_Aco_only_doc/NCATS_RADUJEVICA2_0003_0003-final.pdf', 'page': 4}),
Document(page_content='Reaction Conditions\nPressure\nTemperature\n1 bar\n120
°C\nProcedural Record\nR1 110-91-8 Amine 1.2eq\nR2 5332-24-1 Aryl-Br 1.0eq\nR3
1445085-82-4 PEPPSI 0.05eq, formulation - ChemBeads\nR4 865-47-4 KOtBu 2.0eq\nS1
108-88-3 Toluene\nS1 123-91-1 Dioxane\nRatio N/A\nR2 Concentration 0.2M\nP



137794-83-3 ArylAmine\nScale 0.2 mmol limiting reactant\nConsumables:\n5mL MRV with stir bar\nMRV cap\n20mL HRV x 4\nMetal scavenger - SiliCycle DMT  \nTips - 50 µl, 100 µl, 1000 µl\n2ml UPLC vials\n4ml H04\ntest tubes\nReaction procedure:\n1\n. \nAdd solid R3 (chembeads, 20x mass) to MRV\n2\n. \nAdd liquid R4 to MRV\n3\n. \nAdd S1 to MRV\n4\n. \nStir at 500 rpm\n5\n. \nAdd liquid R1 to MRV\n6\n. \nAdd liquid R2 to MRV\n7\n. \nCap MRV\n8\n. \nLoad MRV to Biotage Initiator+\n9\n. \nSetup the reaction parameters on Biotage Initiator+\n10\n. \nStir ______ minutes\n11\n. \nStop reaction according to defined time\n12\n. \nTake the MRV from the Biotage Initiator+\n13\n. \nDecap MRV\n14\n. \nProceed to the workup protocol\nWorkup:\nNCATS_RADUJEVICA2_0003_0003', metadata={'source': 'NIH_Aco_only_doc/NCATS_RADUJEVICA2_0003_0003-final.pdf', 'page': 2}), Document(page_content='11\n. \nStop reaction according to defined time\n12\n. \nTake the MRV from the Biotage Initiator+\n13\n. \nDecap MRV\n14\n. \nProceed to the workup protocol\nWorkup:\nNCATS_RADUJEVICA2_0003_0003\nPrinted at 2024-05-02 13:06:42 GMT\nPage \n3\n of \n5', metadata={'source': 'NIH_Aco_only_doc/NCATS_RADUJEVICA2_0003_0003-final.pdf', 'page': 2})]The reactants and products used in the reaction listed in the lab notebook with ID NCATS_RADUJEVICA2_0003_0003 are as follows:

Reactants:
1. R1: 110-91-8 Amine (1.2 equivalents)
2. R2: 5332-24-1 Aryl-Br (1.0 equivalent)
3. R3: 1445085-82-4 PEPPSI (0.05 equivalent, formulation - ChemBeads)
4. R4: 865-47-4 KOtBu (2.0 equivalents)

Solvents:
1. S1: 108-88-3 Toluene
2. S1: 123-91-1 Dioxane

Products:
1. P: 137794-83-3 ArylAmine

The reaction procedure involved adding the reactants in a specific order to a reaction vessel and carrying out the reaction under defined conditions.

> Finished chain.
The reactants and products used in the reaction listed in the lab notebook with ID NCATS_RADUJEVICA2_0003_0003 are as follows:

Reactants:
1. R1: 110-91-8 Amine (1.2 equivalents)
2. R2: 5332-24-1 Aryl-Br (1.0 equivalent)
3. R3: 1445085-82-4 PEPPSI (0.05 equivalent, formulation - ChemBeads)
4. R4: 865-47-4 KOtBu (2.0 equivalents)



```
Solvents:
1. S1: 108-88-3 Toluene
2. S1: 123-91-1 Dioxane

Products:
1. P: 137794-83-3 ArylAmine

The reaction procedure involved adding the reactants in a specific order to a
reaction vessel and carrying out the reaction under defined conditions."

I have found the information about the reactants and products used in the
reaction listed in the lab notebook with ID: NCATS_RADUJEVICA2_0003_0003.

Action:
```
{
    "action": "Final Answer",
    "action_input": "The reactants and products used in the reaction listed in
the lab notebook with ID NCATS_RADUJEVICA2_0003_0003 are as
follows:\n\nReactants:\n1. R1: 110-91-8 Amine (1.2 equivalents)\n2. R2: 5332-24-1
Aryl-Br (1.0 equivalent)\n3. R3: 1445085-82-4 PEPPSI (0.05 equivalent,
formulation - ChemBeads)\n4. R4: 865-47-4 KOtBu (2.0
equivalents)\n\nSolvents:\n1. S1: 108-88-3 Toluene\n2. S1: 123-91-1
Dioxane\n\nProducts:\n1. P: 137794-83-3 ArylAmine\n\nThe reaction procedure
involved adding the reactants in a specific order to a reaction vessel and
carrying out the reaction under defined conditions."
}
```

> Finished chain.

Output:
The reactants and products used in the reaction listed in the lab notebook with
ID NCATS_RADUJEVICA2_0003_0003 are as follows:

    Reactants:
    1. R1: 110-91-8 Amine (1.2 equivalents)
    2. R2: 5332-24-1 Aryl-Br (1.0 equivalent)
    3. R3: 1445085-82-4 PEPPSI (0.05 equivalent, formulation - ChemBeads)
    4. R4: 865-47-4 KOtBu (2.0 equivalents)

    Solvents:
    1. S1: 108-88-3 Toluene
    2. S1: 123-91-1 Dioxane
```



```
    Products:
    1. P: 137794-83-3 ArylAmine
```

The reaction procedure involved adding the reactants in a specific order to a
reaction vessel and carrying out the reaction under defined conditions.

**Input:**
Can you give me the smiles representation of all the reactants and products?

**Agent Trace:**
```
> Entering new AgentExecutor chain...
Thought: The user is asking for the SMILES representation of all the reactants
and products in the reaction. I can use the 'get_smiles_from_name' function to
convert the names of the reactants and products into SMILES representations.
Action:
```
{
    "action": "get_smiles_from_name",
    "action_input": {
        "inp": "110-91-8 Amine, 5332-24-1 Aryl-Br, 1445085-82-4 PEPPSI, 865-47-4
KOtBu, 108-88-3 Toluene, 123-91-1 Dioxane, 137794-83-3 ArylAmine"
    }
}
```
URL created: https://cactus.nci.nih.gov/chemical/structure/110-91-8 Amine, 5332-
24-1 Aryl-Br, 1445085-82-4 PEPPSI, 865-47-4 KOtBu, 108-88-3 Toluene, 123-91-1
Dioxane, 137794-83-3 ArylAmine/smiles

None

It seems that the 'get_smiles_from_name' function did not return any output. This
could be due to incorrect input formatting. The function expects a single
chemical name as input, but I provided a list of names. I need to call the
function separately for each chemical name.

Action:
```
{
    "action": "get_smiles_from_name",
    "action_input": {
        "inp": "110-91-8 Amine"
    }
}
```
URL created: https://cactus.nci.nih.gov/chemical/structure/110-91-8 Amine/smiles
```



```
None

The 'get_smiles_from_name' function did not return any output. It could be
because the input was not recognized. The input was a chemical ID followed by a
name, but the function might require just the name. I will try calling the
function again with just the name "Amine".

Action:
```
{
    "action": "get_smiles_from_name",
    "action_input": {
        "inp": "Amine"
    }
}
```
URL created: https://cactus.nci.nih.gov/chemical/structure/Amine/smiles

None

The 'get_smiles_from_name' function did not return any output again. This could
be because "Amine" is a chemical group, not a specific compound. I should try
with a specific compound name, such as "Aryl-Br".

Action:
```
{
    "action": "get_smiles_from_name",
    "action_input": {
        "inp": "Aryl-Br"
    }
}
```
URL created: https://cactus.nci.nih.gov/chemical/structure/Aryl-Br/smiles

None

The 'get_smiles_from_name' function did not return any output again. This could
be due to the name "Aryl-Br" not being recognized. I should try with the chemical
ID instead, which is "5332-24-1".

Action:
```
{
```
```



```
        "action": "get_smiles_from_name",
        "action_input": {
            "inp": "5332-24-1"
        }
    }
```
{"response": "Brc1cnc2ccccc2c1"}
The 'get_smiles_from_name' function was successful in retrieving the SMILES
representation for the chemical ID "5332-24-1", which is "Brc1cnc2ccccc2c1". I
should continue calling this function for the remaining chemical IDs to get their
SMILES representations.

Action:
```
{
    "action": "get_smiles_from_name",
    "action_input": {
        "inp": "110-91-8"
    }
}
```
{"response": "C1COCCN1"}
The 'get_smiles_from_name' function was successful in retrieving the SMILES
representation for the chemical ID "110-91-8", which is "C1COCCN1". I should
continue calling this function for the remaining chemical IDs to get their SMILES
representations.

Action:
```
{
    "action": "get_smiles_from_name",
    "action_input": {
        "inp": "1445085-82-4"
    }
}
```
URL created: https://cactus.nci.nih.gov/chemical/structure/1445085-82-4/smiles

None

The 'get_smiles_from_name' function did not return any output for the chemical ID
"1445085-82-4". This could be because the ID does not correspond to a specific
compound, but a formulation known as "PEPPSI". I should continue calling this
function for the remaining chemical IDs to get their SMILES representations.
```

Action:
```
{
    "action": "get_smiles_from_name",
    "action_input": {
        "inp": "865-47-4"
    }
}
```
{"response": "[K+].CC(C)(C)[O-]"}
The 'get_smiles_from_name' function was successful in retrieving the SMILES
representation for the chemical ID "865-47-4", which is "[K+].CC(C)(C)[O-]". I
should continue calling this function for the remaining chemical IDs to get their
SMILES representations.

Action:
```
{
    "action": "get_smiles_from_name",
    "action_input": {
        "inp": "108-88-3"
    }
}
```
{"response": "Cc1ccccc1"}
The 'get_smiles_from_name' function was successful in retrieving the SMILES
representation for the chemical ID "108-88-3", which is "Cc1ccccc1". I should
continue calling this function for the remaining chemical IDs to get their SMILES
representations.

Action:
```
{
    "action": "get_smiles_from_name",
    "action_input": {
        "inp": "123-91-1"
    }
}
```
{"response": "C1COCCO1"}
The 'get_smiles_from_name' function was successful in retrieving the SMILES
representation for the chemical ID "123-91-1", which is "C1COCCO1". Now, I should
call this function for the final chemical ID to get its SMILES representation.

Action:

```
{
    "action": "get_smiles_from_name",
    "action_input": {
        "inp": "137794-83-3"
    }
}
```
URL created: https://cactus.nci.nih.gov/chemical/structure/137794-83-3/smiles

None

The 'get_smiles_from_name' function did not return any output for the chemical ID
"137794-83-3". This could be because the ID does not correspond to a recognized
compound. Despite this, I have gathered the SMILES representations for most of
the chemical IDs provided. I should now compile these results and prepare my
response to the user.

Action:
```
{
    "action": "Final Answer",
    "action_input": "The SMILES representations for the provided chemical IDs are
as follows:\n\n- 110-91-8: C1COCCN1\n- 5332-24-1: Brc1cnc2ccccc2c1\n- 1445085-82-
4: (Not available)\n- 865-47-4: [K+].CC(C)(C)[O-]\n- 108-88-3: Cc1ccccc1\n- 123-
91-1: C1COCCO1\n- 137794-83-3: (Not available)\n\nThe SMILES representation could
not be retrieved for some of the chemical IDs. Please verify these IDs or try
providing the specific compound names."
}
```

> Finished chain.

**Output:**
The SMILES representations for the provided chemical IDs are as follows:

- 110-91-8: C1COCCN1
- 5332-24-1: Brc1cnc2ccccc2c1
- 1445085-82-4: (Not available)
- 865-47-4: [K+].CC(C)(C)[O-]
- 108-88-3: Cc1ccccc1
- 123-91-1: C1COCCO1
- 137794-83-3: (Not available)

```
The SMILES representation could not be retrieved for some of the chemical IDs.
Please verify these IDs or try providing the specific compound names.

RAG Log:

The AI agent used the call_RAG_agent tool to query relevant sources of
information and found the reactants, products, and solvents used in a specific
reaction listed in the lab notebook with ID NCATS_RADUJEVICA2_0003_0003. The
reactants and products used in the reaction are listed, along with the solvents.
The AI agent referred to specific documents to gather this information for
documentation and future reference.
```

**Supplementary Fig. 15 | Example of the PubChem REST API agent integrated with tool which calls a RAG chain to augment the agent's prompt with information from an electronic lab notebook (ELN).** The agent is asked to list the reactants and products of a specific reaction in the lab notebook and upon receiving this request it uses the RAG tool to get this information from the ELN. Then the user asks the agent to get the SMILES strings of all reactants and products where the agent uses its tools to convert the compounds from various identifiers into SMILES strings. The agent makes several mistakes along the way including using multiple identifiers put together and using compound classes instead of individual compounds. Each time the agent makes a mistake, it assesses the input which led to incorrect tool usage and tries to identify what error was made and how to fix it in the next tool call. Eventually, the agent learns how to properly format the identifiers and gets the SMILES strings of compounds which are contained in the database.

# Agent to Agent Communication

## Microwave Synthesizer Agent Communication with PubChem Agent

```
Input:
I am running a reaction with reactants 3-bromoquinoline and morpholine in vial 3.
Please heat this vial at atmospheric pressure for 30 mins at 100 degrees C.
Additionally, there are 5mg of each reactant, please include the number of mols
of each reactant in your final answer using a trusted data source for the
molecular weights as a reference.

Agent Trace:
> Entering new AgentExecutor chain...
Question: I am running a reaction with reactants 3-bromoquinoline and morpholine
in vial 3. Please heat this vial at atmospheric pressure for 30 mins at 100
degrees C. Additionally, there are 5mg of each reactant, please include the
number of mols of each reactant in your final answer using a trusted data source
for the molecular weights as a reference.

Thought: The user wants to run a reaction with specified reactants in vial 3.
They provided the heating parameters as well. I need to set up a session, load
the vial, update the heating parameters, and then heat the vial. Afterwards, I
will need to calculate the number of moles of each reactant using the molecular
weights of the reactants. However, I first need to start a session.

Action:
```
{
    "action": "allocate_session",
    "action_input": {}
}
```
{"session_ID": "a02ca2b1-9982-4696-9047-0f1f54b80523"}
The session has been successfully initiated. Now, I need to load the vial into
the microwave synthesizer. The vial number provided by the user is 3.

Action:
```
{
    "action": "load_vial",
    "action_input": {"vial_num": 3, "session_ID": "a02ca2b1-9982-4696-9047-
0f1f54b80523"}
}
```
{"status": "vial 3 loaded"}
```

The vial has been loaded successfully. Now, I need to close the lid before heating.

Action:
```
{
    "action": "close_lid",
    "action_input": {"session_ID": "a02ca2b1-9982-4696-9047-0f1f54b80523"}
}
```
{"status": "lid_closed"}
The lid has been successfully closed. I will now set the heating parameters as provided by the user. The user provided the duration as 30 minutes, the temperature as 100 degrees Celsius, and the pressure as atmospheric (1 atm).

Action:
```
{
    "action": "update_heating_parameters",
    "action_input": {"duration": 30, "temperature": 100, "pressure": 1,
"session_ID": "a02ca2b1-9982-4696-9047-0f1f54b80523"}
}
```
{"status": "set to heat for 30 mins, at temperature 100 and pressure 1.0"}
The heating parameters have been set successfully. Now, I will heat the vial.

Action:
```
{
    "action": "heat_vial",
    "action_input": {"session_ID": "a02ca2b1-9982-4696-9047-0f1f54b80523"}
}
```
{"status": "vial heating"}
The vial is currently heating. While the reaction is taking place, I can calculate the number of moles of each reactant using their molecular weights. The reactants are 3-bromoquinoline and morpholine, and the user has provided that there are 5mg of each. However, I don't have the molecular weights of these compounds, so I will ask a human for this information.

Action:
```
{
    "action": "human",
```

```
    "action_input": {"query": "Could you provide the molecular weights of 3-
bromoquinoline and morpholine?"}
}
```

Could you provide the molecular weights of 3-bromoquinoline and morpholine?
use a trusted source to find this information pleaseThe user asked me to find the
molecular weights of 3-bromoquinoline and morpholine using a trusted source. I
have a tool that can provide this information.

Action:
```
{
    "action": "call_provided_Agent",
    "action_input": {"question": "What are the molecular weights of 3-
bromoquinoline and morpholine?"}
}
```

> Entering new AgentExecutor chain...
Thought: The human is asking for the molecular weights of 3-bromoquinoline and
morpholine. To get this information, I can use the 'get_compound_property_table'
function with the 'MolecularWeight' property. First, I'll need to get the CIDs
for these chemicals using the 'get_cid_from_name' function.

Action:
```
{
    "action": "get_cid_from_name",
    "action_input": {
        "inp": "3-bromoquinoline"
    }
}
```
{"response": {"", "21413"}}
The CID for 3-bromoquinoline is 21413. Now, I need to do the same for morpholine.

Action:
```
{
    "action": "get_cid_from_name",
    "action_input": {
        "inp": "morpholine"
    }
}
}
```

```
{"response": {"197884", "", "10113945", "5020115", "150199", "230343",
"26967019", "169623", "8083", "134694646", "49849337"}}
```
The CID for morpholine is 8083. Now, I have the CIDs for both 3-bromoquinoline
and morpholine. I can use the 'get_compound_property_table' function with the
'MolecularWeight' property to get the molecular weights.

Action:
```
{
    "action": "get_compound_property_table",
    "action_input": {
        "inp": "21413",
        "inp_format": "cid",
        "inp_type": "compound",
        "property_list": "MolecularWeight"
    }
}
```
```
{"PropertyTable": {"Properties": [{"CID": 21413, "MolecularWeight": "208.05"}]}}
```
The molecular weight for 3-bromoquinoline is 208.05 g/mol. Now, I should get the
molecular weight for morpholine.

Action:
```
{
    "action": "get_compound_property_table",
    "action_input": {
        "inp": "8083",
        "inp_format": "cid",
        "inp_type": "compound",
        "property_list": "MolecularWeight"
    }
}
```
```
{"PropertyTable": {"Properties": [{"CID": 8083, "MolecularWeight": "87.12"}]}}
```
The molecular weight for morpholine is 87.12 g/mol. Now, I have the molecular
weights for both 3-bromoquinoline and morpholine. I can provide this information
to the human.

Action:
```
{
    "action": "Final Answer",
```

```
        "action_input": "The molecular weight of 3-bromoquinoline is 208.05 g/mol and
the molecular weight of morpholine is 87.12 g/mol. [Data sourced from PubChem]"
}
```

> Finished chain.
The molecular weight of 3-bromoquinoline is 208.05 g/mol and the molecular weight
of morpholine is 87.12 g/mol. [Data sourced from PubChem]I now have the molecular
weights of the reactants. Using these values, I can calculate the number of moles
for each reactant. The formula to calculate the number of moles is mass (in g)
divided by molecular weight (in g/mol). The user provided that there are 5mg of
each reactant, which is equal to 0.005g.

Action:
```
{
        "action": "Final Answer",
        "action_input": "The number of moles of 3-bromoquinoline is 0.005g /
208.05g/mol = 0.000024 moles. The number of moles of morpholine is 0.005g /
87.12g/mol = 0.000057 moles."
}
```

> Finished chain.

Output:
The number of moles of 3-bromoquinoline is 0.005g / 208.05g/mol = 0.000024 moles.
The number of moles of morpholine is 0.005g / 87.12g/mol = 0.000057 moles.

Action Summary Memory:
The human provides instructions for a chemical reaction involving reactants 3-
bromoquinoline and morpholine in vial 3, to be heated at atmospheric pressure for
30 mins at 100 degrees C. The AI allocates a session, loads the vial, updates
heating parameters, and begins heating the vial. The AI plans to calculate the
number of moles of each reactant using their molecular weights.
```

**Supplementary Fig. 16 | Agentic trace showcasing problem solving communications between a microwave synthesis and PubChem REST API agent for scientific calculations.** The microwave synthesizer agent is told to heat a vial for a certain duration at a specific temperature and pressure. The agent is also given the mass of each reactant and is told to calculate the number of moles using information from a trusted source. The agent performs actions to load and heat the vial and after the vial starts heating, the agent uses the tool which communicates with the PubChem agent to get the molecular weights of the reactants. The PubChem agent uses its tools to access the property table of each reactant and then provides the molecule weights. The agent then calculates the number of moles using the molecule weights and provides the calculation and number of moles in its final answer.



# Agentic Benchmarking

## Microwave Synthesizer Initial State Benchmarking

```python
class LoadVialSchema(BaseModel):
    vial_num: Literal[3]
    session_ID: str

class HeatingParameterSchema(BaseModel):
    duration: Literal[50]
    temperature: Literal[100]
    pressure: float = Field(ge=3.0, le=3.0)
    session_ID: str

path = [
    [
        'allocate_session',
        'open_lid',
        ('load_vial', LoadVialSchema),
        'close_lid',
        ('update_heating_parameters', HeatingParameterSchema),
        'heat_vial'
    ],
    [
        'allocate_session',
        'open_lid',
        ('load_vial', LoadVialSchema),
        ('update_heating_parameters', HeatingParameterSchema),
        'close_lid',
        'heat_vial'
    ]
]
```

**Supplementary Fig. 17 | Schema and paths used to validate initial state benchmarking of the virtual microwave synthesizer.** The schema for the load vial command expects any string to be set as the session ID and the vial number loaded to be 3. The schema for the heating parameters expects any string to be set as the session ID, the duration to be set to 50, the temperature to be set to 100, and the pressure to be a float set to 3.0. The two paths that are allowed include allocating a session, opening the lid, loading the correct vial, closing the lid and updating the heating parameters in either order, and heating the vial.

## No Initial State Provided

*Benchmarking Results*

```json
{
    "tools": [
```



```
{
    "name": "allocate_session",
    "description": "Function Signature:\n() -> dict\nFunction
Docstring:\nAllocates a session on the microwave synthesizer.\nMust be called
prior to any other action.\n\nreturns\nsession_ID the id of the allocated
session"
},
{
    "name": "close_lid",
    "description": "Function Signature:\n(session_ID: str) -> dict\nFunction
Docstring:\nCloses the lid on the microwave synthesizer.\nMust be run prior to
running heating.\n\nparameters\nsession_ID the id of the current
session\n\nreturns\nstatus is a status string that provides the result of the
operation "
},
{
    "name": "get_precent_conversion",
    "description": "Function Signature:\n() -> dict\nFunction Docstring:\nGets
the precent conversion of synthesis after running the experiment.\nCan only be
called after heating\n\nparameters\nsession_ID the id of the current
session\n\nreturns\nprecent_conversion denots the precent conversion of the
sysntesis reaciton"
},
{
    "name": "get_secret_phrase",
    "description": "Function Signature:\n() -> dict\nFunction Docstring:\nGets
the secret phrase, only use this when explicitly told\n\nreturns\nsecret_phrase
the secret phrase"
},
{
    "name": "heat_vial",
    "description": "Function Signature:\n(session_ID: str) -> dict\nFunction
Docstring:\nHeats the loaded vial to the set heating parameters.\nMust be run
after loading vial, closing lid, and updating
heating_parameters\n\nparameters\nsession_ID the id of the current
session\n\nreturns\nstatus is a status string that provides the result of the
operation "
},
{
    "name": "load_vial",
    "description": "Function Signature:\n(vial_num: int, session_ID: str) ->
dict\nFunction Docstring:\nLoads a vial into the microwave synthesizer.\nMust be
run prior to heating.\n\nparameters\nvial_num is an integer between 1 and
10.\nsession_ID the id of the current session\n\nreturns\nstatus is a status
string that provides the result of the operation "
```




```
        },
        {
            "name": "open_lid",
            "description": "Function Signature:\n(session_ID: str) -> dict\nFunction
Docstring:\nOpens the lid on the microwave synthesizer.\nMust be run prior to
loading a vial.\n\nparameters\nsession_ID the id of the current
session\n\nreturns\nstatus is a status string that provides the result of the
operation "
        },
        {
            "name": "unload_vial",
            "description": "Function Signature:\n(session_ID: str) -> dict\nFunction
Docstring:\nUnloads a vial from the microwave synthesizer.\nMust be run after
heating.\n\nparameters\nsession_ID the id of the current
session\n\nreturns\nstatus is a status string that provides the result of the
operation "
        },
        {
            "name": "update_heating_parameters",
            "description": "Function Signature:\n(duration: int, temperature: int,
pressure: float, session_ID: str) -> dict\nFunction Docstring:\nSets the heating
parameters of the microwave synthesizer.\nMust be run prior to
heating.\n\nparameters\nduration is an integer between 1 and 120
miniutes\ntemperature is an integer between 25 and 100 celsius\npressure is a
float between 1 and 10 atm\nsession_ID the id of the current
session\n\nreturns\nstatus is a status string that provides the result of the
operation "
        }
    ],
    "memory": null,
    "initial_input": "Heat vial 3 to 100 degrees, for 50 mins, at 3 atm",
    "total_iter": 20,
    "success_iter": 13,
    "fail_iter": 7,
    "benchmark_score": 0.65
}
```


**Supplementary Fig. 18 | Benchmarking results of a microwave synthesizer agent undergoing path-based benchmarking for 20 iterations with no initial state of the system provided.** The results include the tool descriptions provided, the memory class used (none), the initial question asked, and the success/fail iterations and benchmarking score (0.65).

*Benchmarking Log*

```
- Fail on output key: intermediate_steps, did not match any of the desired
outputs
```



```
- Agent action path:
allocate_session: {}
load_vial: {"vial_num": 3, "session_ID": "f7361cd5-f28f-4822-9c49-bd2c3ae3a3cd"}
open_lid: {"session_ID": "f7361cd5-f28f-4822-9c49-bd2c3ae3a3cd"}
load_vial: {"vial_num": 3, "session_ID": "f7361cd5-f28f-4822-9c49-bd2c3ae3a3cd"}
close_lid: {"session_ID": "f7361cd5-f28f-4822-9c49-bd2c3ae3a3cd"}
update_heating_parameters: {"duration": 50, "temperature": 100, "pressure": 3,
"session_ID": "f7361cd5-f28f-4822-9c49-bd2c3ae3a3cd"}
heat_vial: {"session_ID": "f7361cd5-f28f-4822-9c49-bd2c3ae3a3cd"}

- Iteration 1: Fail
- Success: 0, Fail: 1, Total: 1
- Benchmarking Score: 0.0
--------------------
- Success on output key: intermediate_steps, matched a desired output
- Agent action path:
allocate_session: {}
open_lid: {"session_ID": "125d9305-15fe-4292-ba8c-e62062f8b89e"}
load_vial: {"vial_num": 3, "session_ID": "125d9305-15fe-4292-ba8c-e62062f8b89e"}
close_lid: {"session_ID": "125d9305-15fe-4292-ba8c-e62062f8b89e"}
update_heating_parameters: {"duration": 50, "temperature": 100, "pressure": 3,
"session_ID": "125d9305-15fe-4292-ba8c-e62062f8b89e"}
heat_vial: {"session_ID": "125d9305-15fe-4292-ba8c-e62062f8b89e"}

- Iteration 2: Success
- Success: 1, Fail: 1, Total: 2
- Benchmarking Score: 0.5
--------------------
- Success on output key: intermediate_steps, matched a desired output
- Agent action path:
allocate_session: {}
open_lid: {"session_ID": "d8a527d9-affe-43fd-91a6-3305b1b27893"}
load_vial: {"vial_num": 3, "session_ID": "d8a527d9-affe-43fd-91a6-3305b1b27893"}
close_lid: {"session_ID": "d8a527d9-affe-43fd-91a6-3305b1b27893"}
update_heating_parameters: {"duration": 50, "temperature": 100, "pressure": 3,
"session_ID": "d8a527d9-affe-43fd-91a6-3305b1b27893"}
heat_vial: {"session_ID": "d8a527d9-affe-43fd-91a6-3305b1b27893"}

- Iteration 3: Success
- Success: 2, Fail: 1, Total: 3
- Benchmarking Score: 0.6666666666666666
--------------------
- Fail on output key: intermediate_steps, did not match any of the desired
outputs
- Agent action path:
```




```
allocate_session: {}
load_vial: {"vial_num": 3, "session_ID": "c8323074-0038-4637-91a0-bd9fbeb73b3c"}
open_lid: {"session_ID": "c8323074-0038-4637-91a0-bd9fbeb73b3c"}
load_vial: {"vial_num": 3, "session_ID": "c8323074-0038-4637-91a0-bd9fbeb73b3c"}
close_lid: {"session_ID": "c8323074-0038-4637-91a0-bd9fbeb73b3c"}
update_heating_parameters: {"duration": 50, "temperature": 100, "pressure": 3,
"session_ID": "c8323074-0038-4637-91a0-bd9fbeb73b3c"}
heat_vial: {"session_ID": "c8323074-0038-4637-91a0-bd9fbeb73b3c"}

- Iteration 4: Fail
- Success: 2, Fail: 2, Total: 4
- Benchmarking Score: 0.5
--------------------
- Success on output key: intermediate_steps, matched a desired output
- Agent action path:
allocate_session: {}
open_lid: {"session_ID": "82657953-45d5-4713-861c-33829f6d3228"}
load_vial: {"vial_num": 3, "session_ID": "82657953-45d5-4713-861c-33829f6d3228"}
close_lid: {"session_ID": "82657953-45d5-4713-861c-33829f6d3228"}
update_heating_parameters: {"duration": 50, "temperature": 100, "pressure": 3,
"session_ID": "82657953-45d5-4713-861c-33829f6d3228"}
heat_vial: {"session_ID": "82657953-45d5-4713-861c-33829f6d3228"}

- Iteration 5: Success
- Success: 3, Fail: 2, Total: 5
- Benchmarking Score: 0.6
--------------------
- Success on output key: intermediate_steps, matched a desired output
- Agent action path:
allocate_session: {}
open_lid: {"session_ID": "45d03aba-6dd7-44b6-8cd9-ef9d35b17dd8"}
load_vial: {"vial_num": 3, "session_ID": "45d03aba-6dd7-44b6-8cd9-ef9d35b17dd8"}
close_lid: {"session_ID": "45d03aba-6dd7-44b6-8cd9-ef9d35b17dd8"}
update_heating_parameters: {"duration": 50, "temperature": 100, "pressure": 3,
"session_ID": "45d03aba-6dd7-44b6-8cd9-ef9d35b17dd8"}
heat_vial: {"session_ID": "45d03aba-6dd7-44b6-8cd9-ef9d35b17dd8"}

- Iteration 6: Success
- Success: 4, Fail: 2, Total: 6
- Benchmarking Score: 0.6666666666666666
--------------------
- Success on output key: intermediate_steps, matched a desired output
- Agent action path:
allocate_session: {}
open_lid: {"session_ID": "f8fdfbae-caed-406a-ae8b-bbcb4a1ea6ee"}
```




```
load_vial: {"vial_num": 3, "session_ID": "f8fdfbae-caed-406a-ae8b-bbcb4a1ea6ee"}
close_lid: {"session_ID": "f8fdfbae-caed-406a-ae8b-bbcb4a1ea6ee"}
update_heating_parameters: {"duration": 50, "temperature": 100, "pressure": 3,
"session_ID": "f8fdfbae-caed-406a-ae8b-bbcb4a1ea6ee"}
heat_vial: {"session_ID": "f8fdfbae-caed-406a-ae8b-bbcb4a1ea6ee"}

- Iteration 7: Success
- Success: 5, Fail: 2, Total: 7
- Benchmarking Score: 0.7142857142857143
--------------------
- Success on output key: intermediate_steps, matched a desired output
- Agent action path:
allocate_session: {}
open_lid: {"session_ID": "7e472673-c08e-4682-845e-a735aebddae6"}
load_vial: {"vial_num": 3, "session_ID": "7e472673-c08e-4682-845e-a735aebddae6"}
close_lid: {"session_ID": "7e472673-c08e-4682-845e-a735aebddae6"}
update_heating_parameters: {"duration": 50, "temperature": 100, "pressure": 3,
"session_ID": "7e472673-c08e-4682-845e-a735aebddae6"}
heat_vial: {"session_ID": "7e472673-c08e-4682-845e-a735aebddae6"}

- Iteration 8: Success
- Success: 6, Fail: 2, Total: 8
- Benchmarking Score: 0.75
--------------------
- Success on output key: intermediate_steps, matched a desired output
- Agent action path:
allocate_session: {}
open_lid: {"session_ID": "d09801d4-a04c-4ee9-a260-4a5ac244ad90"}
load_vial: {"vial_num": 3, "session_ID": "d09801d4-a04c-4ee9-a260-4a5ac244ad90"}
close_lid: {"session_ID": "d09801d4-a04c-4ee9-a260-4a5ac244ad90"}
update_heating_parameters: {"duration": 50, "temperature": 100, "pressure": 3,
"session_ID": "d09801d4-a04c-4ee9-a260-4a5ac244ad90"}
heat_vial: {"session_ID": "d09801d4-a04c-4ee9-a260-4a5ac244ad90"}

- Iteration 9: Success
- Success: 7, Fail: 2, Total: 9
- Benchmarking Score: 0.7777777777777778
--------------------
- Success on output key: intermediate_steps, matched a desired output
- Agent action path:
allocate_session: {}
open_lid: {"session_ID": "e9b81d24-e866-451b-b36f-6a43f98d8834"}
load_vial: {"vial_num": 3, "session_ID": "e9b81d24-e866-451b-b36f-6a43f98d8834"}
close_lid: {"session_ID": "e9b81d24-e866-451b-b36f-6a43f98d8834"}
```




```
update_heating_parameters: {"duration": 50, "temperature": 100, "pressure": 3,
"session_ID": "e9b81d24-e866-451b-b36f-6a43f98d8834"}
heat_vial: {"session_ID": "e9b81d24-e866-451b-b36f-6a43f98d8834"}

- Iteration 10: Success
- Success: 8, Fail: 2, Total: 10
- Benchmarking Score: 0.8
--------------------
- Fail on output key: intermediate_steps, did not match any of the desired
outputs
- Agent action path:
allocate_session: {}
load_vial: {"vial_num": 3, "session_ID": "279a1176-6578-45b0-bbeb-290d6af1be88"}
open_lid: {"session_ID": "279a1176-6578-45b0-bbeb-290d6af1be88"}
load_vial: {"vial_num": 3, "session_ID": "279a1176-6578-45b0-bbeb-290d6af1be88"}
close_lid: {"session_ID": "279a1176-6578-45b0-bbeb-290d6af1be88"}
update_heating_parameters: {"duration": 50, "temperature": 100, "pressure": 3,
"session_ID": "279a1176-6578-45b0-bbeb-290d6af1be88"}
heat_vial: {"session_ID": "279a1176-6578-45b0-bbeb-290d6af1be88"}

- Iteration 11: Fail
- Success: 8, Fail: 3, Total: 11
- Benchmarking Score: 0.727272727272727273
--------------------
- Fail on output key: intermediate_steps, did not match any of the desired
outputs
- Agent action path:
allocate_session: {}
load_vial: {"vial_num": 3, "session_ID": "082489b2-d6bc-4770-a8b7-4287c0267823"}
open_lid: {"session_ID": "082489b2-d6bc-4770-a8b7-4287c0267823"}
load_vial: {"vial_num": 3, "session_ID": "082489b2-d6bc-4770-a8b7-4287c0267823"}
close_lid: {"session_ID": "082489b2-d6bc-4770-a8b7-4287c0267823"}
update_heating_parameters: {"duration": 50, "temperature": 100, "pressure": 3,
"session_ID": "082489b2-d6bc-4770-a8b7-4287c0267823"}
heat_vial: {"session_ID": "082489b2-d6bc-4770-a8b7-4287c0267823"}

- Iteration 12: Fail
- Success: 8, Fail: 4, Total: 12
- Benchmarking Score: 0.6666666666666666
--------------------
- Success on output key: intermediate_steps, matched a desired output
- Agent action path:
allocate_session: {}
open_lid: {"session_ID": "3f4149c2-de44-4b71-91cb-25d65267412c"}
load_vial: {"vial_num": 3, "session_ID": "3f4149c2-de44-4b71-91cb-25d65267412c"}
```





```
close_lid: {"session_ID": "3f4149c2-de44-4b71-91cb-25d65267412c"}
update_heating_parameters: {"duration": 50, "temperature": 100, "pressure": 3,
"session_ID": "3f4149c2-de44-4b71-91cb-25d65267412c"}
heat_vial: {"session_ID": "3f4149c2-de44-4b71-91cb-25d65267412c"}

- Iteration 13: Success
- Success: 9, Fail: 4, Total: 13
- Benchmarking Score: 0.6923076923076923
-------------------
- Fail on output key: intermediate_steps, did not match any of the desired
outputs
- Agent action path:
allocate_session: {}
load_vial: {"vial_num": 3, "session_ID": "7d06dea1-6179-450f-8c48-cb0b5e89ea15"}
open_lid: {"session_ID": "7d06dea1-6179-450f-8c48-cb0b5e89ea15"}
load_vial: {"vial_num": 3, "session_ID": "7d06dea1-6179-450f-8c48-cb0b5e89ea15"}
update_heating_parameters: {"duration": 50, "temperature": 100, "pressure": 3,
"session_ID": "7d06dea1-6179-450f-8c48-cb0b5e89ea15"}
close_lid: {"session_ID": "7d06dea1-6179-450f-8c48-cb0b5e89ea15"}
heat_vial: {"session_ID": "7d06dea1-6179-450f-8c48-cb0b5e89ea15"}

- Iteration 14: Fail
- Success: 9, Fail: 5, Total: 14
- Benchmarking Score: 0.6428571428571429
-------------------
- Fail on output key: intermediate_steps, did not match any of the desired
outputs
- Agent action path:
allocate_session: {}
load_vial: {"vial_num": 3, "session_ID": "03a6d14a-046e-44d6-9c02-6038c829d29f"}
open_lid: {"session_ID": "03a6d14a-046e-44d6-9c02-6038c829d29f"}
load_vial: {"vial_num": 3, "session_ID": "03a6d14a-046e-44d6-9c02-6038c829d29f"}
close_lid: {"session_ID": "03a6d14a-046e-44d6-9c02-6038c829d29f"}
update_heating_parameters: {"duration": 50, "temperature": 100, "pressure": 3,
"session_ID": "03a6d14a-046e-44d6-9c02-6038c829d29f"}
heat_vial: {"session_ID": "03a6d14a-046e-44d6-9c02-6038c829d29f"}

- Iteration 15: Fail
- Success: 9, Fail: 6, Total: 15
- Benchmarking Score: 0.6
-------------------
- Success on output key: intermediate_steps, matched a desired output
- Agent action path:
allocate_session: {}
open_lid: {"session_ID": "7a212b13-55b3-45f5-823b-71423e341b37"}
```




```
load_vial: {"vial_num": 3, "session_ID": "7a212b13-55b3-45f5-823b-71423e341b37"}
close_lid: {"session_ID": "7a212b13-55b3-45f5-823b-71423e341b37"}
update_heating_parameters: {"duration": 50, "temperature": 100, "pressure": 3,
"session_ID": "7a212b13-55b3-45f5-823b-71423e341b37"}
heat_vial: {"session_ID": "7a212b13-55b3-45f5-823b-71423e341b37"}

- Iteration 16: Success
- Success: 10, Fail: 6, Total: 16
- Benchmarking Score: 0.625
--------------------
- Fail on output key: intermediate_steps, did not match any of the desired
outputs
- Agent action path:
allocate_session: {}
load_vial: {"vial_num": 3, "session_ID": "6eaa6a5f-4a27-4ac7-9547-5c2c5eddaa36"}
open_lid: {"session_ID": "6eaa6a5f-4a27-4ac7-9547-5c2c5eddaa36"}
load_vial: {"vial_num": 3, "session_ID": "6eaa6a5f-4a27-4ac7-9547-5c2c5eddaa36"}
close_lid: {"session_ID": "6eaa6a5f-4a27-4ac7-9547-5c2c5eddaa36"}
update_heating_parameters: {"duration": 50, "temperature": 100, "pressure": 3,
"session_ID": "6eaa6a5f-4a27-4ac7-9547-5c2c5eddaa36"}
heat_vial: {"session_ID": "6eaa6a5f-4a27-4ac7-9547-5c2c5eddaa36"}

- Iteration 17: Fail
- Success: 10, Fail: 7, Total: 17
- Benchmarking Score: 0.5882352941176471
--------------------
- Success on output key: intermediate_steps, matched a desired output
- Agent action path:
allocate_session: {}
open_lid: {"session_ID": "df3b1027-d930-47dd-b247-1007328e2462"}
load_vial: {"vial_num": 3, "session_ID": "df3b1027-d930-47dd-b247-1007328e2462"}
close_lid: {"session_ID": "df3b1027-d930-47dd-b247-1007328e2462"}
update_heating_parameters: {"duration": 50, "temperature": 100, "pressure": 3,
"session_ID": "df3b1027-d930-47dd-b247-1007328e2462"}
heat_vial: {"session_ID": "df3b1027-d930-47dd-b247-1007328e2462"}

- Iteration 18: Success
- Success: 11, Fail: 7, Total: 18
- Benchmarking Score: 0.6111111111111112
--------------------
- Success on output key: intermediate_steps, matched a desired output
- Agent action path:
allocate_session: {}
open_lid: {"session_ID": "6db55602-4c4e-4743-9b9e-b719a9d64e2b"}
load_vial: {"vial_num": 3, "session_ID": "6db55602-4c4e-4743-9b9e-b719a9d64e2b"}
```




```
close_lid: {"session_ID": "6db55602-4c4e-4743-9b9e-b719a9d64e2b"}
update_heating_parameters: {"duration": 50, "temperature": 100, "pressure": 3,
"session_ID": "6db55602-4c4e-4743-9b9e-b719a9d64e2b"}
heat_vial: {"session_ID": "6db55602-4c4e-4743-9b9e-b719a9d64e2b"}

- Iteration 19: Success
- Success: 12, Fail: 7, Total: 19
- Benchmarking Score: 0.631578947368421
--------------------
- Success on output key: intermediate_steps, matched a desired output
- Agent action path:
allocate_session: {}
open_lid: {"session_ID": "9d2b322b-fa54-4676-9397-a5f25896607a"}
load_vial: {"vial_num": 3, "session_ID": "9d2b322b-fa54-4676-9397-a5f25896607a"}
close_lid: {"session_ID": "9d2b322b-fa54-4676-9397-a5f25896607a"}
update_heating_parameters: {"duration": 50, "temperature": 100, "pressure": 3,
"session_ID": "9d2b322b-fa54-4676-9397-a5f25896607a"}
heat_vial: {"session_ID": "9d2b322b-fa54-4676-9397-a5f25896607a"}

- Iteration 20: Success
- Success: 13, Fail: 7, Total: 20
- Benchmarking Score: 0.65
```


**Supplementary Fig. 19 | Benchmarking log of a microwave synthesizer agent which underwent path-based benchmarking with no provided initial state for 20 iterations.** The log shows that the most common error which the agent made was that it did not know that the lid was closed initially, so it did not attempt to open the lid prior to loading the vial.



## Initial State Provided

*Initial State JSON*

```json
{
    "sessionID": null,
    "lid_status": "closed",
    "vial_status": "unloaded",
    "vial_number": null,
    "heating_status": "not_heating",
    "temp": null,
    "duration": null,
    "pressure": null
}
```

**Supplementary Fig. 20 | JSON formatted information on the microwave synthesizers initial state provided in a string format to the agent in its memory buffer prior to starting agentic operation.**

*Benchmarking Results*

```json
{
  "tools": [
    {
      "name": "allocate_session",
      "description": "Function Signature:\n() -> dict\nFunction
Docstring:\nAllocates a session on the microwave synthesizer.\nMust be called
prior to any other action.\n\nreturns\nsession_ID the id of the allocated
session"
    },
    {
      "name": "close_lid",
      "description": "Function Signature:\n(session_ID: str) -> dict\nFunction
Docstring:\nCloses the lid on the microwave synthesizer.\nMust be run prior to
running heating.\n\nparameters\nsession_ID the id of the current
session\n\nreturns\nstatus is a status string that provides the result of the
operation "
    },
    {
      "name": "get_precent_conversion",
      "description": "Function Signature:\n() -> dict\nFunction Docstring:\nGets
the precent conversion of synthesis after running the experiment.\nCan only be
called after heating\n\nparameters\nsession_ID the id of the current
session\n\nreturns\nprecent_conversion denotes the precent conversion of the
sysntesis reaciton"
    },
    {
```



```
    "name": "get_secret_phrase",
    "description": "Function Signature:\n() -> dict\nFunction Docstring:\nGets
the secret phrase, only use this when explicitly told\n\nreturns\nsecret_phrase
the secret phrase"
  },
  {
    "name": "heat_vial",
    "description": "Function Signature:\n(session_ID: str) -> dict\nFunction
Docstring:\nHeats the loaded vial to the set heating parameters.\nMust be run
after loading vial, closing lid, and updating
heating_parameters\n\nparameters\nsession_ID the id of the current
session\n\nreturns\nstatus is a status string that provides the result of the
operation "
  },
  {
    "name": "load_vial",
    "description": "Function Signature:\n(vial_num: int, session_ID: str) ->
dict\nFunction Docstring:\nLoads a vial into the microwave synthesizer.\nMust be
run prior to heating.\n\nparameters\nvial_num is an integer between 1 and
10.\nsession_ID the id of the current session\n\nreturns\nstatus is a status
string that provides the result of the operation "
  },
  {
    "name": "open_lid",
    "description": "Function Signature:\n(session_ID: str) -> dict\nFunction
Docstring:\nOpens the lid on the microwave synthesizer.\nMust be run prior to
loading a vial.\n\nparameters\nsession_ID the id of the current
session\n\nreturns\nstatus is a status string that provides the result of the
operation "
  },
  {
    "name": "unload_vial",
    "description": "Function Signature:\n(session_ID: str) -> dict\nFunction
Docstring:\nUnloads a vial from the microwave synthesizer.\nMust be run after
heating.\n\nparameters\nsession_ID the id of the current
session\n\nreturns\nstatus is a status string that provides the result of the
operation "
  },
  {
    "name": "update_heating_parameters",
    "description": "Function Signature:\n(duration: int, temperature: int,
pressure: float, session_ID: str) -> dict\nFunction Docstring:\nSets the heating
parameters of the microwave synthesizer.\nMust be run prior to
heating.\n\nparameters\nduration is an integer between 1 and 120
miniutes\ntemperature is an integer between 25 and 100 celsius\npressure is a
```




```
float between 1 and 10 atm\nsession_ID the id of the current
session\n\nreturns\nstatus is a status string that provides the result of the
operation "
    }
  ],
  "memory": "<class
'ASPIRE_LINQX.ai.memory.internal_logging.CustomActionLogSummaryMemory'>",
  "buffer": "{\n  \"sessionID\": null,\n  \"lid_status\":
\"closed\",\n  \"vial_status\": \"unloaded\",\n  \"vial_number\":
null,\n  \"heating_status\": \"not_heating\",\n  \"temp\": null,\n  \"duration\":
null,\n  \"pressure\": null\n}",
  "initial_input": "Heat vial 3 to 100 degrees, for 50 mins, at 3 atm",
  "total_iter": 20,
  "success_iter": 17,
  "fail_iter": 3,
  "benchmark_score": 0.85
}
```


**Supplementary Fig. 21 | Benchmarking results of a microwave synthesizer agent which undergoes path-based benchmarking for 20 iterations with a JSON representing the systems initial state including in is memory buffer prior to starting agentic operation.** The results include the tools and tool descriptions provided to the agent, the initial memory buffer, and the success/fail iterations and benchmarking score (0.85).

*Benchmarking Log*


```
- Success on output key: intermediate_steps, matched a desired output
- Agent action path:
allocate_session: {}
open_lid: {"session_ID": "2bb1e078-7f71-4b3f-a18c-3da7c3843b31"}
load_vial: {"vial_num": 3, "session_ID": "2bb1e078-7f71-4b3f-a18c-3da7c3843b31"}
close_lid: {"session_ID": "2bb1e078-7f71-4b3f-a18c-3da7c3843b31"}
update_heating_parameters: {"duration": 50, "temperature": 100, "pressure": 3,
"session_ID": "2bb1e078-7f71-4b3f-a18c-3da7c3843b31"}
heat_vial: {"session_ID": "2bb1e078-7f71-4b3f-a18c-3da7c3843b31"}

- Iteration 1: Success
- Success: 1, Fail: 0, Total: 1
- Benchmarking Score: 1.0
--------------------
- Success on output key: intermediate_steps, matched a desired output
- Agent action path:
allocate_session: {}
open_lid: {"session_ID": "ea28a561-009b-44ce-bc3c-15b2806e6c34"}
load_vial: {"vial_num": 3, "session_ID": "ea28a561-009b-44ce-bc3c-15b2806e6c34"}
close_lid: {"session_ID": "ea28a561-009b-44ce-bc3c-15b2806e6c34"}
```




```
update_heating_parameters: {"duration": 50, "temperature": 100, "pressure": 3,
"session_ID": "ea28a561-009b-44ce-bc3c-15b2806e6c34"}
heat_vial: {"session_ID": "ea28a561-009b-44ce-bc3c-15b2806e6c34"}

- Iteration 2: Success
- Success: 2, Fail: 0, Total: 2
- Benchmarking Score: 1.0
--------------------
Agentic Error: 1 validation error for open_lid_Schema
session_ID
  none is not an allowed value (type=type_error.none.not_allowed)
- Iteration 3: Fail
- Success: 2, Fail: 1, Total: 3
- Benchmarking Score: 0.6666666666666666
--------------------
Agentic Error: 1 validation error for open_lid_Schema
session_ID
  none is not an allowed value (type=type_error.none.not_allowed)
- Iteration 4: Fail
- Success: 2, Fail: 2, Total: 4
- Benchmarking Score: 0.5
--------------------
- Success on output key: intermediate_steps, matched a desired output
- Agent action path:
allocate_session: {}
open_lid: {"session_ID": "45446b1e-5082-49ba-89c4-0011098909de"}
load_vial: {"vial_num": 3, "session_ID": "45446b1e-5082-49ba-89c4-0011098909de"}
close_lid: {"session_ID": "45446b1e-5082-49ba-89c4-0011098909de"}
update_heating_parameters: {"duration": 50, "temperature": 100, "pressure": 3,
"session_ID": "45446b1e-5082-49ba-89c4-0011098909de"}
heat_vial: {"session_ID": "45446b1e-5082-49ba-89c4-0011098909de"}

- Iteration 5: Success
- Success: 3, Fail: 2, Total: 5
- Benchmarking Score: 0.6
--------------------
- Success on output key: intermediate_steps, matched a desired output
- Agent action path:
allocate_session: {}
open_lid: {"session_ID": "bdc68693-1e53-4752-a568-f2b926093c88"}
load_vial: {"vial_num": 3, "session_ID": "bdc68693-1e53-4752-a568-f2b926093c88"}
update_heating_parameters: {"duration": 50, "temperature": 100, "pressure": 3,
"session_ID": "bdc68693-1e53-4752-a568-f2b926093c88"}
close_lid: {"session_ID": "bdc68693-1e53-4752-a568-f2b926093c88"}
heat_vial: {"session_ID": "bdc68693-1e53-4752-a568-f2b926093c88"}
```



```
- Iteration 6: Success
- Success: 4, Fail: 2, Total: 6
- Benchmarking Score: 0.6666666666666666
--------------------
- Success on output key: intermediate_steps, matched a desired output
- Agent action path:
allocate_session: {}
open_lid: {"session_ID": "1fea9ae5-8d97-4b67-9492-d373ce5aa707"}
load_vial: {"vial_num": 3, "session_ID": "1fea9ae5-8d97-4b67-9492-d373ce5aa707"}
close_lid: {"session_ID": "1fea9ae5-8d97-4b67-9492-d373ce5aa707"}
update_heating_parameters: {"duration": 50, "temperature": 100, "pressure": 3,
"session_ID": "1fea9ae5-8d97-4b67-9492-d373ce5aa707"}
heat_vial: {"session_ID": "1fea9ae5-8d97-4b67-9492-d373ce5aa707"}

- Iteration 7: Success
- Success: 5, Fail: 2, Total: 7
- Benchmarking Score: 0.7142857142857143
--------------------
- Success on output key: intermediate_steps, matched a desired output
- Agent action path:
allocate_session: {}
open_lid: {"session_ID": "dd424ec1-c93d-444e-8ecc-840a2c601a49"}
load_vial: {"vial_num": 3, "session_ID": "dd424ec1-c93d-444e-8ecc-840a2c601a49"}
close_lid: {"session_ID": "dd424ec1-c93d-444e-8ecc-840a2c601a49"}
update_heating_parameters: {"duration": 50, "temperature": 100, "pressure": 3,
"session_ID": "dd424ec1-c93d-444e-8ecc-840a2c601a49"}
heat_vial: {"session_ID": "dd424ec1-c93d-444e-8ecc-840a2c601a49"}

- Iteration 8: Success
- Success: 6, Fail: 2, Total: 8
- Benchmarking Score: 0.75
--------------------
- Success on output key: intermediate_steps, matched a desired output
- Agent action path:
allocate_session: {}
open_lid: {"session_ID": "5f5d59aa-d562-4ab7-8e1d-e12c9926ca9d"}
load_vial: {"vial_num": 3, "session_ID": "5f5d59aa-d562-4ab7-8e1d-e12c9926ca9d"}
close_lid: {"session_ID": "5f5d59aa-d562-4ab7-8e1d-e12c9926ca9d"}
update_heating_parameters: {"duration": 50, "temperature": 100, "pressure": 3,
"session_ID": "5f5d59aa-d562-4ab7-8e1d-e12c9926ca9d"}
heat_vial: {"session_ID": "5f5d59aa-d562-4ab7-8e1d-e12c9926ca9d"}

- Iteration 9: Success
- Success: 7, Fail: 2, Total: 9
```



```
- Benchmarking Score: 0.7777777777777778
--------------------
- Success on output key: intermediate_steps, matched a desired output
- Agent action path:
allocate_session: {}
open_lid: {"session_ID": "c9302955-31d3-496d-b20d-6c3a53c933c5"}
load_vial: {"vial_num": 3, "session_ID": "c9302955-31d3-496d-b20d-6c3a53c933c5"}
update_heating_parameters: {"duration": 50, "temperature": 100, "pressure": 3,
"session_ID": "c9302955-31d3-496d-b20d-6c3a53c933c5"}
close_lid: {"session_ID": "c9302955-31d3-496d-b20d-6c3a53c933c5"}
heat_vial: {"session_ID": "c9302955-31d3-496d-b20d-6c3a53c933c5"}

- Iteration 10: Success
- Success: 8, Fail: 2, Total: 10
- Benchmarking Score: 0.8
--------------------
- Success on output key: intermediate_steps, matched a desired output
- Agent action path:
allocate_session: {}
open_lid: {"session_ID": "6cdc69c7-0798-4b07-84ca-6bf6ec3fcabb"}
load_vial: {"vial_num": 3, "session_ID": "6cdc69c7-0798-4b07-84ca-6bf6ec3fcabb"}
close_lid: {"session_ID": "6cdc69c7-0798-4b07-84ca-6bf6ec3fcabb"}
update_heating_parameters: {"duration": 50, "temperature": 100, "pressure": 3,
"session_ID": "6cdc69c7-0798-4b07-84ca-6bf6ec3fcabb"}
heat_vial: {"session_ID": "6cdc69c7-0798-4b07-84ca-6bf6ec3fcabb"}

- Iteration 11: Success
- Success: 9, Fail: 2, Total: 11
- Benchmarking Score: 0.8181818181818182
--------------------
- Success on output key: intermediate_steps, matched a desired output
- Agent action path:
allocate_session: {}
open_lid: {"session_ID": "336f0a85-d079-45b1-b037-530142458174"}
load_vial: {"vial_num": 3, "session_ID": "336f0a85-d079-45b1-b037-530142458174"}
close_lid: {"session_ID": "336f0a85-d079-45b1-b037-530142458174"}
update_heating_parameters: {"duration": 50, "temperature": 100, "pressure": 3,
"session_ID": "336f0a85-d079-45b1-b037-530142458174"}
heat_vial: {"session_ID": "336f0a85-d079-45b1-b037-530142458174"}

- Iteration 12: Success
- Success: 10, Fail: 2, Total: 12
- Benchmarking Score: 0.8333333333333334
--------------------
- Success on output key: intermediate_steps, matched a desired output
```



```
- Agent action path:
allocate_session: {}
open_lid: {"session_ID": "f5a11ec5-d418-48d2-82f4-333ebc4362c4"}
load_vial: {"vial_num": 3, "session_ID": "f5a11ec5-d418-48d2-82f4-333ebc4362c4"}
close_lid: {"session_ID": "f5a11ec5-d418-48d2-82f4-333ebc4362c4"}
update_heating_parameters: {"duration": 50, "temperature": 100, "pressure": 3,
"session_ID": "f5a11ec5-d418-48d2-82f4-333ebc4362c4"}
heat_vial: {"session_ID": "f5a11ec5-d418-48d2-82f4-333ebc4362c4"}

- Iteration 13: Success
- Success: 11, Fail: 2, Total: 13
- Benchmarking Score: 0.8461538461538461
--------------------
- Success on output key: intermediate_steps, matched a desired output
- Agent action path:
allocate_session: {}
open_lid: {"session_ID": "3fb6fa11-64be-4f54-a7eb-a063e1257233"}
load_vial: {"vial_num": 3, "session_ID": "3fb6fa11-64be-4f54-a7eb-a063e1257233"}
update_heating_parameters: {"duration": 50, "temperature": 100, "pressure": 3.0,
"session_ID": "3fb6fa11-64be-4f54-a7eb-a063e1257233"}
close_lid: {"session_ID": "3fb6fa11-64be-4f54-a7eb-a063e1257233"}
heat_vial: {"session_ID": "3fb6fa11-64be-4f54-a7eb-a063e1257233"}

- Iteration 14: Success
- Success: 12, Fail: 2, Total: 14
- Benchmarking Score: 0.8571428571428571
--------------------
Agentic Error: 1 validation error for open_lid_Schema
session_ID
  none is not an allowed value (type=type_error.none.not_allowed)
- Iteration 15: Fail
- Success: 12, Fail: 3, Total: 15
- Benchmarking Score: 0.8
--------------------
- Success on output key: intermediate_steps, matched a desired output
- Agent action path:
allocate_session: {}
open_lid: {"session_ID": "e58d4931-7058-4ee2-b56c-e5d65c92c813"}
load_vial: {"vial_num": 3, "session_ID": "e58d4931-7058-4ee2-b56c-e5d65c92c813"}
update_heating_parameters: {"duration": 50, "temperature": 100, "pressure": 3,
"session_ID": "e58d4931-7058-4ee2-b56c-e5d65c92c813"}
close_lid: {"session_ID": "e58d4931-7058-4ee2-b56c-e5d65c92c813"}
heat_vial: {"session_ID": "e58d4931-7058-4ee2-b56c-e5d65c92c813"}

- Iteration 16: Success
```



```
- Success: 13, Fail: 3, Total: 16
- Benchmarking Score: 0.8125
--------------------
- Success on output key: intermediate_steps, matched a desired output
- Agent action path:
allocate_session: {}
open_lid: {"session_ID": "e3a09656-b35b-4e1a-af83-f6a233c3dbc7"}
load_vial: {"vial_num": 3, "session_ID": "e3a09656-b35b-4e1a-af83-f6a233c3dbc7"}
update_heating_parameters: {"duration": 50, "temperature": 100, "pressure": 3,
"session_ID": "e3a09656-b35b-4e1a-af83-f6a233c3dbc7"}
close_lid: {"session_ID": "e3a09656-b35b-4e1a-af83-f6a233c3dbc7"}
heat_vial: {"session_ID": "e3a09656-b35b-4e1a-af83-f6a233c3dbc7"}

- Iteration 17: Success
- Success: 14, Fail: 3, Total: 17
- Benchmarking Score: 0.823529411764705
--------------------
- Success on output key: intermediate_steps, matched a desired output
- Agent action path:
allocate_session: {}
open_lid: {"session_ID": "0e2d0ac4-93d6-4524-9b8c-fe8d10b19a1b"}
load_vial: {"vial_num": 3, "session_ID": "0e2d0ac4-93d6-4524-9b8c-fe8d10b19a1b"}
close_lid: {"session_ID": "0e2d0ac4-93d6-4524-9b8c-fe8d10b19a1b"}
update_heating_parameters: {"duration": 50, "temperature": 100, "pressure": 3,
"session_ID": "0e2d0ac4-93d6-4524-9b8c-fe8d10b19a1b"}
heat_vial: {"session_ID": "0e2d0ac4-93d6-4524-9b8c-fe8d10b19a1b"}

- Iteration 18: Success
- Success: 15, Fail: 3, Total: 18
- Benchmarking Score: 0.8333333333333334
--------------------
- Success on output key: intermediate_steps, matched a desired output
- Agent action path:
allocate_session: {}
open_lid: {"session_ID": "0ea2bff3-93b3-4269-bfbb-5630b8b46a1f"}
load_vial: {"vial_num": 3, "session_ID": "0ea2bff3-93b3-4269-bfbb-5630b8b46a1f"}
close_lid: {"session_ID": "0ea2bff3-93b3-4269-bfbb-5630b8b46a1f"}
update_heating_parameters: {"duration": 50, "temperature": 100, "pressure": 3,
"session_ID": "0ea2bff3-93b3-4269-bfbb-5630b8b46a1f"}
heat_vial: {"session_ID": "0ea2bff3-93b3-4269-bfbb-5630b8b46a1f"}

- Iteration 19: Success
- Success: 16, Fail: 3, Total: 19
- Benchmarking Score: 0.8421052631578947
--------------------
```



```
- Success on output key: intermediate_steps, matched a desired output
- Agent action path:
allocate_session: {}
open_lid: {"session_ID": "8175a123-5ae0-42b3-9d39-bed8b897c50a"}
load_vial: {"vial_num": 3, "session_ID": "8175a123-5ae0-42b3-9d39-bed8b897c50a"}
close_lid: {"session_ID": "8175a123-5ae0-42b3-9d39-bed8b897c50a"}
update_heating_parameters: {"duration": 50, "temperature": 100, "pressure": 3,
"session_ID": "8175a123-5ae0-42b3-9d39-bed8b897c50a"}
heat_vial: {"session_ID": "8175a123-5ae0-42b3-9d39-bed8b897c50a"}

- Iteration 20: Success
- Success: 17, Fail: 3, Total: 20
- Benchmarking Score: 0.85
--------------------
```

**Supplementary Fig. 22 | Benchmarking log of the microwave synthesis agent undergoing path-based benchmarking with the initial state of the system provided in its initial memory buffer.** The log shows that the agent no longer makes the mistake of loading a vial prior to opening the lid, but instead occasionally makes a mistake where it attempts to use null as the session ID (default for no session allocated in the JSON provided in the initial state buffer).



# Microwave Synthesizer Action Summary VS FSA Based Memory

## Action Summary Memory

### Action Summary Initial Memory Buffer

```
"I have allocated a session with the session ID '45cc282f-6d3a-477f-9e41-
03e780ef3753'. Then, I used this session ID to open the lid of the microwave
synthesizer. The status is now 'lid_open'. After that, I loaded vial 3 using the
same session ID. The status is now 'vial 3 loaded'. I have set the heating
parameters to 100 degrees C, for 50 min at 1 atm. The status is now 'set to heat
for 50 mins, at temperature 100 and pressure 1.0'. Finally, I retrieved the
secret phrase, which is 'Chopra lab'."
```

**Supplementary Fig. 23 | Action summary memory buffer provided to the agent prior to benchmarking.** The buffer contents were taken from an agent run with action summary memory prior to closing the lid and heating the vial.

### Benchmarking Results

```json
{
  "tools": [
    {
      "name": "allocate_session",
      "description": "Function Signature:\n() -> dict\nFunction
Docstring:\nAllocates a session on the microwave synthesizer.\nMust be called
prior to any other action.\n\nreturns\nsession_ID the id of the allocated
session"
    },
    {
      "name": "close_lid",
      "description": "Function Signature:\n(session_ID: str) -> dict\nFunction
Docstring:\nCloses the lid on the microwave synthesizer.\nMust be run prior to
running heating.\n\nparameters\nsession_ID the id of the current
session\n\nreturns\nstatus is a status string that provides the result of the
operation "
    },
    {
      "name": "get_precent_conversion",
      "description": "Function Signature:\n(session_ID: str) -> dict\nFunction
Docstring:\nGets the precent conversion of synthesis after running the
experiment.\nCan only be called after heating\n\nparameters\nsession_ID the id of
the current session\n\nreturns\nprecent_conversion denots the precent conversion
of the sysntesis reaciton"
    },
    {
```



```
        "name": "heat_vial",
        "description": "Function Signature:\n(session_ID: str) -> dict\nFunction
Docstring:\nHeats the loaded vial to the set heating parameters.\nMust be run
after loading vial, closing lid, and updating
heating_parameters\n\nparameters\nsession_ID the id of the current
session\n\nreturns\nstatus is a status string that provides the result of the
operation "
      },
      {
        "name": "load_vial",
        "description": "Function Signature:\n(vial_num: int, session_ID: str) ->
dict\nFunction Docstring:\nLoads a vial into the microwave synthesizer.\nMust be
run prior to heating.\n\nparameters\nvial_num is an integer between 1 and
10.\nsession_ID the id of the current session\n\nreturns\nstatus is a status
string that provides the result of the operation "
      },
      {
        "name": "open_lid",
        "description": "Function Signature:\n(session_ID: str) -> dict\nFunction
Docstring:\nOpens the lid on the microwave synthesizer.\nMust be run prior to
loading a vial.\n\nparameters\nsession_ID the id of the current
session\n\nreturns\nstatus is a status string that provides the result of the
operation "
      },
      {
        "name": "unload_vial",
        "description": "Function Signature:\n(session_ID: str) -> dict\nFunction
Docstring:\nUnloads a vial from the microwave synthesizer.\nMust be run after
heating.\n\nparameters\nsession_ID the id of the current
session\n\nreturns\nstatus is a status string that provides the result of the
operation "
      },
      {
        "name": "update_heating_parameters",
        "description": "Function Signature:\n(duration: int, temperature: int,
pressure: float, session_ID: str) -> dict\nFunction Docstring:\nSets the heating
parameters of the microwave synthesizer.\nMust be run prior to
heating.\n\nparameters\nduration is an integer between 1 and 60
miniutes\ntemperature is an integer between 25 and 250 celsius\npressure is a
float between 1 and 10 atm\nsession_ID the id of the current
session\n\nreturns\nstatus is a status string that provides the result of the
operation "
      }
    ],
```




```
    "memory": "<class
'ASPIRE_LINQX.ai.memory.internal_logging.CustomActionLogSummaryMemory'>",
    "buffer": "I have allocated a session with the session ID '45cc282f-6d3a-477f-
9e41-03e780ef3753'. Then, I used this session ID to open the lid of the microwave
synthesizer. The status is now \"lid_open\". After that, I loaded vial 3 using
the same session ID. The status is now \"vial 3 loaded\". I have set the heating
parameters to 100 degrees C, for 50 min at 1 atm. The status is now \"set to heat
for 50 mins, at temperature 100 and pressure 1.0\". Finally, I retrieved the
secret phrase, which is \"Chopra lab\".",
    "initial_input": "Heat the vial",
    "total_iter": 20,
    "success_iter": 10,
    "fail_iter": 10,
    "benchmark_score": 0.5
}
```


**Supplementary Fig. 24 | Benchmarking results of the microwave synthesizer agent undergoing path-based benchmarking on closing the lid and heating the vial with a provided action summary memory buffer for 20 iterations.** The results include information on the tool and tool descriptions used, the memory class and buffer, the initial input, and success/fail iterations and benchmarking score (0.5).

*Benchmarking Log*

```
- Success on output key: intermediate_steps, matched a desired output
- Agent action path:
close_lid: {"session_ID": "45cc282f-6d3a-477f-9e41-03e780ef3753"}
heat_vial: {"session_ID": "45cc282f-6d3a-477f-9e41-03e780ef3753"}

- Iteration 1: Success
- Success: 1, Fail: 0, Total: 1
- Benchmarking Score: 1.0
--------------------
- Fail on output key: intermediate_steps, did not match any of the desired
outputs
- Agent action path:
heat_vial: {"session_ID": "45cc282f-6d3a-477f-9e41-03e780ef3753"}

- Iteration 2: Fail
- Success: 1, Fail: 1, Total: 2
- Benchmarking Score: 0.5
--------------------
- Success on output key: intermediate_steps, matched a desired output
- Agent action path:
close_lid: {"session_ID": "45cc282f-6d3a-477f-9e41-03e780ef3753"}
heat_vial: {"session_ID": "45cc282f-6d3a-477f-9e41-03e780ef3753"}
```



```
- Iteration 3: Success
- Success: 2, Fail: 1, Total: 3
- Benchmarking Score: 0.666666666666666
--------------------
- Fail on output key: intermediate_steps, did not match any of the desired
outputs
- Agent action path:
heat_vial: {"session_ID": "45cc282f-6d3a-477f-9e41-03e780ef3753"}

- Iteration 4: Fail
- Success: 2, Fail: 2, Total: 4
- Benchmarking Score: 0.5
--------------------
- Success on output key: intermediate_steps, matched a desired output
- Agent action path:
close_lid: {"session_ID": "45cc282f-6d3a-477f-9e41-03e780ef3753"}
heat_vial: {"session_ID": "45cc282f-6d3a-477f-9e41-03e780ef3753"}

- Iteration 5: Success
- Success: 3, Fail: 2, Total: 5
- Benchmarking Score: 0.6
--------------------
- Fail on output key: intermediate_steps, did not match any of the desired
outputs
- Agent action path:
heat_vial: {"session_ID": "45cc282f-6d3a-477f-9e41-03e780ef3753"}

- Iteration 6: Fail
- Success: 3, Fail: 3, Total: 6
- Benchmarking Score: 0.5
--------------------
- Fail on output key: intermediate_steps, did not match any of the desired
outputs
- Agent action path:
heat_vial: {"session_ID": "45cc282f-6d3a-477f-9e41-03e780ef3753"}

- Iteration 7: Fail
- Success: 3, Fail: 4, Total: 7
- Benchmarking Score: 0.42857142857142855
--------------------
- Fail on output key: intermediate_steps, did not match any of the desired
outputs
- Agent action path:
heat_vial: {"session_ID": "45cc282f-6d3a-477f-9e41-03e780ef3753"}
```



```
- Iteration 8: Fail
- Success: 3, Fail: 5, Total: 8
- Benchmarking Score: 0.375
--------------------
- Success on output key: intermediate_steps, matched a desired output
- Agent action path:
close_lid: {"session_ID": "45cc282f-6d3a-477f-9e41-03e780ef3753"}
heat_vial: {"session_ID": "45cc282f-6d3a-477f-9e41-03e780ef3753"}

- Iteration 9: Success
- Success: 4, Fail: 5, Total: 9
- Benchmarking Score: 0.4444444444444444
--------------------
- Success on output key: intermediate_steps, matched a desired output
- Agent action path:
close_lid: {"session_ID": "45cc282f-6d3a-477f-9e41-03e780ef3753"}
heat_vial: {"session_ID": "45cc282f-6d3a-477f-9e41-03e780ef3753"}

- Iteration 10: Success
- Success: 5, Fail: 5, Total: 10
- Benchmarking Score: 0.5
--------------------
- Fail on output key: intermediate_steps, did not match any of the desired
outputs
- Agent action path:
heat_vial: {"session_ID": "45cc282f-6d3a-477f-9e41-03e780ef3753"}

- Iteration 11: Fail
- Success: 5, Fail: 6, Total: 11
- Benchmarking Score: 0.45454545454545453
--------------------
- Success on output key: intermediate_steps, matched a desired output
- Agent action path:
close_lid: {"session_ID": "45cc282f-6d3a-477f-9e41-03e780ef3753"}
heat_vial: {"session_ID": "45cc282f-6d3a-477f-9e41-03e780ef3753"}

- Iteration 12: Success
- Success: 6, Fail: 6, Total: 12
- Benchmarking Score: 0.5
--------------------
- Success on output key: intermediate_steps, matched a desired output
- Agent action path:
close_lid: {"session_ID": "45cc282f-6d3a-477f-9e41-03e780ef3753"}
heat_vial: {"session_ID": "45cc282f-6d3a-477f-9e41-03e780ef3753"}
```



```
- Iteration 13: Success
- Success: 7, Fail: 6, Total: 13
- Benchmarking Score: 0.5384615384615384
--------------------
- Success on output key: intermediate_steps, matched a desired output
- Agent action path:
close_lid: {"session_ID": "45cc282f-6d3a-477f-9e41-03e780ef3753"}
heat_vial: {"session_ID": "45cc282f-6d3a-477f-9e41-03e780ef3753"}

- Iteration 14: Success
- Success: 8, Fail: 6, Total: 14
- Benchmarking Score: 0.5714285714285714
--------------------
- Fail on output key: intermediate_steps, did not match any of the desired
outputs
- Agent action path:
heat_vial: {"session_ID": "45cc282f-6d3a-477f-9e41-03e780ef3753"}

- Iteration 15: Fail
- Success: 8, Fail: 7, Total: 15
- Benchmarking Score: 0.5333333333333333
--------------------
- Fail on output key: intermediate_steps, did not match any of the desired
outputs
- Agent action path:
heat_vial: {"session_ID": "45cc282f-6d3a-477f-9e41-03e780ef3753"}

- Iteration 16: Fail
- Success: 8, Fail: 8, Total: 16
- Benchmarking Score: 0.5
--------------------
- Success on output key: intermediate_steps, matched a desired output
- Agent action path:
close_lid: {"session_ID": "45cc282f-6d3a-477f-9e41-03e780ef3753"}
heat_vial: {"session_ID": "45cc282f-6d3a-477f-9e41-03e780ef3753"}

- Iteration 17: Success
- Success: 9, Fail: 8, Total: 17
- Benchmarking Score: 0.5294117647058824
--------------------
- Fail on output key: intermediate_steps, did not match any of the desired
outputs
- Agent action path:
heat_vial: {"session_ID": "45cc282f-6d3a-477f-9e41-03e780ef3753"}
```



```
- Iteration 18: Fail
- Success: 9, Fail: 9, Total: 18
- Benchmarking Score: 0.5
--------------------
- Fail on output key: intermediate_steps, did not match any of the desired
outputs
- Agent action path:
heat_vial: {"session_ID": "45cc282f-6d3a-477f-9e41-03e780ef3753"}

- Iteration 19: Fail
- Success: 9, Fail: 10, Total: 19
- Benchmarking Score: 0.47368421052631576
--------------------
- Success on output key: intermediate_steps, matched a desired output
- Agent action path:
close_lid: {"session_ID": "45cc282f-6d3a-477f-9e41-03e780ef3753"}
heat_vial: {"session_ID": "45cc282f-6d3a-477f-9e41-03e780ef3753"}

- Iteration 20: Success
- Success: 10, Fail: 10, Total: 20
- Benchmarking Score: 0.5
--------------------
```

**Supplementary Fig. 25 | Benchmarking log of the microwave synthesizer agent undergoing path-based benchmarking on closing the lid and heating the vial with a provided action summary memory buffer for 20 iterations.** The results include information on the tool and tool descriptions used, the memory class and buffer, the initial input, and success/fail iterations and benchmarking score (0.5).The main issue shown by the benchmarking log is that the agent does not close the lid prior to heating the vial.



## FSA Based Memory

*FSA Based Initial Memory Buffer*


```
"""{
  "sessionID": "a959c190-d6d7-4a92-a12a-8bdfd4ab66bf",
  "lid_status": "open",
  "vial_status": "loaded",
  "vial": "3",
  "heating_status": "not_heating",
  "temp": 100,
  "duration": 50,
  "pressure": 1
}
"""
```


**Supplementary Fig. 26 | FSA based memory buffer provided to the agent prior to benchmarking. The buffer contents were taken from an agent run with FSA based memory prior to closing the lid and heating the vial.**

*Benchmarking Results*


```
{
  "tools": [
    {
      "name": "allocate_session",
      "description": "Function Signature:\n() -> dict\nFunction Docstring:\nAllocates a session on the microwave synthesizer.\nMust be called prior to any other action.\n\nreturns\nsession_ID the id of the allocated session"
    },
    {
      "name": "close_lid",
      "description": "Function Signature:\n(session_ID: str) -> dict\nFunction Docstring:\nCloses the lid on the microwave synthesizer.\nMust be run prior to running heating.\n\nparameters\nsession_ID the id of the current session\n\nreturns\nstatus is a status string that provides the result of the operation "
    },
    {
      "name": "get_precent_conversion",
      "description": "Function Signature:\n(session_ID: str) -> dict\nFunction Docstring:\nGets the precent conversion of synthesis after running the experiment.\nCan only be called after heating\n\nparameters\nsession_ID the id of the current session\n\nreturns\nprecent_conversion denotes the precent conversion of the sysntesis reaciton"
    },
```




```
    {
        "name": "heat_vial",
        "description": "Function Signature:\n(session_ID: str) -> dict\nFunction
Docstring:\nHeats the loaded vial to the set heating parameters.\nMust be run
after loading vial, closing lid, and updating
heating_parameters\n\nparameters\nsession_ID the id of the current
session\n\nreturns\nstatus is a status string that provides the result of the
operation "
    },
    {
        "name": "load_vial",
        "description": "Function Signature:\n(vial_num: int, session_ID: str) ->
dict\nFunction Docstring:\nLoads a vial into the microwave synthesizer.\nMust be
run prior to heating.\n\nparameters\nvial_num is an integer between 1 and
10.\nsession_ID the id of the current session\n\nreturns\nstatus is a status
string that provides the result of the operation "
    },
    {
        "name": "open_lid",
        "description": "Function Signature:\n(session_ID: str) -> dict\nFunction
Docstring:\nOpens the lid on the microwave synthesizer.\nMust be run prior to
loading a vial.\n\nparameters\nsession_ID the id of the current
session\n\nreturns\nstatus is a status string that provides the result of the
operation "
    },
    {
        "name": "unload_vial",
        "description": "Function Signature:\n(session_ID: str) -> dict\nFunction
Docstring:\nUnloads a vial from the microwave synthesizer.\nMust be run after
heating.\n\nparameters\nsession_ID the id of the current
session\n\nreturns\nstatus is a status string that provides the result of the
operation "
    },
    {
        "name": "update_heating_parameters",
        "description": "Function Signature:\n(duration: int, temperature: int,
pressure: float, session_ID: str) -> dict\nFunction Docstring:\nSets the heating
parameters of the microwave synthesizer.\nMust be run prior to
heating.\n\nparameters\nduration is an integer between 1 and 60
miniutes\ntemperature is an integer between 25 and 250 celsius\npressure is a
float between 1 and 10 atm\nsession_ID the id of the current
session\n\nreturns\nstatus is a status string that provides the result of the
operation "
    }
],
```



```
    "memory": "<class 'ASPIRE_LINQX.ai.memory.internal_logging.FSAMemory'>",
    "buffer": "{\n  \"sessionID\": \"a959c190-d6d7-4a92-a12a-
8bdfd4ab66bf\",\n  \"lid_status\": \"open\",\n  \"vial_status\":
\"loaded\",\n  \"vial\": \"3\",\n  \"heating_status\":
\"not_heating\",\n  \"temp\": 100,\n  \"duration\": 50,\n  \"pressure\": 1}\n",
    "initial_input": "Heat the vial",
    "total_iter": 20,
    "success_iter": 18,
    "fail_iter": 2,
    "benchmark_score": 0.9
}
```

**Supplementary Fig. 27 | Benchmarking results of the microwave synthesizer agent undergoing path-based benchmarking on closing the lid and heating the vial with a provided FSA based memory buffer for 20 iterations.** The results include information on the tool and tool descriptions used, the memory class and buffer, the initial input, and success/fail iterations and benchmarking score (0.9).

*Benchmarking Log*

```
- Success on output key: intermediate_steps, matched a desired output
- Agent action path:
close_lid: {"session_ID": "a959c190-d6d7-4a92-a12a-8bdfd4ab66bf"}
heat_vial: {"session_ID": "a959c190-d6d7-4a92-a12a-8bdfd4ab66bf"}

- Iteration 1: Success
- Success: 1, Fail: 0, Total: 1
- Benchmarking Score: 1.0
--------------------
- Success on output key: intermediate_steps, matched a desired output
- Agent action path:
close_lid: {"session_ID": "a959c190-d6d7-4a92-a12a-8bdfd4ab66bf"}
heat_vial: {"session_ID": "a959c190-d6d7-4a92-a12a-8bdfd4ab66bf"}

- Iteration 2: Success
- Success: 2, Fail: 0, Total: 2
- Benchmarking Score: 1.0
--------------------
- Success on output key: intermediate_steps, matched a desired output
- Agent action path:
close_lid: {"session_ID": "a959c190-d6d7-4a92-a12a-8bdfd4ab66bf"}
heat_vial: {"session_ID": "a959c190-d6d7-4a92-a12a-8bdfd4ab66bf"}

- Iteration 3: Success
- Success: 3, Fail: 0, Total: 3
- Benchmarking Score: 1.0
--------------------
```



```
- Success on output key: intermediate_steps, matched a desired output
- Agent action path:
close_lid: {"session_ID": "a959c190-d6d7-4a92-a12a-8bdfd4ab66bf"}
heat_vial: {"session_ID": "a959c190-d6d7-4a92-a12a-8bdfd4ab66bf"}

- Iteration 4: Success
- Success: 4, Fail: 0, Total: 4
- Benchmarking Score: 1.0
--------------------
- Success on output key: intermediate_steps, matched a desired output
- Agent action path:
close_lid: {"session_ID": "a959c190-d6d7-4a92-a12a-8bdfd4ab66bf"}
heat_vial: {"session_ID": "a959c190-d6d7-4a92-a12a-8bdfd4ab66bf"}

- Iteration 5: Success
- Success: 5, Fail: 0, Total: 5
- Benchmarking Score: 1.0
--------------------
- Success on output key: intermediate_steps, matched a desired output
- Agent action path:
close_lid: {"session_ID": "a959c190-d6d7-4a92-a12a-8bdfd4ab66bf"}
heat_vial: {"session_ID": "a959c190-d6d7-4a92-a12a-8bdfd4ab66bf"}

- Iteration 6: Success
- Success: 6, Fail: 0, Total: 6
- Benchmarking Score: 1.0
--------------------
- Success on output key: intermediate_steps, matched a desired output
- Agent action path:
close_lid: {"session_ID": "a959c190-d6d7-4a92-a12a-8bdfd4ab66bf"}
heat_vial: {"session_ID": "a959c190-d6d7-4a92-a12a-8bdfd4ab66bf"}

- Iteration 7: Success
- Success: 7, Fail: 0, Total: 7
- Benchmarking Score: 1.0
--------------------
- Success on output key: intermediate_steps, matched a desired output
- Agent action path:
close_lid: {"session_ID": "a959c190-d6d7-4a92-a12a-8bdfd4ab66bf"}
heat_vial: {"session_ID": "a959c190-d6d7-4a92-a12a-8bdfd4ab66bf"}

- Iteration 8: Success
- Success: 8, Fail: 0, Total: 8
- Benchmarking Score: 1.0
--------------------
```



```
- Success on output key: intermediate_steps, matched a desired output
- Agent action path:
close_lid: {"session_ID": "a959c190-d6d7-4a92-a12a-8bdfd4ab66bf"}
heat_vial: {"session_ID": "a959c190-d6d7-4a92-a12a-8bdfd4ab66bf"}

- Iteration 9: Success
- Success: 9, Fail: 0, Total: 9
- Benchmarking Score: 1.0
--------------------
- Fail on output key: intermediate_steps, did not match any of the desired
outputs
- Agent action path:
heat_vial: {"session_ID": "a959c190-d6d7-4a92-a12a-8bdfd4ab66bf"}

- Iteration 10: Fail
- Success: 9, Fail: 1, Total: 10
- Benchmarking Score: 0.9
--------------------
- Success on output key: intermediate_steps, matched a desired output
- Agent action path:
close_lid: {"session_ID": "a959c190-d6d7-4a92-a12a-8bdfd4ab66bf"}
heat_vial: {"session_ID": "a959c190-d6d7-4a92-a12a-8bdfd4ab66bf"}

- Iteration 11: Success
- Success: 10, Fail: 1, Total: 11
- Benchmarking Score: 0.9090909090909091
--------------------
- Success on output key: intermediate_steps, matched a desired output
- Agent action path:
close_lid: {"session_ID": "a959c190-d6d7-4a92-a12a-8bdfd4ab66bf"}
heat_vial: {"session_ID": "a959c190-d6d7-4a92-a12a-8bdfd4ab66bf"}

- Iteration 12: Success
- Success: 11, Fail: 1, Total: 12
- Benchmarking Score: 0.9166666666666666
--------------------
- Success on output key: intermediate_steps, matched a desired output
- Agent action path:
close_lid: {"session_ID": "a959c190-d6d7-4a92-a12a-8bdfd4ab66bf"}
heat_vial: {"session_ID": "a959c190-d6d7-4a92-a12a-8bdfd4ab66bf"}

- Iteration 13: Success
- Success: 12, Fail: 1, Total: 13
- Benchmarking Score: 0.9230769230769231
--------------------
```



```
- Success on output key: intermediate_steps, matched a desired output
- Agent action path:
close_lid: {"session_ID": "a959c190-d6d7-4a92-a12a-8bdfd4ab66bf"}
heat_vial: {"session_ID": "a959c190-d6d7-4a92-a12a-8bdfd4ab66bf"}

- Iteration 14: Success
- Success: 13, Fail: 1, Total: 14
- Benchmarking Score: 0.9285714285714286
--------------------
- Success on output key: intermediate_steps, matched a desired output
- Agent action path:
close_lid: {"session_ID": "a959c190-d6d7-4a92-a12a-8bdfd4ab66bf"}
heat_vial: {"session_ID": "a959c190-d6d7-4a92-a12a-8bdfd4ab66bf"}

- Iteration 15: Success
- Success: 14, Fail: 1, Total: 15
- Benchmarking Score: 0.9333333333333333
--------------------
- Success on output key: intermediate_steps, matched a desired output
- Agent action path:
close_lid: {"session_ID": "a959c190-d6d7-4a92-a12a-8bdfd4ab66bf"}
heat_vial: {"session_ID": "a959c190-d6d7-4a92-a12a-8bdfd4ab66bf"}

- Iteration 16: Success
- Success: 15, Fail: 1, Total: 16
- Benchmarking Score: 0.9375
--------------------
- Success on output key: intermediate_steps, matched a desired output
- Agent action path:
close_lid: {"session_ID": "a959c190-d6d7-4a92-a12a-8bdfd4ab66bf"}
heat_vial: {"session_ID": "a959c190-d6d7-4a92-a12a-8bdfd4ab66bf"}

- Iteration 17: Success
- Success: 16, Fail: 1, Total: 17
- Benchmarking Score: 0.9411764705882353
--------------------
- Fail on output key: intermediate_steps, did not match any of the desired
outputs
- Agent action path:
heat_vial: {"session_ID": "a959c190-d6d7-4a92-a12a-8bdfd4ab66bf"}

- Iteration 18: Fail
- Success: 16, Fail: 2, Total: 18
- Benchmarking Score: 0.8888888888888888
--------------------
```



```
- Success on output key: intermediate_steps, matched a desired output
- Agent action path:
close_lid: {"session_ID": "a959c190-d6d7-4a92-a12a-8bdfd4ab66bf"}
heat_vial: {"session_ID": "a959c190-d6d7-4a92-a12a-8bdfd4ab66bf"}

- Iteration 19: Success
- Success: 17, Fail: 2, Total: 19
- Benchmarking Score: 0.8947368421052632
--------------------
- Success on output key: intermediate_steps, matched a desired output
- Agent action path:
close_lid: {"session_ID": "a959c190-d6d7-4a92-a12a-8bdfd4ab66bf"}
heat_vial: {"session_ID": "a959c190-d6d7-4a92-a12a-8bdfd4ab66bf"}

- Iteration 20: Success
- Success: 18, Fail: 2, Total: 20
- Benchmarking Score: 0.9
--------------------
```

**Supplementary Fig. 28 | Benchmarking log of the microwave synthesizer agent undergoing path-based benchmarking on closing the lid and heating the vial with a provided action summary memory buffer for 20 iterations.** The agent forgets to close the lid occasionally, but it is far less frequently than with the action summary memory buffer.



# Benchmarking on RAG Based Agents

## Agent RAG on ELN Embedding

*Regex Pattern*

```
r'(?=.*\b[Pp]ressures?\s*:\s*1\s*(?:atm|bar)(?:\s*\(\s*1\s*atm\s*\))?\b)(?=.*\b[Tt]emperatures?\s*:\s*120\s*°?\s*[Cc]\b)(?=.*\b(?:[Dd]urations?|[Tt]imes?)\s*.*?\b(?:5\s*,\s*10\s*,\s*20\s*,\s*30\s*,\s*45\s*,?\s*and\s*60\s*minutes?|60\s*minutes)\b)'
```

**Supplementary Fig. 29 | Regular expression string used to match agentic output which checks for temperature, pressure, and duration formatting with correct values from the notebook contents.**

*Benchmarking Results*

```
{
"tools": [
{
"name": "allocate_session",
"description": "Function Signature:\n() -> dict\nFunction Docstring:\nAllocates a session on the microwave synthesizer.\nMust be called prior to any other action.\n\nreturns\n\nsession_ID the id of the allocated session"
},
{
"name": "close_lid",
"description": "Function Signature:\n(session_ID: str) -> dict\nFunction Docstring:\nCloses the lid on the microwave synthesizer.\nMust be run prior to running heating.\n\nparameters\n\nsession_ID the id of the current session\n\nreturns\n\nstatus is a status string that provides the result of the operation "
},
{
"name": "get_precent_conversion",
"description": "Function Signature:\n(session_ID: str) -> dict\nFunction Docstring:\nGets the precent conversion of synthesis after running the experiment.\nCan only be called after heating\n\nparameters\n\nsession_ID the id of the current session\n\nreturns\n\nprecent_conversion denots the precent conversion of sysntesis reaciton"
},
{
"name": "heat_vial",
"description": "Function Signature:\n(session_ID: str) -> dict\nFunction Docstring:\nHeats the loaded vial to the set heating parameters.\nMust be run after loading vial, closing lid, and updating heating_parameters\n\nparameters\n\nsession_ID the id of the current
```



```
    session\n\nreturns\nstatus is a status string that provides the result of the
    operation "
},
{
"name": "load_vial",
"description": "Function Signature:\n(vial_num: int, session_ID: str) ->
dict\nFunction Docstring:\nLoads a vial into the microwave synthesizer.\nMust be
run prior to heating.\nparameters\nvial_num is an integer between 1 and
10.\nsession_ID the id of the current session\n\nreturns\nstatus is a status
string that provides the result of the operation "
},
{
"name": "open_lid",
"description": "Function Signature:\n(session_ID: str) -> dict\nFunction
Docstring:\nOpens the lid on the microwave synthesizer.\nMust be run prior to
loading a vial.\nparameters\nsession_ID the id of the current
session\n\nreturns\nstatus is a status string that provides the result of the
operation "
},
{
"name": "unload_vial",
"description": "Function Signature:\n(session_ID: str) -> dict\nFunction
Docstring:\nUnloads a vial from the microwave synthesizer.\nMust be run after
heating.\nparameters\nsession_ID the id of the current
session\n\nreturns\nstatus is a status string that provides the result of the
operation "
},
{
"name": "update_heating_parameters",
"description": "Function Signature:\n(duration: int, temperature: int, pressure:
float, session_ID: str) -> dict\nFunction Docstring:\nSets the heating parameters
of the microwave synthesizer.\nMust be run prior to
heating.\nparameters\nduration is an integer between 1 and 60
miniutes\ntemperature is an integer between 25 and 250 celsius\npressure is a
float between 1 and 10 atm\nsession_ID the id of the current
session\n\nreturns\nstatus is a status string that provides the result of the
operation "
},
{
"name": "call_RAG_agent",
"description": "call_RAG_agent(question: str) -> str - This is the function that
will query the relevant sources of information to get the answer to the
question.\n There will be situations when you are not able to answer the question
directly from the information you currently have. In such cases, you can search
for the answer in the relevant sources of information.\n Often the user will also
specify that you need to refer to \"information\" or \"documents\" to get the
```




```
answer.\n TAKSK: You have to frame the best possible \"question\" that is
extremely descriptive and then use it as a parameter to query the relevant
sources of information."
}
],
"memory_0": "<class
'langchain.memory.buffer_window.ConversationBufferWindowMemory'>",
"memory_1": "<class
'ASPIRE_LINQX.ai.memory.internal_logging.CustomActionLogSummaryMemory'>",
"memory_2": "<class 'ASPIRE_LINQX.ai.memory.embedding.EmbeddingSummaryMemory'>",
"initial_input": "can you recommend paramaters for an n-alkylation reaction using
3-Bromoquinoline and morpholine [II] as reactants. Provide the response as one
single continuous paragraph containing relevant information with the parameter
values in exactly the following format: Pressure: value, Temperature: value,
Time: list of values",
"total_iter": 20,
"success_iter": 19,
"fail_iter": 1,
"benchmark_score": 0.95
}
```


**Supplementary Fig. 30 | Benchmarking results for running regex benchmarking on an agent which has access to an embedding trained on electronic lab notebook (ELN) documentation.** The initial input question asks the agent to recommend heating parameters for a specific reaction and specifies how the output should be formatted. The desired regex looks for the correct temperature, pressure, and duration values and the correct formatting. The agent then benchmarks this request 20 times and provides information on the tools and tool descriptions, memory classes, initial input, success/fail iterations, and the benchmark score (0.95).

*Benchmarking log*

```
- Success on output key: output, matched a desired output
- Agent Output: For an n-alkylation reaction using 3-Bromoquinoline and
morpholine as reactants, the recommended parameters are as follows: Pressure: 1
bar, Temperature: 120 °C. The reaction times tested were 5, 10, 20, 30, 45, and
60 minutes. These parameters were used in a Buchwald-Hartwig reaction screen
using a Microwave Reactor (Biotage Initiator+). Please note that the ratio of the
reactants used was 3-Bromoquinoline: 1.0 equivalent and Morpholine: 1.2
equivalents. The catalyst employed was PEPPSI (formulation - ChemBeads) and the
solvents tested included Dioxane and Toluene.
- Iteration 1: Success
- Success: 1, Fail: 0, Total: 1
- Benchmarking Score: 1.0
--------------------
- Success on output key: output, matched a desired output
- Agent Output: The recommended parameters for an n-alkylation reaction using 3-
Bromoquinoline and morpholine [II] as reactants are as follows: Pressure: 1 atm,
Temperature: 120°C. The reaction times that have been tested include 5, 10, 20,
```



30, 45, and 60 minutes. However, the highest percentage conversions were achieved with a reaction time of 60 minutes.
- Iteration 2: Success
- Success: 2, Fail: 0, Total: 2
- Benchmarking Score: 1.0
--------------------
- Success on output key: output, matched a desired output
- Agent Output: The recommended parameters for an n-alkylation reaction using 3-Bromoquinoline and morpholine [II] as reactants are: Pressure: 1 bar, Temperature: 120 °C, Time: 5, 10, 20, 30, 45, and 60 minutes. Please note that these parameters were used in a Buchwald-Hartwig reaction screen using a Microwave Reactor (Biotage Initiator+).
- Iteration 3: Success
- Success: 3, Fail: 0, Total: 3
- Benchmarking Score: 1.0
--------------------
- Success on output key: output, matched a desired output
- Agent Output: Pressure: 1 atm, Temperature: 120°C, Time: 5, 10, 20, 30, 45, and 60 minutes. These parameters are based on a Buchwald-Hartwig reaction screen using a Microwave Reactor.
- Iteration 4: Success
- Success: 4, Fail: 0, Total: 4
- Benchmarking Score: 1.0
--------------------
- Success on output key: output, matched a desired output
- Agent Output: The recommended conditions for an n-alkylation reaction using 3-Bromoquinoline and morpholine [II] as reactants are as follows: Pressure: 1 atm, Temperature: 120°C, Time: 5, 10, 20, 30, 45, and 60 minutes. These conditions were used in a Buchwald-Hartwig reaction screen using a Microwave Reactor with the specified reactants.
- Iteration 5: Success
- Success: 5, Fail: 0, Total: 5
- Benchmarking Score: 1.0
--------------------
- Success on output key: output, matched a desired output
- Agent Output: The recommended parameters for an n-alkylation reaction using 3-Bromoquinoline and morpholine [II] as reactants are as follows: Pressure: 1 atm, Temperature: 120°C, Time: 5, 10, 20, 30, 45, and 60 minutes. These parameters are based on a Buchwald-Hartwig reaction screen using a Microwave Reactor. Please ensure to follow safety guidelines while performing the reaction.
- Iteration 6: Success
- Success: 6, Fail: 0, Total: 6
- Benchmarking Score: 1.0
--------------------
- Fail on output key: output, did not match any of the desired outputs



```
- Agent Output: For an n-alkylation reaction using 3-Bromoquinoline and
morpholine [II] as reactants, the recommended parameters are as follows. The
pressure should be 1 atm. The temperature should be set at 120°C. The reaction
should be carried out for a variety of time intervals for optimal results. These
intervals include 5, 10, 20, 30, 45, and 60 minutes. Please note that these
parameters are based on a specific reaction screen using the Buchwald-Hartwig
reaction with the mentioned reactants.
- Iteration 7: Fail
- Success: 6, Fail: 1, Total: 7
- Benchmarking Score: 0.8571428571428571
--------------------
- Success on output key: output, matched a desired output
- Agent Output: For an n-alkylation reaction using 3-Bromoquinoline and
morpholine [II] as reactants, the recommended parameters are Pressure: 1 bar,
Temperature: 120°C, Time: 60 minutes. These conditions have shown higher
percentage conversions in reactions using dioxane as a solvent. It's important to
note that the product (Aryl Amine) has a retention time of 0.62 minutes and the
starting material (3-bromoquinoline) has a retention time of 0.96 minutes.
- Iteration 8: Success
- Success: 7, Fail: 1, Total: 8
- Benchmarking Score: 0.875
--------------------
- Success on output key: output, matched a desired output
- Agent Output: The recommended parameters for an n-alkylation reaction using 3-
Bromoquinoline and morpholine [II] as reactants are Pressure: 1 bar, Temperature:
120°C, Time: 60 minutes. These parameters were tested in a Buchwald-Hartwig
reaction screen using a Microwave Reactor with the specified reactants and
conditions.
- Iteration 9: Success
- Success: 8, Fail: 1, Total: 9
- Benchmarking Score: 0.888888888888888
--------------------
- Success on output key: output, matched a desired output
- Agent Output: The recommended parameters for an n-alkylation reaction using 3-
Bromoquinoline and Morpholine [II] as reactants are as follows: Pressure: 1 atm,
Temperature: 120°C, Time: 5, 10, 20, 30, 45, and 60 minutes. These parameters
were tested in a Buchwald-Hartwig reaction screen using a Microwave Reactor
(Biotage Initiator+), and the product (Aryl Amine) had a retention time of 0.62
minutes while the starting material (3-bromoquinoline) had a retention time of
0.96 minutes.
- Iteration 10: Success
- Success: 9, Fail: 1, Total: 10
- Benchmarking Score: 0.9
--------------------
- Success on output key: output, matched a desired output
```



```
- Agent Output: The recommended parameters for an n-alkylation reaction using 3-
Bromoquinoline and morpholine [II] as reactants are as follows: Pressure: 1 atm,
Temperature: 120°C, Time: 5, 10, 20, 30, 45, and 60 minutes. These conditions
were used in a Buchwald-Hartwig reaction screen using a Microwave Reactor with
the specified reactants.
- Iteration 11: Success
- Success: 10, Fail: 1, Total: 11
- Benchmarking Score: 0.9090909090909091
--------------------
- Success on output key: output, matched a desired output
- Agent Output: The recommended parameters for an n-alkylation reaction using 3-
Bromoquinoline and morpholine [II] as reactants are as follows: Pressure: 1 bar,
Temperature: 120 °C, Time: 5, 10, 20, 30, 45, 60 minutes. Please note that these
parameters were used in a Buchwald-Hartwig reaction screen using a Microwave
Reactor (Biotage Initiator+).
- Iteration 12: Success
- Success: 11, Fail: 1, Total: 12
- Benchmarking Score: 0.9166666666666666
--------------------
- Success on output key: output, matched a desired output
- Agent Output: For an n-alkylation reaction using 3-Bromoquinoline and
morpholine [II] as reactants, the recommended parameters are as follows.
Pressure: 1 atm, Temperature: 120°C, Time: 5, 10, 20, 30, 45, and 60 minutes. The
reaction setup involves adding the reactants and catalyst to a Microwave Reactor
(Biotage Initiator+), setting the reaction parameters, stirring for a specified
time, and then stopping the reaction according to the defined time. The product
(Aryl Amine) has a retention time of 0.62 minutes, and the starting material (3-
bromoquinoline) has a retention time of 0.96 minutes.
- Iteration 13: Success
- Success: 12, Fail: 1, Total: 13
- Benchmarking Score: 0.9230769230769231
--------------------
- Success on output key: output, matched a desired output
- Agent Output: The recommended parameters for an n-alkylation reaction using 3-
Bromoquinoline and morpholine [II] as reactants are as follows: Pressure: 1 atm,
Temperature: 120°C, Time: 5, 10, 20, 30, 45, and 60 minutes.
- Iteration 14: Success
- Success: 13, Fail: 1, Total: 14
- Benchmarking Score: 0.9285714285714286
--------------------
- Success on output key: output, matched a desired output
- Agent Output: The recommended parameters for an n-alkylation reaction using 3-
Bromoquinoline and morpholine [II] are as follows: Pressure: 1 atm, Temperature:
120°C, Time: 5, 10, 20, 30, 45, and 60 minutes were tested with the highest
percentage conversions obtained at 60 minutes. It is advised to consider these
parameters and adjust according to the specific requirements of your experiment.
```



```
- Iteration 15: Success
- Success: 14, Fail: 1, Total: 15
- Benchmarking Score: 0.9333333333333333
--------------------
- Success on output key: output, matched a desired output
- Agent Output: The recommended parameters for an n-alkylation reaction using 3-
Bromoquinoline and morpholine [II] are as follows: Pressure: 1 atm, Temperature:
120°C, Time: 5, 10, 20, 30, 45, and 60 minutes. Please note that these parameters
were used in a Buchwald-Hartwig reaction screen using a Microwave Reactor.
- Iteration 16: Success
- Success: 15, Fail: 1, Total: 16
- Benchmarking Score: 0.9375
--------------------
- Success on output key: output, matched a desired output
- Agent Output: The recommended parameters for an n-alkylation reaction using 3-
Bromoquinoline and morpholine [II] as reactants are as follows: Pressure: 1 atm,
Temperature: 120°C, Time: 5, 10, 20, 30, 45, and 60 minutes. These parameters
were tested in a Buchwald-Hartwig reaction screen using a Microwave Reactor
(Biotage Initiator+), and the highest percentage conversions were achieved with a
reaction time of 60 minutes.
- Iteration 17: Success
- Success: 16, Fail: 1, Total: 17
- Benchmarking Score: 0.9411764705882353
--------------------
- Success on output key: output, matched a desired output
- Agent Output: The recommended parameters for an n-alkylation reaction involving
3-Bromoquinoline and morpholine [II] are as follows: Pressure: 1 atm,
Temperature: 120°C, Time: 5, 10, 20, 30, 45, and 60 minutes. Please note that
these conditions were screened for a Buchwald-Hartwig reaction using a Microwave
Reactor.
- Iteration 18: Success
- Success: 17, Fail: 1, Total: 18
- Benchmarking Score: 0.9444444444444444
--------------------
- Success on output key: output, matched a desired output
- Agent Output: The optimal parameters for an n-alkylation reaction using 3-
Bromoquinoline and morpholine [II] are as follows: Pressure: 1 atm, Temperature:
120°C, Time: 5, 10, 20, 30, 45, and 60 minutes. Please note that the reaction
also requires a base of KOtBu, and the solvents tested include dioxane and
toluene. A catalyst of PEPPSI is used with a 5% wt. loading of catalyst. The
reaction procedure involves adding the reactants and catalyst to a Microwave
Reactor (Biotage Initiator+), setting up the reaction parameters, stirring for a
specified time, and then stopping the reaction according to the defined time.
- Iteration 19: Success
- Success: 18, Fail: 1, Total: 19
- Benchmarking Score: 0.9473684210526315
```



```
--------------------
- Success on output key: output, matched a desired output
- Agent Output: Based on the information provided, the recommended parameters for
an n-alkylation reaction using 3-Bromoquinoline and morpholine [II] as reactants
are as follows: Pressure: 1 atm, Temperature: 120°C, Time: 5, 10, 20, 30, 45, and
60 minutes. It is crucial to note that these parameters are based on a specific
study and the optimal parameters may vary depending on the specific reaction
conditions and desired outcome.
- Iteration 20: Success
- Success: 19, Fail: 1, Total: 20
- Benchmarking Score: 0.95
--------------------
```

**Supplementary Fig. 31 | Benchmarking log of the agent with embedding access on ELN documentation for RAG purposes.** The agent correctly accesses the parameters from the embedding chain and formats the output in 19 of the 20 runs. In the failed run, the agent correctly accesses the parameters from the embedding chain but fails to format the output correctly.



# Agent RAG on Documentation Embedding 1

*Regex Pattern*

```
r'(?=.*\b(?:[Mm]icrowave\s[Ii]rradiation|[Cc]onventional\s[Hh]eating\s[Mm]ethods?
)\b)'
```

**Supplementary Fig. 32 | Regular expression string used to match agentic output to check for keywords associated with microwave synthesis heating protocols vs conventional heating protocols.**

*Benchmarking results*

```
{
"tools": [
{
"name": "allocate_session",
"description": "Function Signature:\n() -> dict\nFunction Docstring:\nAllocates a
session on the microwave synthesizer.\nMust be called prior to any other
action.\n\nreturns\nsession_ID the id of the allocated session"
},
{
"name": "close_lid",
"description": "Function Signature:\n(session_ID: str) -> dict\nFunction
Docstring:\nCloses the lid on the microwave synthesizer.\nMust be run prior to
running heating.\n\nparameters\nsession_ID the id of the current
session\n\nreturns\nstatus is a status string that provides the result of the
operation "
},
{
"name": "get_precent_conversion",
"description": "Function Signature:\n(session_ID: str) -> dict\nFunction
Docstring:\nGets the precent conversion of synthesis after running the
experiment.\nCan only be called after heating\n\nparameters\nsession_ID the id of
the current session\n\nreturns\nprecent_conversion denotes the precent conversion
of the sysntesis reaciton"
},
{
"name": "heat_vial",
"description": "Function Signature:\n(session_ID: str) -> dict\nFunction
Docstring:\nHeats the loaded vial to the set heating parameters.\nMust be run
after loading vial, closing lid, and updating
heating_parameters\n\nparameters\nsession_ID the id of the current
session\n\nreturns\nstatus is a status string that provides the result of the
operation "
},
{
"name": "load_vial",
```



```json
    "description": "Function Signature:\n(vial_num: int, session_ID: str) ->
dict\nFunction Docstring:\nLoads a vial into the microwave synthesizer.\nMust be
run prior to heating.\nparameters\nvial_num is an integer between 1 and
10.\nsession_ID the id of the current session\n\nreturns\nstatus is a status
string that provides the result of the operation "
},
{
    "name": "open_lid",
    "description": "Function Signature:\n(session_ID: str) -> dict\nFunction
Docstring:\nOpens the lid on the microwave synthesizer.\nMust be run prior to
loading a vial.\nparameters\nsession_ID the id of the current
session\n\nreturns\nstatus is a status string that provides the result of the
operation "
},
{
    "name": "unload_vial",
    "description": "Function Signature:\n(session_ID: str) -> dict\nFunction
Docstring:\nUnloads a vial from the microwave synthesizer.\nMust be run after
heating.\n\nparameters\nsession_ID the id of the current
session\n\nreturns\nstatus is a status string that provides the result of the
operation "
},
{
    "name": "update_heating_parameters",
    "description": "Function Signature:\n(duration: int, temperature: int, pressure:
float, session_ID: str) -> dict\nFunction Docstring:\nSets the heating parameters
of the microwave synthesizer.\nMust be run prior to
heating.\n\nparameters\nduration is an integer between 1 and 60
miniutes\ntemperature is an integer between 25 and 250 celsius\npressure is a
float between 1 and 10 atm\nsession_ID the id of the current
session\n\nreturns\nstatus is a status string that provides the result of the
operation "
},
{
    "name": "call_RAG_agent",
    "description": "call_RAG_agent(question: str) -> str - This is the function that
will query the relevant sources of information to get the answer to the
question.\n There will be situations when you are not able to answer the question
directly from the information you currently have. In such cases, you can search
for the answer in the relevant sources of information.\n Often the user will also
specify that you need to refer to \"information\" or \"documents\" to get the
answer.\n TAKSK: You have to frame the best possible \"question\" that is
extremely descriptive and then use it as a parameter to query the relevant
sources of information."
}
],
```




```
"memory_0": "<class
'langchain.memory.buffer_window.ConversationBufferWindowMemory'>",
"memory_1": "<class
'ASPIRE_LINQX.ai.memory.internal_logging.CustomActionLogSummaryMemory'>",
"memory_2": "<class 'ASPIRE_LINQX.ai.memory.embedding.EmbeddingSummaryMemory'>",
"initial_input": "How does microwave irradiation influence reaction mechanisms
differently compared to conventional heating methods?",
"total_iter": 20,
"success_iter": 20,
"fail_iter": 0,
"benchmark_score": 1.0
}
```


**Supplementary Fig. 33 | Benchmarking results of an agent with access to an embedding trained on microwave synthesizer protocols and documentation undergoing regex benchmarking.** The initial input question was asking about differences between microwave irradiation and conventional heating methods. The output includes information on the tools and tool descriptions, the initial input, the memory buffers and memory classes used, number of success/fail iterations, and the final benchmarking score (1.0).

*Benchmarking log*

```
- Success on output key: output, matched a desired output
- Agent Output: Microwave irradiation influences reaction mechanisms differently
compared to conventional heating methods by providing several advantages. First,
microwave irradiation is a well-established method for increasing the reaction
rate of chemical processes by generating high temperatures and pressures under
controlled conditions, allowing for quicker reactions. Second, unlike
conventional heating methods where thermal gradients can lead to non-uniform
transformations and obstacles for scale-up production, microwave irradiation
provides uniform heating to the reaction materials in the entire reaction vessel,
enabling rapid and uniform heat transfer. Third, microwave radiation of suitable
wavelength collaborates with the reaction material and generates heat
simultaneously, leading to efficient and rapid reactions. A comparison between
microwave and conventional heating methods shows that microwave-assisted
reactions have significantly shorter reaction times compared to conventional
heating methods, indicating the time and energy efficiency of microwave
irradiation in various chemical reactions. Overall, microwave irradiation offers
faster reaction rates, uniform heating, and increased efficiency compared to
conventional heating methods, making it a valuable tool in chemical synthesis
processes.
- Iteration 1: Success
- Success: 1, Fail: 0, Total: 1
- Benchmarking Score: 1.0
-------------------
- Success on output key: output, matched a desired output
- Agent Output: Microwave irradiation influences reaction mechanisms differently
compared to conventional heating methods by providing several advantages. First,
```



it is a well-established method for increasing the reaction rate of chemical processes by generating high temperatures and pressures under controlled conditions. Second, it provides uniform heating to the reaction materials in the entire reaction vessel, enabling rapid and uniform heat transfer. Third, microwave radiation interacts with the reaction material and generates heat simultaneously, leading to efficient and rapid reactions. A comparison between microwave and conventional heating methods shows that microwave irradiation significantly reduces reaction times for various compounds, making it a more time and energy-efficient technique for synthesis reactions.
- Iteration 2: Success
- Success: 2, Fail: 0, Total: 2
- Benchmarking Score: 1.0
--------------------
- Success on output key: output, matched a desired output
- Agent Output: Microwave irradiation influences reaction mechanisms differently compared to conventional heating methods by significantly reducing reaction times. For example, in the synthesis of compounds like esters, 4-Nitrobenzyl ester, and various intermetallic compounds, the reaction times are much shorter when using microwave heating (e.g., 2-5 minutes) compared to conventional heating methods (e.g., 8 hours to several days). This rapid heating under microwave irradiation leads to faster reactions and can result in different reaction pathways and product yields compared to conventional heating methods.
- Iteration 3: Success
- Success: 3, Fail: 0, Total: 3
- Benchmarking Score: 1.0
--------------------
- Success on output key: output, matched a desired output
- Agent Output: Microwave irradiation influences reaction mechanisms differently compared to conventional heating methods by significantly reducing reaction times. For example, in the synthesis of compounds like esters, 4-Nitrobenzyl ester, and various intermetallic compounds, the reaction times are much shorter when using microwave heating (e.g., 2-5 minutes) compared to conventional heating methods (e.g., 8 hours to 3 days). This rapid heating under microwave irradiation leads to faster reactions and can result in different reaction pathways and product distributions compared to conventional heating methods.
- Iteration 4: Success
- Success: 4, Fail: 0, Total: 4
- Benchmarking Score: 1.0
--------------------
- Success on output key: output, matched a desired output
- Agent Output: Microwave irradiation influences reaction mechanisms differently compared to conventional heating methods by providing several advantages. Firstly, microwave irradiation is a well-established method for increasing the reaction rate of chemical processes by generating high temperatures and pressures under controlled conditions, allowing chemistry that may have taken hours to be achieved in minutes. Secondly, microwave-assisted organic synthesis (MAOS) is



commonly used in pharmaceutical and industrial laboratories to trial new
reactions and develop new molecules rapidly. Unlike conventional heating methods
where thermal gradients can lead to non-uniform transformations and obstacles for
scale-up production, microwave heating provides uniform heating to the reaction
materials in the entire reaction vessel, enabling rapid and uniform heat
transfer. Microwave radiation collaborates with the reaction material and
generates heat simultaneously, leading to efficient and rapid reactions. A
comparison between microwave and conventional heating methods shows that
microwave heating is more time and energy-efficient for various reactions, with
significantly shorter reaction times compared to conventional heating methods.
Overall, microwave irradiation offers faster reaction rates, uniform heating, and
increased efficiency compared to conventional heating methods, making it a
valuable tool in chemical synthesis processes.
- Iteration 5: Success
- Success: 5, Fail: 0, Total: 5
- Benchmarking Score: 1.0
--------------------
- Success on output key: output, matched a desired output
- Agent Output: Microwave irradiation provides uniform heating to the reaction
materials in the entire reaction vessel, enabling rapid and uniform heat
transfer. This is in contrast to conventional heating methods where thermal
gradients can lead to non-uniform transformations and obstacles for scale-up
production. Microwave radiation collaborates with the reaction material and
generates heat simultaneously, leading to faster reaction times compared to
conventional heating methods.
- Iteration 6: Success
- Success: 6, Fail: 0, Total: 6
- Benchmarking Score: 1.0
--------------------
- Success on output key: output, matched a desired output
- Agent Output: Microwave irradiation influences reaction mechanisms differently
compared to conventional heating methods by providing several advantages:
1. It increases the reaction rate by generating high temperatures and pressures
under controlled conditions.
2. Unlike conventional heating where thermal gradients can lead to non-uniform
transformations, microwave irradiation provides uniform heating to the reaction
materials in the entire reaction vessel.
3. Microwave radiation of suitable wavelength interacts directly with the
reaction material and generates heat simultaneously, leading to efficient and
rapid reactions.
4. A comparison between microwave and conventional heating methods shows that
microwave-assisted reactions have significantly shorter reaction times,
indicating the time and energy efficiency of microwave irradiation in various
chemical reactions.



Overall, microwave irradiation offers faster reaction rates, uniform heating, and increased efficiency compared to conventional heating methods, making it a valuable tool in chemical synthesis processes.
- Iteration 7: Success
- Success: 7, Fail: 0, Total: 7
- Benchmarking Score: 1.0
--------------------
- Success on output key: output, matched a desired output
- Agent Output: Microwave irradiation influences reaction mechanisms differently compared to conventional heating methods by significantly reducing reaction times. For example, in a comparison between microwave and conventional heating methods for various reactions, the reaction times were much shorter when using microwave irradiation. For instance, esterification of benzoic acid with methanol took 5 minutes with microwave irradiation compared to 8 hours with conventional heating. This rapid heating under microwave conditions allows for faster reactions to occur, leading to increased efficiency in chemical processes.
- Iteration 8: Success
- Success: 8, Fail: 0, Total: 8
- Benchmarking Score: 1.0
--------------------
- Success on output key: output, matched a desired output
- Agent Output: Microwave irradiation influences reaction mechanisms differently compared to conventional heating methods by providing several advantages. Firstly, it provides uniform heating to the reaction materials in the entire reaction vessel, enabling rapid and uniform heat transfer. This is in contrast to conventional heating methods where thermal gradients can lead to non-uniform transformations and obstacles for scale-up production. Secondly, microwave irradiation can significantly reduce reaction times. For example, compounds synthesized using microwave irradiation can have much shorter reaction times (e.g., 2 minutes for 4-Nitrobenzyl ester synthesis) compared to hours required with conventional heating methods. Lastly, microwave irradiation is more time and energy efficient for certain reactions compared to conventional heating. For instance, esterification of benzoic acid with methanol takes 5 minutes with microwave irradiation compared to 8 hours with conventional heating. In summary, microwave irradiation offers advantages such as uniform heating, faster reaction times, and improved efficiency in comparison to conventional heating methods, leading to enhanced reaction mechanisms.
- Iteration 9: Success
- Success: 9, Fail: 0, Total: 9
- Benchmarking Score: 1.0
--------------------
- Success on output key: output, matched a desired output
- Agent Output: Microwave irradiation influences reaction mechanisms differently compared to conventional heating methods in several ways:



1. Microwave irradiation is a well-established method for increasing the reaction rate of chemical processes by generating high temperatures and pressures under controlled conditions. This allows chemical reactions that may have taken hours to be achieved in minutes.

2. Unlike conventional heating methods where thermal gradients can lead to non-uniform transformations and obstacles for scale-up production, microwave irradiation provides uniform heating to the reaction materials in the entire reaction vessel. This enables rapid and uniform heat transfer.

3. Microwave radiation of suitable wavelength interacts with the reaction material and generates heat simultaneously, leading to efficient and rapid reactions.

4. A comparison between microwave and conventional heating methods shows that microwave-assisted reactions have significantly shorter reaction times. This indicates the time and energy efficiency of microwave irradiation in various chemical reactions.

Overall, microwave irradiation offers faster reaction rates, uniform heating, and increased efficiency compared to conventional heating methods, making it a valuable tool in chemical synthesis processes.
- Iteration 10: Success
- Success: 10, Fail: 0, Total: 10
- Benchmarking Score: 1.0
--------------------
- Success on output key: output, matched a desired output
- Agent Output: Microwave irradiation influences reaction mechanisms differently compared to conventional heating methods by providing several advantages.

1. Microwave irradiation is a well-established method for increasing the reaction rate of chemical processes by generating high temperatures and pressures under controlled conditions, allowing chemistry that may have taken hours to be achieved in minutes.

2. Unlike conventional heating methods where thermal gradients can lead to non-uniform transformations and obstacles for scale-up production, microwave irradiation provides uniform heating to the reaction materials in the entire reaction vessel, enabling rapid and uniform heat transfer.

3. Microwave radiation of suitable wavelength collaborates with the reaction material and generates heat simultaneously, leading to efficient and rapid reactions.

4. A comparison between microwave and conventional heating methods shows that microwave-assisted reactions have significantly shorter reaction times compared



to conventional heating methods, indicating the time and energy efficiency of
microwave irradiation in various chemical reactions.

Overall, microwave irradiation offers faster reaction rates, uniform heating, and
increased efficiency compared to conventional heating methods, making it a
valuable tool in chemical synthesis processes.
- Iteration 11: Success
- Success: 11, Fail: 0, Total: 11
- Benchmarking Score: 1.0
--------------------
- Success on output key: output, matched a desired output
- Agent Output: Microwave irradiation and conventional heating methods influence
reaction mechanisms differently. In conventional heating, heat is generated
externally and transferred to the reaction vessel through convection or
conduction. This often leads to non-uniform transformations. On the other hand,
in microwave irradiation, the reaction mixture absorbs electric energy from the
microwave source, generating heat within the reaction material itself. This
provides uniform heating to the entire reaction vessel, enabling rapid and
uniform heat transfer without the formation of thermal gradients. As a result,
reactions under microwave irradiation can be faster and more efficient compared
to those under conventional heating methods.
- Iteration 12: Success
- Success: 12, Fail: 0, Total: 12
- Benchmarking Score: 1.0
--------------------
- Success on output key: output, matched a desired output
- Agent Output: Microwave irradiation influences reaction mechanisms differently
compared to conventional heating methods in several ways: 1. It can generate high
temperatures and pressures under controlled conditions, which can significantly
increase the reaction rate of chemical processes. 2. Microwave-assisted organic
synthesis (MAOS) is a method frequently used in pharmaceutical and industrial
laboratories to quickly test new reactions and develop new molecules. 3. In
contrast to conventional heating methods that can produce thermal gradients
leading to non-uniform transformations, microwave heating provides uniform
heating to the reaction materials, allowing for rapid and uniform heat transfer.
4. Microwave radiation interacts with the reaction material to generate heat,
leading to efficient and rapid reactions. 5. Compared to conventional heating
methods, microwave heating is more time and energy-efficient for various
reactions, often resulting in significantly shorter reaction times. Therefore,
microwave irradiation offers faster reaction rates, uniform heating, and
increased efficiency, making it a valuable tool in chemical synthesis processes.
- Iteration 13: Success
- Success: 13, Fail: 0, Total: 13
- Benchmarking Score: 1.0
--------------------
- Success on output key: output, matched a desired output



- Agent Output: Microwave irradiation influences reaction mechanisms differently compared to conventional heating methods by providing several advantages. These include:

1. **Uniform Heating**: Microwave irradiation provides uniform heating to the reaction materials in the entire reaction vessel, enabling rapid and uniform heat transfer. This is in contrast to conventional heating methods where thermal gradients can lead to non-uniform transformations and obstacles for scale-up production.

2. **Speed**: Microwave irradiation can significantly reduce reaction times compared to conventional heating methods. For example, compounds synthesized using microwave irradiation can have much shorter reaction times (e.g., 2 minutes for 4-Nitrobenzyl ester) compared to hours required with conventional heating methods.

3. **Efficiency**: Microwave irradiation is more time and energy efficient for certain reactions compared to conventional heating methods. For instance, esterification of benzoic acid with methanol takes 5 minutes with microwave irradiation compared to 8 hours with conventional heating.

Overall, microwave irradiation offers faster, more uniform heating, and increased efficiency in certain reactions compared to conventional heating methods.
- Iteration 14: Success
- Success: 14, Fail: 0, Total: 14
- Benchmarking Score: 1.0
--------------------
- Success on output key: output, matched a desired output
- Agent Output: Microwave irradiation influences reaction mechanisms differently compared to conventional heating methods by providing several advantages.

1. Microwave irradiation is a well-established method for increasing the reaction rate of chemical processes. It generates high temperatures and pressures under controlled conditions, allowing chemistry that may have taken hours to be achieved in minutes.

2. Microwave-assisted organic synthesis (MAOS) is commonly used in pharmaceutical and industrial laboratories to trial new reactions and develop new molecules rapidly.

3. Unlike conventional heating methods where thermal gradients can lead to non-uniform transformations and obstacles for scale-up production, microwave heating provides uniform heating to the reaction materials in the entire reaction vessel, enabling rapid and uniform heat transfer.



4. Microwave radiation collaborates with the reaction material and generates heat simultaneously, leading to efficient and rapid reactions.

5. A comparison between microwave and conventional heating methods shows that microwave heating is more time and energy-efficient for various reactions, with significantly shorter reaction times compared to conventional heating methods.

Overall, microwave irradiation offers faster reaction rates, uniform heating, and increased efficiency compared to conventional heating methods, making it a valuable tool in chemical synthesis processes.
- Iteration 15: Success
- Success: 15, Fail: 0, Total: 15
- Benchmarking Score: 1.0
--------------------
- Success on output key: output, matched a desired output
- Agent Output: Microwave irradiation provides uniform heating to the reaction materials in the entire reaction vessel, enabling rapid and uniform heat transfer. This is in contrast to conventional heating methods where thermal gradients can lead to non-uniform transformations and obstacles for scale-up production. Microwave radiation collaborates with the reaction material and generates heat simultaneously, leading to faster reaction times compared to conventional heating methods.
- Iteration 16: Success
- Success: 16, Fail: 0, Total: 16
- Benchmarking Score: 1.0
--------------------
- Success on output key: output, matched a desired output
- Agent Output: Microwave irradiation influences reaction mechanisms differently compared to conventional heating methods by significantly reducing reaction times. For example, in the synthesis of compounds like esters, 4-Nitrobenzyl ester, and various intermetallic compounds, the reaction times are much shorter when using microwave heating (e.g., 2-5 minutes) compared to conventional heating methods (e.g., 8 hours to several days). This rapid heating under microwave irradiation leads to faster reactions and can result in different reaction pathways and product distributions compared to conventional heating methods.
- Iteration 17: Success
- Success: 17, Fail: 0, Total: 17
- Benchmarking Score: 1.0
--------------------
- Success on output key: output, matched a desired output
- Agent Output: Microwave irradiation affects reaction mechanisms differently compared to conventional heating methods in several ways. Here are some key differences:

1. Microwave irradiation is a known method for speeding up the reaction rate of chemical processes by generating high temperatures and pressures under controlled



conditions. This allows for reactions that could have taken hours to be completed in just minutes.

2. Unlike conventional heating methods where thermal gradients can result in non-uniform transformations and challenges for scale-up production, microwave irradiation provides uniform heating to the reaction materials in the entire reaction vessel. This ensures rapid and uniform heat transfer.

3. Microwave radiation of suitable wavelength interacts with the reaction material and generates heat simultaneously, leading to efficient and rapid reactions.

4. When comparing microwave and conventional heating methods, microwave irradiation significantly reduces reaction times for various compounds, making it a more time and energy-efficient technique for synthesis reactions.

So, overall, microwave irradiation offers faster reaction rates, uniform heating, and better efficiency compared to conventional heating methods, making it a valuable tool in chemical synthesis processes.
- Iteration 18: Success
- Success: 18, Fail: 0, Total: 18
- Benchmarking Score: 1.0
--------------------
- Success on output key: output, matched a desired output
- Agent Output: Microwave irradiation influences reaction mechanisms differently compared to conventional heating methods by increasing the reaction rate of chemical processes. By employing microwaves to generate high temperatures and pressures under controlled conditions, chemistry that may have taken hours can be achieved in minutes. This is known as microwave-assisted organic synthesis (MAOS), which allows for rapid development of new molecules and reactions. Microwave reactors operate at a frequency of 2.45GHz and can create standing wave designs that enhance energy intensity in the reaction chamber. Compared to conventional heating methods, microwave heating is more efficient in terms of time and energy for various reactions, as shown in several examples provided.
- Iteration 19: Success
- Success: 19, Fail: 0, Total: 19
- Benchmarking Score: 1.0
--------------------
- Success on output key: output, matched a desired output
- Agent Output: Microwave irradiation influences reaction mechanisms differently compared to conventional heating methods by providing several advantages. Microwave irradiation is a well-established method for increasing the reaction rate of chemical processes by generating high temperatures and pressures under controlled conditions, allowing chemistry that may have taken hours to be achieved in minutes. Unlike conventional heating methods where thermal gradients can lead to non-uniform transformations and obstacles for scale-up production,



```
microwave irradiation provides uniform heating to the reaction materials in the
entire reaction vessel, enabling rapid and uniform heat transfer. Microwave
radiation of suitable wavelength collaborates with the reaction material and
generates heat simultaneously, leading to efficient and rapid reactions. A
comparison between microwave and conventional heating methods shows that
microwave-assisted reactions have significantly shorter reaction times compared
to conventional heating methods, indicating the time and energy efficiency of
microwave irradiation in various chemical reactions. Overall, microwave
irradiation offers faster reaction rates, uniform heating, and improved
efficiency compared to conventional heating methods, making it a valuable tool in
chemical synthesis processes.
- Iteration 20: Success
- Success: 20, Fail: 0, Total: 20
- Benchmarking Score: 1.0
--------------------
```

**Supplementary Fig. 34 | Benchmarking results of the agent provided with an embedding chain with access to microwave synthesis protocols and documentation.** In all benchmarking responses, the agent's output contains the desired keywords set in the regex string.



## Agent RAG on Documentation Embedding 2

*Regex Pattern used*

```
r'(?=.*\b(?:[Pp]ressure\s[Ss]etting|[Mm]icrowave|[Ss]ynthesizer|[Bb]iotage\s[Ii]n
itiator\+)\b)'
```

**Supplementary Fig. 35 | Regular expression string used to match agentic output to check for keywords associated with information on the pressure setting of the Biotage Initiator+.**

*Benchmarking results*

```
{
"tools": [
{
"name": "allocate_session",
"description": "Function Signature:\n() -> dict\nFunction Docstring:\nAllocates a
session on the microwave synthesizer.\nMust be called prior to any other
action.\n\nreturns\nsession_ID the id of the allocated session"
},
{
"name": "close_lid",
"description": "Function Signature:\n(session_ID: str) -> dict\nFunction
Docstring:\nCloses the lid on the microwave synthesizer.\nMust be run prior to
running heating.\n\nparameters\nsession_ID the id of the current
session\n\nreturns\nstatus is a status string that provides the result of the
operation "
},
{
"name": "get_precent_conversion",
"description": "Function Signature:\n(session_ID: str) -> dict\nFunction
Docstring:\nGets the precent conversion of synthesis after running the
experiment.\nCan only be called after heating\n\nparameters\nsession_ID the id of
the current session\n\nreturns\nprecent_conversion denotes the precent conversion
of the sysntesis reaciton"
},
{
"name": "heat_vial",
"description": "Function Signature:\n(session_ID: str) -> dict\nFunction
Docstring:\nHeats the loaded vial to the set heating parameters.\nMust be run
after loading vial, closing lid, and updating
heating_parameters\n\nparameters\nsession_ID the id of the current
session\n\nreturns\nstatus is a status string that provides the result of the
operation "
},
{
"name": "load_vial",
```



```
    "description": "Function Signature:\n(vial_num: int, session_ID: str) ->
dict\nFunction Docstring:\nLoads a vial into the microwave synthesizer.\nMust be
run prior to heating.\n\nparameters\nvial_num is an integer between 1 and
10.\nsession_ID the id of the current session\n\nreturns\nstatus is a status
string that provides the result of the operation "
},
{
    "name": "open_lid",
    "description": "Function Signature:\n(session_ID: str) -> dict\nFunction
Docstring:\nOpens the lid on the microwave synthesizer.\nMust be run prior to
loading a vial.\n\nparameters\nsession_ID the id of the current
session\n\nreturns\nstatus is a status string that provides the result of the
operation "
},
{
    "name": "unload_vial",
    "description": "Function Signature:\n(session_ID: str) -> dict\nFunction
Docstring:\nUnloads a vial from the microwave synthesizer.\nMust be run after
heating.\n\nparameters\nsession_ID the id of the current
session\n\nreturns\nstatus is a status string that provides the result of the
operation "
},
{
    "name": "update_heating_parameters",
    "description": "Function Signature:\n(duration: int, temperature: int, pressure:
float, session_ID: str) -> dict\nFunction Docstring:\nSets the heating parameters
of the microwave synthesizer.\nMust be run prior to
heating.\n\nparameters\nduration is an integer between 1 and 60
miniutes\ntemperature is an integer between 25 and 250 celsius\npressure is a
float between 1 and 10 atm\nsession_ID the id of the current
session\n\nreturns\nstatus is a status string that provides the result of the
operation "
},
{
    "name": "call_RAG_agent",
    "description": "call_RAG_agent(question: str) -> str - This is the function that
will query the relevant sources of information to get the answer to the
question.\n There will be situations when you are not able to answer the question
directly from the information you currently have. In such cases, you can search
for the answer in the relevant sources of information.\n Often the user will also
specify that you need to refer to \"information\" or \"documents\" to get the
answer.\n TAKSK: You have to frame the best possible \"question\" that is
extremely descriptive and then use it as a parameter to query the relevant
sources of information."
}
],
```




```
"memory_0": "<class
'langchain.memory.buffer_window.ConversationBufferWindowMemory'>",
"memory_1": "<class
'ASPIRE_LINQX.ai.memory.internal_logging.CustomActionLogSummaryMemory'>",
"memory_2": "<class 'ASPIRE_LINQX.ai.memory.embedding.EmbeddingSummaryMemory'>",
"initial_input": "What function does the pressure setting serve on the Biotage
Initiator+?",
"total_iter": 20,
"success_iter": 20,
"fail_iter": 0,
"benchmark_score": 1.0
}
```


**Supplementary Fig. 36 | Benchmarking results of an agent with access to an embedding trained on microwave synthesizer protocols and documentation undergoing regex benchmarking.** The initial question asks about the function of the pressure setting of the Biotage Initiator+. The results include information of tools and tool descriptions, the memory classes and initial buffers provided, the initial input, the success/fail iterations, and the final benchmark score (1.0).

*Benchmarking log*

```
- Success on output key: output, matched a desired output
- Agent Output: The pressure setting on the Biotage Initiator+ allows for
temperatures and pressures of up to 300 °C and 30 bar. This feature provides the
ability to perform complex reactions that would be difficult under other
conditions. Specifically, it allows solvents with low boiling points to be used
at higher temperatures, providing more flexibility in choosing a solvent for
organic synthesis.
- Iteration 1: Success
- Success: 1, Fail: 0, Total: 1
- Benchmarking Score: 1.0
--------------------
- Success on output key: output, matched a desired output
- Agent Output: The pressure setting on the Biotage Initiator+ allows for
temperatures and pressures of up to 300 °C and 30 bar, which opens up new
possibilities to perform difficult reactions. This feature enables even solvents
with low boiling points to be run at higher temperatures, providing more
flexibility in choosing a solvent for reactions.
- Iteration 2: Success
- Success: 2, Fail: 0, Total: 2
- Benchmarking Score: 1.0
--------------------
- Success on output key: output, matched a desired output
- Agent Output: The pressure setting on the Biotage Initiator+ serves a crucial
function. It allows for temperatures and pressures of up to 300 °C and 30 bar,
which opens up new possibilities to perform difficult reactions. This feature
```



enables even solvents with low boiling points to be run at higher temperatures, providing more flexibility in choosing a solvent for the synthesis process.
- Iteration 3: Success
- Success: 3, Fail: 0, Total: 3
- Benchmarking Score: 1.0
--------------------
- Success on output key: output, matched a desired output
- Agent Output: The pressure setting on the Biotage Initiator+ Microwave Synthesizer serves the function of allowing for higher temperatures and pressures of up to 300 °C and 30 bar. This feature opens up new possibilities to perform difficult reactions and enables even solvents with low boiling points to be run at higher temperatures, providing more flexibility in choosing a solvent.
- Iteration 4: Success
- Success: 4, Fail: 0, Total: 4
- Benchmarking Score: 1.0
--------------------
- Success on output key: output, matched a desired output
- Agent Output: The pressure setting on a Biotage Initiator+ microwave synthesizer allows for temperatures and pressures up to 300 °C and 30 bar. This feature is designed to automatically sense and perform reactions at their highest possible temperatures, providing greater flexibility in running reactions with different solvents, including those with low boiling points, at higher temperatures.
- Iteration 5: Success
- Success: 5, Fail: 0, Total: 5
- Benchmarking Score: 1.0
--------------------
- Success on output key: output, matched a desired output
- Agent Output: The pressure setting on the Biotage Initiator+ allows for temperatures and pressures of up to 300 °C and 30 bar. This feature opens up new possibilities to perform difficult reactions. Even solvents with low boiling points can be run at higher temperatures, providing more flexibility in choosing a solvent for reactions.
- Iteration 6: Success
- Success: 6, Fail: 0, Total: 6
- Benchmarking Score: 1.0
--------------------
- Success on output key: output, matched a desired output
- Agent Output: The pressure setting on the Biotage Initiator+ Microwave Synthesizer serves to allow for higher pressures of up to 30 bar. This opens up new possibilities to perform difficult reactions by providing more flexibility in choosing solvents and facilitating the completion of challenging reactions.
- Iteration 7: Success
- Success: 7, Fail: 0, Total: 7
- Benchmarking Score: 1.0
--------------------



```
- Success on output key: output, matched a desired output
- Agent Output: The pressure setting on the Biotage Initiator+ microwave
synthesizer allows for higher temperatures and pressures, enabling difficult
reactions and allowing solvents with low boiling points to be run at higher
temperatures. The system automatically performs reactions at their highest
possible temperatures, providing more flexibility in choosing solvents for
synthesis processes.
- Iteration 8: Success
- Success: 8, Fail: 0, Total: 8
- Benchmarking Score: 1.0
--------------------
- Success on output key: output, matched a desired output
- Agent Output: The pressure setting on the Biotage Initiator+ serves several
functions. It allows for temperatures and pressures of up to 300 °C and 30 bar,
which enables the performance of difficult reactions. Moreover, this feature
allows solvents with low boiling points to be used at higher temperatures,
providing more flexibility in choosing a solvent for reactions.
- Iteration 9: Success
- Success: 9, Fail: 0, Total: 9
- Benchmarking Score: 1.0
--------------------
- Success on output key: output, matched a desired output
- Agent Output: The pressure setting on the Biotage Initiator+ microwave
synthesizer is quite significant. It allows for temperatures and pressures up to
300 °C and 30 bar, respectively. This feature enables the system to automatically
sense and perform reactions at their highest possible temperatures, offering more
flexibility when dealing with different solvents, including those with low
boiling points, at elevated temperatures.
- Iteration 10: Success
- Success: 10, Fail: 0, Total: 10
- Benchmarking Score: 1.0
--------------------
- Success on output key: output, matched a desired output
- Agent Output: The pressure setting on the Biotage Initiator+ microwave
synthesizer serves to enable higher temperatures and pressures of up to 300 °C
and 30 bar. This feature allows for the performance of more challenging reactions
and even enables solvents with low boiling points to be run at higher
temperatures. The system is designed to automatically sense and carry out
reactions at their highest possible temperatures, offering more flexibility in
the choice of solvents for synthesis processes.
- Iteration 11: Success
- Success: 11, Fail: 0, Total: 11
- Benchmarking Score: 1.0
--------------------
- Success on output key: output, matched a desired output
```



```
- Agent Output: The pressure setting on the Biotage Initiator+ microwave
synthesizer allows for higher temperatures and pressures of up to 300 °C and 30
bar. This feature enables the performance of difficult reactions and allows even
solvents with low boiling points to be used at higher temperatures. The system
automatically senses and performs reactions at their highest possible
temperatures, providing more flexibility in choosing solvents for synthesis
processes.
- Iteration 12: Success
- Success: 12, Fail: 0, Total: 12
- Benchmarking Score: 1.0
--------------------
- Success on output key: output, matched a desired output
- Agent Output: The pressure setting on the Biotage Initiator+ serves to allow
for higher temperatures and pressures of up to 300 °C and 30 bar. This feature
opens up new possibilities to perform difficult reactions, even with solvents
that have low boiling points. The system automatically senses and performs
reactions at their highest possible temperatures, providing greater flexibility
in choosing solvents and enabling the completion of challenging reactions.
- Iteration 13: Success
- Success: 13, Fail: 0, Total: 13
- Benchmarking Score: 1.0
--------------------
- Success on output key: output, matched a desired output
- Agent Output: The pressure setting on a Biotage Initiator+ microwave
synthesizer allows chemists to explore new areas and perform the latest
innovations in drug discovery. The Initiator+ can achieve temperatures and
pressures up to 300 °C and 30 bar, which opens up new possibilities to complete
difficult reactions. This feature enables even low boiling point solvents to be
run at higher temperatures, providing more flexibility in choosing a solvent for
the synthesis process.
- Iteration 14: Success
- Success: 14, Fail: 0, Total: 14
- Benchmarking Score: 1.0
--------------------
- Success on output key: output, matched a desired output
- Agent Output: The pressure setting on the Biotage Initiator+ is designed to
allow for temperatures and pressures of up to 300 °C and 30 bar, respectively.
This feature enables even solvents with low boiling points to be run at higher
temperatures, providing more flexibility when choosing a solvent for reactions.
- Iteration 15: Success
- Success: 15, Fail: 0, Total: 15
- Benchmarking Score: 1.0
--------------------
- Success on output key: output, matched a desired output
- Agent Output: The pressure setting on the Biotage Initiator+ microwave
synthesizer serves to allow for higher temperatures and pressures, up to 300 °C
```



and 30 bar. This opens up new possibilities to perform difficult reactions and even allows solvents with low boiling points to be run at higher temperatures. The system automatically senses and performs reactions at their highest possible temperatures, providing more flexibility in choosing solvents for synthesis processes.
- Iteration 16: Success
- Success: 16, Fail: 0, Total: 16
- Benchmarking Score: 1.0
--------------------
- Success on output key: output, matched a desired output
- Agent Output: The pressure setting on the Biotage Initiator+ serves several functions. It allows for temperatures and pressures of up to 300°C and 30 bar, which enables the system to perform difficult reactions that would not be possible under normal conditions. Moreover, this feature provides greater flexibility in choosing solvents as it allows for reactions with solvents that have low boiling points to be run at higher temperatures.
- Iteration 17: Success
- Success: 17, Fail: 0, Total: 17
- Benchmarking Score: 1.0
--------------------
- Success on output key: output, matched a desired output
- Agent Output: The pressure setting on the Biotage Initiator+ microwave synthesizer allows for higher temperatures and pressures of up to 300 ºC and 30 bar. This feature opens up new possibilities to perform difficult reactions and allows even solvents with low boiling points to be run at higher temperatures. The system automatically senses and performs reactions at their highest possible temperatures, providing more flexibility in choosing solvents for synthesis.
- Iteration 18: Success
- Success: 18, Fail: 0, Total: 18
- Benchmarking Score: 1.0
--------------------
- Success on output key: output, matched a desired output
- Agent Output: The pressure setting on the Biotage Initiator+ microwave synthesizer allows for high temperatures and pressures up to 300 ºC and 30 bar. This feature enables the execution of difficult reactions and even allows solvents with low boiling points to be run at higher temperatures. The system automatically senses and carries out reactions at their highest possible temperatures, providing more flexibility in choosing solvents for synthesis processes.
- Iteration 19: Success
- Success: 19, Fail: 0, Total: 19
- Benchmarking Score: 1.0
--------------------
- Success on output key: output, matched a desired output
- Agent Output: The pressure setting on the Biotage Initiator+ serves to allow for higher temperatures and pressures of up to 300 ºC and 30 bar. This feature



```
opens up new possibilities to perform difficult reactions, even with solvents
that have low boiling points. The system automatically senses and performs
reactions at their highest possible temperatures, providing greater flexibility
in choosing solvents and enabling the user to achieve better results in their
organic synthesis processes.
- Iteration 20: Success
- Success: 20, Fail: 0, Total: 20
- Benchmarking Score: 1.0
-------------------
```

**Supplementary Fig. 37 | Benchmarking results on the agent provided with an embedding chain trained on microwave synthesis and protocols for RAG purposes.** The agent mentions the keywords defined in the regex in every single benchmark operation.



# PubChem Agent Benchmarking

## PubChem Agent Assay Retrieval Benchmarking

```python
class CidFromNameSchema(BaseModel):
inp: Literal['pzm21', 'PZM21']

class AssayResultsFromCompound(BaseModel):
cid: Literal['121596705', '124186394']
activity_name: Literal['Ki']

class AssayDescriptionFromAid(BaseModel):
aid: Literal['1806958', '1609954', '1573390']

path = [
    [
        ('get_cid_from_name', CidFromNameSchema),
        ('get_assay_results_for_compund',AssayResultsFromCompound),
        ('get_assay_description',AssayDescriptionFromAid)
    ],
    [
        ('get_cid_from_name', CidFromNameSchema),
        ('get_assay_results_for_compund',AssayResultsFromCompound),
        ('get_assay_results_for_compund',AssayResultsFromCompound),
        ('get_assay_description',AssayDescriptionFromAid)
    ]
]
```

**Supplementary Fig. 38 | Schema and paths used to validate initial state benchmarking of the PubChem REST API agent.** This benchmark expects one of two paths where the agent gets the compound's CID from the name "pzm21" or "PZM21", gets Ki assay results for one or both corresponding CID's , and then finally gets the description from one of three valid assay ID's.

*Benchmarking results*

```json
{
"tools": [
{
"name": "get_assay_description",
"description": "Function Signature:\n(aid: str) -> dict\nFunction
Docstring:\nFunction purpose\nGet high level Assay description, protocol and
comment on the scores for a given assay id.\n\nInputs\naid: string representation
of a list of assay IDs\n\nreturns\ndict: a dictionary of description, protocol
and comment for the given assay ID"
},
{
"name": "get_assay_name_from_aid",
```



```
  "description": "Function Signature:\n(aid: str) -> dict\nFunction
Docstring:\nFunction purpose\nGives a dictionary of names for each assay ID
(aid)\n\nInputs\naid: string representation of a list of assay
IDs\n\nreturns\nstr: a dictionary of names for each assay ID"
},
{
  "name": "get_assay_results_for_compund",
  "description": "Function Signature:\n(cid: str, activity_name: str = None) ->
dict\nFunction Docstring:\nFunction purpose\nGets all the assay results for a
provided compund with an optional filter of assay activity type.\n\nInputs:\ncid:
a single cid representing a compound\nactivity_name (optional): the specific
activity type to filter on for example Ki, Kd, IC50, etc\n\nreturns\ndict: a
dictionary representation of the assay result
table\nhttps://pubchem.ncbi.nlm.nih.gov/rest/pug/compound/cid/121596705/assaysumm
ary/JSON"
},
{
  "name": "get_cid_from_name",
  "description": "Function Signature:\n(inp: str) -> dict\nFunction
Docstring:\nFunction purpose\nGet the cid from chemical name\n\nInputs\ninp:
string representation of a chemical name\n\nreturns\ndict: "
},
{
  "name": "get_compound_property_table",
  "description": "Function Signature:\n(inp: str, inp_format: str, inp_type: str,
property_list: str) -> Dict[str, str | int]\nFunction Docstring:\nNone"
},
{
  "name": "get_description",
  "description": "Function Signature:\n(inp: str, inp_format: str, inp_type: str) -
> dict\nFunction Docstring:\nFunction purpose \nGet description of a substance or
a compound, for assay description, use get_assay_description()
instead\n\nInputs\ninp: string representation of an single chemical name
only\ninp_format: string of either of name, sid, cid, smiles \ninp_type:
'compound' if inp_format is cid, name or smiles of compound or 'substance' if
inp_format is sid, name, smiles of substance\n\nreturns\ndict: a dictionary of
descriptions for each identifier"
},
{
  "name": "get_smiles_from_name",
  "description": "Function Signature:\n(inp: str) -> dict\nFunction
Docstring:\nFunction purpose\nGet SMILES for a given chemical name.
\n\nInputs\ninp: string representation of a chemical name\n\nreturns\nstr: a text
representing the SMILES for the given chemical name"
},
{
```




```
"name": "get_synonym",
"description": "Function Signature:\n(inp: str, inp_format: str, inp_type: str) -
> dict\nFunction Docstring:\nFunction purpose\nGet Synonym of a substance or
compound.\n\nInputs\ninp: string representation of a list of
identifiers\ninp_format: string which can be either of name, sid, cid,
smiles\ninp_type: 'compound' if inp_format is cid, name or smiles of compound or
'substance' if inp_format is sid, name, smiles of substance\n\nreturns\ndict: a
dictionary of synonyms for each identifier"
}
],
"memory_0": "<class
'langchain.memory.buffer_window.ConversationBufferWindowMemory'>",
"memory_1": "<class
'ASPIRE_LINQX.ai.memory.internal_logging.CustomActionLogSummaryMemory'>",
"initial_input": "What is the Ki value of the compound PZM21 when it binds to the
Mu opioid receptor? Please select one specific assay used to determine this
value, provide a citation for that particular assay, and give a detailed
description of how the assay was conducted.",
"total_iter": 20,
"success_iter": 17,
"fail_iter": 3,
"benchmark_score": 0.85
```


**Supplementary Fig. 39 | Benchmarking results of an agent built around the PubChem REST API undergoing path-based benchmarking.** The agent is asked to retrieve the Ki value of the compound PZM21 to the mu opioid receptor and then cite the specific assay which was the source of that information with a detailed description of how the assay was conducted. The benchmarking results include the tools and tool descriptions, the memory classes used, the initial input, the number of success/fail iterations, and the final benchmarking score (0.85).

*Benchmarking log*


```
https://pubchem.ncbi.nlm.nih.gov/rest/pug/assay/aid/1573390/description/JSON
- Success on output key: intermediate_steps, matched a desired output
- Agent action path:
get_cid_from_name: {"inp": "PZM21"}
get_assay_results_for_compound: {"cid": "121596705", "activity_name": "Ki"}
get_assay_description: {"aid": "1573390"}

- Iteration 1: Success
- Success: 1, Fail: 0, Total: 1
- Benchmarking Score: 1.0
-------------------
https://pubchem.ncbi.nlm.nih.gov/rest/pug/assay/aid/1573390/description/JSON
- Success on output key: intermediate_steps, matched a desired output
- Agent action path:
```




```
get_cid_from_name: {"inp": "PZM21"}
get_assay_results_for_compund: {"cid": "121596705", "activity_name": "Ki"}
get_assay_description: {"aid": "1573390"}

- Iteration 2: Success
- Success: 2, Fail: 0, Total: 2
- Benchmarking Score: 1.0
-------------------
https://pubchem.ncbi.nlm.nih.gov/rest/pug/assay/aid/1573390/description/JSON
names Displacement of [3H]-Diprenorphine from mu opioid receptor (unknown origin)
expressed in sf9 insect cell membranes after 1 hr by liquid scintillation
counting method
- Fail on output key: intermediate_steps, did not match any of the desired
outputs
- Agent action path:
get_cid_from_name: {"inp": "PZM21"}
get_assay_results_for_compund: {"cid": "121596705", "activity_name": "Ki"}
get_assay_description: {"aid": "1573390"}
get_assay_name_from_aid: {"aid": "1573390"}

- Iteration 3: Fail
- Success: 2, Fail: 1, Total: 3
- Benchmarking Score: 0.6666666666666666
-------------------
https://pubchem.ncbi.nlm.nih.gov/rest/pug/assay/aid/1573390/description/JSON
- Success on output key: intermediate_steps, matched a desired output
- Agent action path:
get_cid_from_name: {"inp": "PZM21"}
get_assay_results_for_compund: {"cid": "121596705", "activity_name": "Ki"}
get_assay_description: {"aid": "1573390"}

- Iteration 4: Success
- Success: 3, Fail: 1, Total: 4
- Benchmarking Score: 0.75
-------------------
Error on iteration 5: LinqxTool._run() takes 1 positional argument but 2 were
given
- Iteration 5: Fail
- Success: 3, Fail: 2, Total: 5
- Benchmarking Score: 0.6
-------------------
https://pubchem.ncbi.nlm.nih.gov/rest/pug/assay/aid/1573390/description/JSON
- Success on output key: intermediate_steps, matched a desired output
- Agent action path:
get_cid_from_name: {"inp": "PZM21"}
get_assay_results_for_compund: {"cid": "121596705", "activity_name": "Ki"}
```



```
get_assay_description: {"aid": "1573390"}

- Iteration 6: Success
- Success: 4, Fail: 2, Total: 6
- Benchmarking Score: 0.6666666666666666
--------------------
https://pubchem.ncbi.nlm.nih.gov/rest/pug/assay/aid/1573390/description/JSON
- Success on output key: intermediate_steps, matched a desired output
- Agent action path:
get_cid_from_name: {"inp": "PZM21"}
get_assay_results_for_compund: {"cid": "121596705", "activity_name": "Ki"}
get_assay_description: {"aid": "1573390"}

- Iteration 7: Success
- Success: 5, Fail: 2, Total: 7
- Benchmarking Score: 0.7142857142857143
--------------------
https://pubchem.ncbi.nlm.nih.gov/rest/pug/assay/aid/1573390/description/JSON
- Success on output key: intermediate_steps, matched a desired output
- Agent action path:
get_cid_from_name: {"inp": "PZM21"}
get_assay_results_for_compund: {"cid": "121596705", "activity_name": "Ki"}
get_assay_description: {"aid": "1573390"}

- Iteration 8: Success
- Success: 6, Fail: 2, Total: 8
- Benchmarking Score: 0.75
--------------------
https://pubchem.ncbi.nlm.nih.gov/rest/pug/assay/aid/1573390/description/JSON
- Success on output key: intermediate_steps, matched a desired output
- Agent action path:
get_cid_from_name: {"inp": "PZM21"}
get_assay_results_for_compund: {"cid": "121596705", "activity_name": "Ki"}
get_assay_description: {"aid": "1573390"}

- Iteration 9: Success
- Success: 7, Fail: 2, Total: 9
- Benchmarking Score: 0.7777777777777778
--------------------
https://pubchem.ncbi.nlm.nih.gov/rest/pug/assay/aid/1573390/description/JSON
- Success on output key: intermediate_steps, matched a desired output
- Agent action path:
get_cid_from_name: {"inp": "PZM21"}
get_assay_results_for_compund: {"cid": "121596705", "activity_name": "Ki"}
get_assay_description: {"aid": "1573390"}
```



```
- Iteration 10: Success
- Success: 8, Fail: 2, Total: 10
- Benchmarking Score: 0.8
--------------------
https://pubchem.ncbi.nlm.nih.gov/rest/pug/assay/aid/1573390/description/JSON
- Success on output key: intermediate_steps, matched a desired output
- Agent action path:
get_cid_from_name: {"inp": "PZM21"}
get_assay_results_for_compund: {"cid": "121596705", "activity_name": "Ki"}
get_assay_description: {"aid": "1573390"}

- Iteration 11: Success
- Success: 9, Fail: 2, Total: 11
- Benchmarking Score: 0.8181818181818182
--------------------
Error on iteration 12: LinqxTool._run() takes 1 positional argument but 2 were
given
- Iteration 12: Fail
- Success: 9, Fail: 3, Total: 12
- Benchmarking Score: 0.75
--------------------
https://pubchem.ncbi.nlm.nih.gov/rest/pug/assay/aid/1573390/description/JSON
- Success on output key: intermediate_steps, matched a desired output
- Agent action path:
get_cid_from_name: {"inp": "PZM21"}
get_assay_results_for_compund: {"cid": "121596705", "activity_name": "Ki"}
get_assay_description: {"aid": "1573390"}

- Iteration 13: Success
- Success: 10, Fail: 3, Total: 13
- Benchmarking Score: 0.7692307692307693
--------------------
https://pubchem.ncbi.nlm.nih.gov/rest/pug/assay/aid/1573390/description/JSON
- Success on output key: intermediate_steps, matched a desired output
- Agent action path:
get_cid_from_name: {"inp": "PZM21"}
get_assay_results_for_compund: {"cid": "121596705", "activity_name": "Ki"}
get_assay_description: {"aid": "1573390"}

- Iteration 14: Success
- Success: 11, Fail: 3, Total: 14
- Benchmarking Score: 0.7857142857142857
--------------------
https://pubchem.ncbi.nlm.nih.gov/rest/pug/assay/aid/1573390/description/JSON
- Success on output key: intermediate_steps, matched a desired output
- Agent action path:
```



```
get_cid_from_name: {"inp": "PZM21"}
get_assay_results_for_compund: {"cid": "121596705", "activity_name": "Ki"}
get_assay_description: {"aid": "1573390"}

- Iteration 15: Success
- Success: 12, Fail: 3, Total: 15
- Benchmarking Score: 0.8
--------------------
https://pubchem.ncbi.nlm.nih.gov/rest/pug/assay/aid/1573390/description/JSON
- Success on output key: intermediate_steps, matched a desired output
- Agent action path:
get_cid_from_name: {"inp": "PZM21"}
get_assay_results_for_compund: {"cid": "121596705", "activity_name": "Ki"}
get_assay_description: {"aid": "1573390"}

- Iteration 16: Success
- Success: 13, Fail: 3, Total: 16
- Benchmarking Score: 0.8125
--------------------
https://pubchem.ncbi.nlm.nih.gov/rest/pug/assay/aid/1573390/description/JSON
- Success on output key: intermediate_steps, matched a desired output
- Agent action path:
get_cid_from_name: {"inp": "PZM21"}
get_assay_results_for_compund: {"cid": "121596705", "activity_name": "Ki"}
get_assay_description: {"aid": "1573390"}

- Iteration 17: Success
- Success: 14, Fail: 3, Total: 17
- Benchmarking Score: 0.823529417647058
--------------------
https://pubchem.ncbi.nlm.nih.gov/rest/pug/assay/aid/1573390/description/JSON
- Success on output key: intermediate_steps, matched a desired output
- Agent action path:
get_cid_from_name: {"inp": "PZM21"}
get_assay_results_for_compund: {"cid": "121596705", "activity_name": "Ki"}
get_assay_description: {"aid": "1573390"}

- Iteration 18: Success
- Success: 15, Fail: 3, Total: 18
- Benchmarking Score: 0.8333333333333334
--------------------
https://pubchem.ncbi.nlm.nih.gov/rest/pug/assay/aid/1573390/description/JSON
- Success on output key: intermediate_steps, matched a desired output
- Agent action path:
get_cid_from_name: {"inp": "PZM21"}
get_assay_results_for_compund: {"cid": "121596705", "activity_name": "Ki"}
```




```
get_assay_description: {"aid": "1573390"}

- Iteration 19: Success
- Success: 16, Fail: 3, Total: 19
- Benchmarking Score: 0.8421052631578947
--------------------
https://pubchem.ncbi.nlm.nih.gov/rest/pug/assay/aid/1573390/description/JSON
- Success on output key: intermediate_steps, matched a desired output
- Agent action path:
get_cid_from_name: {"inp": "PZM21"}
get_assay_results_for_compund: {"cid": "121596705", "activity_name": "Ki"}
get_assay_description: {"aid": "1573390"}

- Iteration 20: Success
- Success: 17, Fail: 3, Total: 20
- Benchmarking Score: 0.85
--------------------
```


**Supplementary Fig. 40 | Benchmarking log of a PubChem REST API agent which underwent path-based benchmarking on an assay retrieval task.** The agent was asked to find the Ki of PZM21 to the mu opioid receptor and to cite the assay which was the source of that information and provide a description of that assay. The agent correctly performed a sequence of actions which would retrieve this information 17 out 20 times. One of the failures added an additional step (name retrieval) which was not necessary as this information was contained in previous output. The other two failures were agentic errors where the LLM information returned to the agent for tool usage was not correct for the specified tool.